\documentclass[twocolumn,showpacs,superscriptaddress,preprintnumbers,amsmath,amssymb,longbibliography]{revtex4-1}
\usepackage[bookmarks=true,colorlinks=true,citecolor=blue,linkcolor=blue,urlcolor=magenta]{hyperref}

\usepackage{amssymb}% http://ctan.org/pkg/amssymb
\usepackage{fdsymbol}
\usepackage{latexsym}
\usepackage{gensymb} 

\usepackage{xfrac}
\usepackage{calc}
\usepackage{graphicx}
\usepackage[tight,footnotesize]{subfigure}
\usepackage{color}
\usepackage{longtable}%for tables that spans more than one page 
\usepackage{dashrule}%used for dashed lines in tables
\usepackage{dcolumn}% Align table columns on decimal point
\usepackage{multirow}%
\usepackage{rotating}%to turn words in the table cells

\usepackage{float}
\usepackage{perpage}
\usepackage{morefloats}%The package will allocate more “float skeletons” than LaTeX does by default

\DeclareFontFamily{OT1}{pzc}{}
\DeclareFontShape{OT1}{pzc}{m}{it}{<-> s * [0.900] pzcmi7t}{}
\DeclareMathAlphabet{\mathpzc}{OT1}{pzc}{m}{it}

\newcommand\scalemath[2]{\scalebox{#1}{\mbox{\ensuremath{\displaystyle #2}}}}
\begin{document}
%\preprint{AIP/DM}
\preprint{PRD}

%\title[]{Modification of the Cosmic Rays flux by the Super Heavy Dark Matter decay}
\title[]{Perturbative width of open rigid strings}

\author{A. S. Bakry}\email[]{ahmed.bakry@mail.com}
\affiliation{Institute of Modern Physics, Chinese Academy of Sciences, Gansu 730000, China }

\author{M. A. Deliyergiyev}\email[]{maksym.deliyergiyev@unige.ch}
\affiliation{Department of Nuclear and Particle Physics, University of Geneva, CH-1211 Geneva, Switzerland}
\affiliation{Institute of Physics, Jan Kochanowski University, 25-406 Kielce, Poland}

\author{A. A. Galal}
\affiliation{Department of Physics, Al Azhar University, Cairo 11651, Egypt}
\author{M. N.\ Khalil}
\affiliation{Department of Mathematics, Bergische Universit\"at Wuppertal, D-42097 Wuppertal, Germany}
\affiliation{Department of Physics, University of Ferrara, Ferrara 44121, Italy}
\affiliation{Computation-based Science Research Center, Cyprus Institute, Nicosia 2121, Cyprus}

\date{\today}

\begin{abstract}
We present a perturbative calculation of the energy profile width of rigid strings up to two loops in D dimensions. The perturbative expansion of the extrinsic-curvature term, signifying the rigidity/smoothness of the string in Polyakov-Kleinert action, is taken around the free Nambu-Goto string. The mean-square width of the string field is derived for open strings with the Dirichlet boundary condition. We compare the broadening of the smooth Polyakov-Kleinert string to the lattice Monte Carlo data of the QCD flux tube just before the deconfinement point and find a good match at the intermediate and large color source separation.	
\end{abstract}

%\pacs{12.38.Gc, 12.38.Lg, 12.38.Aw}%PACS, the Physics and Astronomy %Classification Scheme.
%Use showkeys class option if keyword%display desired
\maketitle

\section{Introduction}
\label{sec:intro}

The confinement of quarks is an outstanding feature of quantum chromodynamics (QCD). An analytic proof of color confinement in any non-Abelian gauge theory remains a far-reaching goal. Nevertheless the confinement property can be probed numerically in the simulations of the bound state of infinitely heavy quark-antiquark ($Q\bar{Q}$) which have shown the linearly rising property of the static potential~\cite{1,2,3,4,5,6,7,8,9}.

The linear increase in the $Q\bar{Q}$ potential is compatible with the creation of a stringlike object that connects the color sources. In a string model~\cite{Luscherfr,Luscher:2002qv} of confinement, the chromoelectric flux tubes are squeezed into a narrow region forming a stringlike tube by the dual Meissner effect~\cite{tHooft,'tHooft:1999au,Mandelstam76,Bali1996,Carmona:2001ja,DiGiacomo:1999a,DiGiacomo:1999b}. String formation is observed in many strongly correlated systems ~\cite{PhysRevB.78.024510,2007arXiv0709.1042K,Nielsen197345,Lo:2005xt} and is often characterized by a roughening phase in which oscillations transverse to the classical world sheet result in universal effects. These effects are detectable in numerical simulations of lattice gauge theories (LGTs)~\cite{Juge:2002br,HariDass2008273,Caselle:2016mqu,caselle-2002,Pennanen:1997qm,Brandt:2016xsp}. The L\"uscher term is the first universal correction to the linearly rising $Q\bar{Q}$ potential, and its analog in the baryonic three-quark system is the L\"uscher-like term~\cite{Jahn2004700,deForcrand:2005vv}.

The confining string is an idealized one-dimension bosonic string providing a low-energy effective description of the confining flux-tube~\cite{Luscherfr,Luscher:2002qv,PhysRevLett.67.1681} that is valid at distance scales larger than the width of the tube~\cite{Caselle:2012rp}. The stringlike object breaks the rotational and translational symmetry of the Yang-Mills vacuum, leading to the generation of spontaneous Goldstone modes~\cite{GODDARD1973109,Low:2001bw}. 
Study of the collective modes profile of the free Nambu-Goto (NG) string has pointed out another universal feature in conjunction with the L\"uscher term of the string potential, which is the logarithmic broadening of the width of the energy profile with the increase of the string length~\cite{Luscher:1980iy}. The broadening of the free string is expected to turn into a linear pattern before deconfinement is reached from below ~\cite{allais,Gliozzi:2010zv,Caselle:2010zs,Gliozzi:2010jh} at large color separations.

The energy profile between the confined color charges and the characteristic broadening is a key source for delving into the physics of the confining force from the first principles. The energy-density profile is a more challenging quantity to measure on the lattice compared to the $Q\bar{Q}$ potential. The exponentially decaying field correlators involve an additional term of the field strength tensor~\cite{PhysRevD.82.094503,Bakry:2010sp,Bicudo:2017uyy,deForcrand:2005vv} that brings about substantial numerical effort in certain gauge groups.

The expansion of computational resources and the constant enhancement of resolution power have opened up a plethora of prospects to examine the energy profile features of confined bound states. 
Many lattice simulations of different gauge groups, for example, have unambiguously confirmed the predicted logarithmic broadening of confined strings and their linear behavior at finite temperatures and large distances ~\cite{Caselle:1995fh,Bonati:2011nt,Caselle:2021eir,HASENBUSCH1994124,Caselle:2006dv,Bringoltz:2008nd,Athenodorou:2008cj,Juge:2002br,HariDass:2006pq,Giudice:2006hw,Bonati:2021vbc,Luscher:2004ib,Pepe:2010na}.

Nevertheless, the analysis of the string's fine structure in the lattice numerical data for the broadening profile reveals substantial deviations~\cite{PhysRevD.82.094503,Bakry:2010sp,Bicudo:2017uyy} from the free string NG model in the intermediate length scale at high temperatures. The excited spectrum~\cite{Luscher:2004ib} and the finite temperature string tension ~\cite{Pisarski,PhysRevD.85.077501,Kac} obtained from the Gaussian partition function of the free NG action shows a similar disagreement with the numerical data ~\cite{Juge:2002br,Kac,Bakry:2010sp,Bakry:2017utr} for large color separation.

Higher-order corrections to the NG action ~\cite{PhysRevD.27.2944,Aharony:2010cx,Billo:2012da} have been extensively investigated numerically to establish their relevance in each gauge model~\cite{Caselle:2004jq,Caselle:2004er,Pepe:2010na,Gliozzi:2010zv}. Indeed, there is no evidence that all orders of power expansion are universal~\cite{Alvarez:1981kc,Arvis:1983fp}.

Different model-dependent effective strings~\cite{Caselle:2004jq} may be identified in the numerical simulations of peculiar gauge groups. For instance, neither Abelian $Z(2)$ nor non-Abelian $SU(2)$ and $SU(3)$ confining gauge models have any universal properties in common, other than the L\"uscher term, at the next-to-leading-order operator~\cite{Caselle:2004jq}. The 3D percolation model ~\cite{Giudice:2009di} has likewise a term that deviates from the universal behavior. At next-to-leading order (NLO), deviations from the NG string are detected in the simulations of non-Abelian $SU(3)$ Yang-Mills theory in 4D ~\cite{Bakry:2017utr,Bakry:2018kpn,Bakry:2020ebo,Bakry:2020flt}.

Along with the boundary corrections~\cite{Billo:2012da,Bakry:2020ebo,Brandt:2010bw,Bakry} and the global geometrical characteristics of the flux tubes~\cite{Bakry:2020akg,Dubovsky}, to gain insight into structure of strings ~\cite{Ambjorn:2014rwa}, one is tempted to investigate likewise the putative physical implications of the local geometries~\cite{POLYAKOV19,Kleinert:1986bk} in the string action ~\cite{Caselle:2014eka,Brandt:2017yzw,Brandt:2021kvt}.

Simulations of the ground state of the Y-shaped flux tubes~\cite{Okiharu:2003vt,Bissey:2006bz,Okiharu:2004ve,Okiharu:2004wy,Bicudo:2011hk} point to absence of acute or sharp angles among strings. The ground state of three-quark potential is minimized while tuning (in Wilson loop overlap formalism) a system of spatial Y-shaped gauge links with equal angles of $120^{\circ}$ at the junction, as shown by Bissey et al.~\cite{Bissey:2006bz}. This obviously follows from a tension that prefers $120^{\circ}$ angles at the junctions to balance the three forces along the string (even at high temperature~\cite{Bakry:2011kga, Bakry:2014gea, Bakry:2016uwt}). On the other-hand, L and T-shaped strings decouple by the evolution of Euclidean time~\cite{Bissey:2006bz} indicating that other angles among strings correspond to higher excited states. These geometrical observations highlight the interconnection between energy of the string's configurations and the steepness of both the fluctuations and self-intersection along one single string.

The interpretation of the string's stiffness may be relevant to the string's vortex line picture ~\cite{Langfeld:2002vy}, or other self-interactions of the string ~\cite{PhysRevD.51.1842}. The consideration of geometric-dependent quantities in the string action could show some pertinence to the characteristics of the confining flux tubes, in particular at intermediate distances and high temperatures or for the excited states.

To realize the rigidity property mathematically the string action is considered using a Poincare-invariant term proportional to the extrinsic-curvature of the string's world sheet. Some years ago, Polyakov~\cite{POLYAKOV19} and Kleinert~\cite{Kleinert:1986bk}  proposed this intriguing generalization to stabilize the NG string action. In this model, the area, as well as the geometrical layout of the embedded sheet in space-time, affect the surface representation of the Polyakov-Kleinert (PK) string. That is, following the NG action, the bosonic free-string action is equipped with an additional term of extrinsic-curvature as a next-order operator~\cite{POLYAKOV19,Kleinert:1986bk}.

The PK string theory, like the ordinary NG string, yields a QCD model of smooth flux sheets across long-distance scales, but with abruptly creased surfaces filtered out. This is handled by the extrinsic-curvature term's leading parameter, which promotes smooth string configurations over those that are sharply curved. 

The model is relevant to QCD main features with ultraviolet (UV) freedom and infrared (IR) confinement properties ~\cite{POLYAKOV19,Kleinert:1986bk} and is consistent with glueball formation ~\cite{Kleinert:1988hz,Viswanathan:1987et}, in addition to a real $Q\bar{Q}$ potential~\cite{German:1989vk,Nesterenko1992,Ambjorn:2014rwa} and a possible tachyonic-free spectrum above some critical coupling~\cite{Viswanathan:1987et}.

The adiabatic and nonrelativistic limit of the Isgur-Paton model~\cite{Isgur}, inspired from the Hamiltonian LGT formulation in the strong coupling, corresponds to glueballs conjectured as rings of glue held together by a string tension equal to the $Q\bar{Q}$ potential, with ``rapid'' transverse oscillations and ``slow'' radial excitation. The stiff glueball model, on the other hand, hypothesizes glueballs as a superposition of flux rings generated by a closed chain of gluons with certain momenta~\cite{Caselle:2013qpa,Johnson:2000qz,Dubovsky:2016cog}. The partition function is that of closed smooth string on genus-one Riemann surfaces in Euclidean space, where the strings have extrinsic-curvature and spontaneously generated tension capable of forming glueballs. The PK model's parameters have a physical origin in QCD as a consequence of the gluonic degrees of freedom's complex dynamics. Even in the absence of quarks, QCD enables the development of a flux tube with a single length scale entering both the string tension and the thickness of the tube~\cite{Kleinert:1988hz,Viswanathan:1987et}.

With these attractive features, the theory induced a research effort to reveal the relevant properties such as its perturbative stability in critical dimensions~\cite{PhysRevLett.58.1300}, the dynamical generation of the string tension~\cite{Kleinert:1988vq}, and the exact potential in the large dimension limit~\cite{Braaten:1987gq,Kleinert:1989re}. This is in addition to the relevant thermodynamic characteristics such as the deconfinement transition point~\cite{Kleinert:1987kv}, partition function~\cite{Elizalde:1993af}, free energy and string tension at finite temperature~\cite{Viswanathan:1988ad,German:1991tc,Nesterenko:1997ku}.

Recently, rigid-string effects have been reported in the numerical simulations of the confining potential Abelian $U(1)$ compact gauge group~\cite{Caselle:2014eka,Caselle:2016mqu}. The rigidity effects in the confining potential are significant in reducing the deviations from the NG string for temperatures just before deconfinement in the quenched QCD ~\cite{Bakry:2017utr,Bakry:2018kpn,Bakry:2020ebo,Bakry:2020flt}. Manifestation of stiffness in the IR region of non-Abelian $SU(N)$ gauge theories in 3D has been reported in Ref.~\cite{Brandt:2017yzw}.

For many QCD processes tractable to stringy portrayals, disclosing the characteristics of the flux tube model beyond the Nambu-Goto arrangement could be demanding. This is especially true for high-energy processes~\cite{Caselle:2015tza,GIDDINGS198955} like the Hagedorn spectrum and glueballs thermodynamics~\cite{Caselle:2013qpa,Johnson:2000qz}, string fireballs~\cite{Kalaydzhyan:2014tfa} and mesonic spectroscopy~\cite{Bali:2013fpo,Kalashnikova:2002zz,Grach:2008ij,GIDDINGS198955}.

The energy chart features of the smooth flux sheets remained ambiguous despite substantial discussion of stiffness at the level of static quark potential~\cite{Kleinert:1988hz,Viswanathan:1987et,German:1989vk,Nesterenko1992,Ambjorn:2014rwa,Kleinert:1988vq,Braaten:1987gq,Kleinert:1989re,German:1991tc,Nesterenko:1997ku}. This, together with the resurgent interest in examining smooth strings on the lattice~\cite{Caselle:2014eka,Caselle:2016mqu,Brandt:2017yzw}, calls for extending the discussion of the string's rigidity effects on the characteristics of the energy field within hadronic bound states.

The goal of this study is to lay out an analytical form for the smooth string mean-square width that is practically viable for numerical models. The calculations described here are for an open string on a cylinder with the Dirichlet boundary condition, which is compatible with energy fields in static mesonic setups. In light of recent lattice studies~\cite{Caselle:2014eka,Caselle:2016mqu,Brandt:2017yzw} that show the delicate manifestation of smooth strings in the IR region of Abelian and non-Abelian gauge models, the mean-square width corrections due to the geometric term in PK action are perturbatively evaluated around the free NG solution.

In Sec.\ref{sec:WidthRigidStringLO} and \ref{sec:IntTerm_contribution}, we consider the perturbative expansion of the extrinsic-curvature term and evaluate the width of the leading- and next-to-leading order approximations expanded beyond the free NG string width, respectively. In Sec.\ref{sec:NumericalDiscussion} we discuss the behavior of the solution with the corresponding fits to lattice Monte Carlo data.
We provide the concluding remarks in Sec.\ref{sec:Conclusions}.

%%%%%%%%%%%%%%%%%%%%%%%%%%%%%%%%%%%%%%%%%%%%%%%%%%%%%%%%%    
%%%%%%%%%%%%%%%%%%%%%%%%%%%%%%%%%%%%%%%%%%%%%%%%%%%%%%%%%  
\section{Width of Rigid String in Leading-order Approximation}
\label{sec:WidthRigidStringLO}

Several physical IR properties of the string's world sheet are in agreement with what is expected for a QCD flux tube at low temperatures. In the IR limit, the dynamics of the flux tube follow toward a massless and free-string theory; for large enough color source separation, an effective field theory with infrared action can be prescribed as 
\begin{equation}
	\begin{split}
		S[\mathbf{X}]=&S_{\rm{cl}}+S_{0}[\mathbf{X}]\\
		=&\sigma R~L_{T}+\dfrac{1}{2} \int d\zeta_{0} \int d\zeta \left(\dfrac{\partial \mathbf{X}}{\partial \zeta_{\alpha}} \cdot \dfrac{\partial \mathbf{X}}{\partial \zeta_{\alpha}}\right),
		\label{LOaction}
	\end{split}        
\end{equation}
where $S_{cl}$ is the classical configuration or perimeter-area term, the coordinates $\zeta_{0}$ and $\zeta$ parameterize the world sheet ($\alpha=0,1$), 
the vector $X^{\mu}(\zeta_{0},\zeta)$ in the transverse gauge describes the fluctuations of the two-dimensional bosonic world sheet relative to the surface of minimal-area/the classical configuration with $(\mu=1,2,...,d-2)$ and the action $S_{0}[\mathbf{X}]$ is known as the Gaussian term. The above action $S[\mathbf{X}]$, Eq.(\ref{LOaction}), is referred to as the free bosonic string action.

The free bosonic string theory can be obtained from the truncated perturbative expansion of the NG string action. The NG string action is the simplest possible action that is proportional to the world sheet area
\begin{equation}
	S^{\rm{NG}}=\frac{1}{2} \int d^{2}\zeta \sqrt{g} g^{\alpha\beta} \frac{\partial \mathbf{X}}{\partial \zeta_{\alpha}} \cdot \frac{\partial \mathbf{X}}{\partial \zeta_{\beta}},           
\end{equation}
where $g^{\alpha\beta}$ is the intrinsic metric induced on the two-dimensional world sheet, parametrized by $X_{\mu}(\zeta)$ embedded in the background $R^{D}$.
\begin{equation}
	g^{\alpha\beta}=\frac{\partial \mathbf{X}}{\partial \zeta_{\alpha}} \cdot \frac{\partial \mathbf{X}}{\partial \zeta_{\beta}},~~~~
	(\alpha,\beta=1,2,....,D-2),~~~~g={\rm{det}}(g^{\alpha\beta}).
\end{equation}

The Dirichlet boundary condition, corresponding to fixed displacement vector at the ends $\zeta=0$ and $\zeta=R$, and periodic boundary condition in time $\zeta_{0}$ with period $L_T=1/T$, proportional to inverse temperature scale. For string tension $\sigma$ and classical area of an open-string world sheet given by $A=L_{T} R$ the NG action $S^{\rm{NG}}$ can be expanded  ~\cite{Giudice:2009di} in a natural dimensionless parameter $\left(\frac{1}{\sigma A}\right)$ such that
%\begin{minipage}{1.45\linewidth}
%\centering	
\begin{equation}
	\begin{split}	
		S^{\rm{NG}}[\mathbf{X}] & =  S_{\rm{cl}}+S_{0}[\mathbf{X}]+S_{1}[\mathbf{X}]+...\\
		& = S_{\rm{cl}}+S_{0}[\mathbf{X}]\\		
		&\scalemath{0.85}{ +\dfrac{1}{\sigma A}  \int \int d\zeta_{0}  d\zeta \Bigg[ \left(\dfrac{\partial \mathbf{X}}{\partial \zeta_{\alpha}} \cdot \dfrac{\partial \mathbf{X}}{\partial \zeta_{\alpha}}\right)^2 +\left(\dfrac{\partial \mathbf{X}}{\partial \zeta_{\alpha}} \cdot \dfrac{\partial \mathbf{X}}{\partial \zeta_{\beta}}\right)^2 \Bigg]}\\	
		&+\mathcal{O}\bigg(\dfrac{1}{(\sigma A)^2}\bigg).
	\end{split}
	\label{NG_action}
\end{equation}
%\end{minipage}

The above expression is an effective low-energy expansion of the string action with truncation that holds in the long string limit. Switching from the physical units ~\cite{Luscher:2004ib} of the bosonic fields to dimensionless form $\frac{\mathbf{X}}{\sqrt{\sigma}} \rightarrow \mathbf{X}$ the leading-order (LO) and the NLO terms of NG action after field redefinition can be written as
\begin{equation}
	\begin{split}
		S^{\rm{NG}}_{\ell o}[\mathbf{X}] & =  \sigma R~L_{T}+\frac{\sigma}{2} \int d\zeta_{0} \int d\zeta \left(\frac{\partial \mathbf{X}}{\partial \zeta_{\alpha}} \cdot \frac{\partial \mathbf{X}}{\partial \zeta_{\alpha}}\right),\\
		S^{\rm{NG}}_{n \ell o}[\mathbf{X}]& = \sigma \int d\zeta_{0} \int d\zeta \left[\left(\frac{\partial \mathbf{X}}{\partial \zeta_{\alpha}} \cdot \frac{\partial \mathbf{X}}{\partial \zeta_{\alpha}}\right)^2 +\left(\frac{\partial \mathbf{X}}{\partial \zeta_{\alpha}} \cdot \frac{\partial \mathbf{X}}{\partial \zeta_{\beta}}\right)^2\right].
		\label{NG_LO_NLO_action}
	\end{split}	
\end{equation}
The action of the PK bosonic string with the extrinsic-curvature term reads as
\begin{align}
	S^{\rm{PK}}[\mathbf{X}] = & S^{\rm{NG}}[\mathbf{X}] + S^{\rm{Ex}},\\
	= &  S^{\rm{NG}}[\mathbf{X}]+ \alpha_{r}\int d^{2}\zeta \sqrt{g}\Delta \mathbf{X} \cdot \Delta \mathbf{X}, \nonumber
	\label{Stiff}    
\end{align}
with the parameter $\alpha_{r}$ referring to the rigidity factor and the $\triangle$ operator given by
\begin{equation}
	\triangle \equiv (1/\sqrt{g}) \partial_{\alpha} ( \sqrt{g} g^{\alpha \beta} \partial_{\beta} ).  
\end{equation}

The second  term proportional to the extrinsic-curvature circumvents the undesirable features of the NG action such as ghost, tachyons, and an imaginary static potential between quarks when the distance becomes smaller than some critical distance
\cite{POLYAKOV19,Kleinert:1986bk,Kleinert:1988hz,Viswanathan:1988ad,German:1989vk,Nesterenko1992,Ambjorn:2014rwa}.

A similar argument can be used to expand the terms proportional to the extrinsic-curvature 
\begin{align}  
	\label{Pert}	
	S^{\rm{Ext}}[\mathbf{X}]&= \left(S^{(1)}+S^{(2)}\right)+....  \\
	&\scalemath{0.75}{  =\alpha_{r} \int_{0}^{L_T} d\zeta_0 \int_{0}^{R} d\zeta\Bigg[\left(\frac{\partial^{2} \mathbf{X}} {\partial \zeta^{2}} \right)^2 + \left(\frac{\partial^{2} \mathbf{X}} {\partial \zeta_{0}^{2}}  \right)^2 } \nonumber\\
	&\scalemath{0.75}{ +\frac{\sigma}{8}\left( \frac{\partial \mathbf{X}} {\partial \zeta_{\alpha}} \right)^4 -\frac{\sigma}{4} \left( \frac{\partial \mathbf{X}} {\partial \zeta_{\alpha}} \cdot \frac{\partial \mathbf{X}} {\partial \zeta_{\beta}}  \right)^2 + 2 \alpha_{r} \left(\frac{\partial^{2} \mathbf{X}} {\partial \zeta_{\alpha}^2 }\right)^2 \left(\frac{\partial \mathbf{X}}{\partial \zeta_{\beta}}\right)^{2} }\nonumber\\
	&\scalemath{0.75}{ - \frac{\alpha_{r}}{2} \left(\frac{\partial \mathbf{X}}{\partial \zeta_{\alpha}} \cdot \frac{\partial^{2} \mathbf{X}} {\partial \zeta_{\beta}^{2}} \right)^2-\alpha_{r} \left( \frac{\partial \mathbf{X}}{\partial \zeta_{\alpha}} \cdot \frac{\partial \mathbf{X}}{\partial \zeta_{\beta}} \right) \left( \frac{\partial^{2} \mathbf{X}} {\partial \zeta_{\alpha} \partial \zeta_{\beta}} \cdot \frac{\partial^{2} \mathbf{X}}{\partial \zeta_{\gamma}^{2}} \right)+..\Bigg] }. \nonumber
\end{align}

The mean-squared width of the string is defined as the second moment of the field with respect to the center of mass of the string $X_0$ and is given by
\begin{equation}
	W^2(\zeta)=\frac{\int  DX( \mathbf{X}(\zeta,\zeta_0)- \mathbf{X_0} )^2 e^{-S^{\rm{PK}}[\mathbf{X}]}} {\int DX e^{-S^{\rm{PK}}[\mathbf{X}]}}. 
	\label{Width2}
\end{equation}
In the truncated form the PK string action reads  
\begin{equation}
	\label{PKaction}  
	S^{\rm{PK}}[\mathbf{X}] \approx S^{\rm{NG}}_{\ell o}[\mathbf{X}]+S^{\rm{Pert}}[\mathbf{X}],
\end{equation}
with the general perturbation series  
\begin{equation}
	\label{Pert1}  
	S^{\rm{Pert}}[\mathbf{X}]= S^{\rm{NG}}_{n\ell o}+...+\left(S^{(1)}+S^{(2)}\right)+....  
\end{equation}

In the following, we calculate the width of the rigid string Eq.\eqref{Width2}, considering the first term  in the perturbative expansion $S^{(1)}$ of the Polyakov extrinsic-curvature term  given by
\begin{equation}
	S^{(1)}= \alpha_{r} \int_{0}^{L_T} d\zeta_0 \int_{0}^{R} d\zeta\left[\left(\frac{\partial^{2} \mathbf{X}} {\partial \zeta^{2}} \right)^2 + \left(\frac{\partial^{2} \mathbf{X}} {\partial \zeta_{0}^{2}}  \right)^2 \right].
	\label{eq:S1_ext}
\end{equation}

In the next section, we evaluate the width at the NLO in the extrinsic-curvature term $S^{(2)}$ given in Eq.\eqref{Pert}. Expanding around the free-string action Eq.\eqref{LOaction}, the squared width of the string ~\cite{Gliozzi:2010zt} is
\begin{align}
	\label{TwoLoopExpansion}	
	W^2(\zeta) & = W_{\rm{NG_{(\ell o)}}}^2(\zeta)- \langle \mathbf{X}^{2}(\zeta,\zeta_{0}) S^{\rm{Pert}} \rangle_0  \\
	& +2 \theta \langle\partial_{\alpha} \mathbf{X}^{2}(\zeta,\zeta_{0})\rangle_0 +  \theta^2 \langle \partial_{\alpha}^{2} \mathbf{X}^{2}(\zeta,\zeta_{0})\rangle_0 \nonumber \\ 
	& -\frac{\theta^2}{L_{T} R} \int d\zeta_{0}~d\zeta~d\zeta^{\prime}_{0}~d\zeta^\prime  \langle  \partial_{\alpha}^{2} \mathbf{X}(\zeta,\zeta_{0})·\partial_{\alpha^\prime}^{2}  \mathbf{X}(\zeta,\zeta^{'}_{0})\rangle_0, \nonumber
\end{align}
where the vacuum expectation value $\langle ...\rangle_0$  is with respect to the free-string partition function, $\theta$ is an effective low-energy parameter, and $L_{T}$ is the length of compactified time direction for the cylindrical boundary condition.

Substituting Eq.\eqref{Pert} in the expansion Eq.\eqref{TwoLoopExpansion}, the mean-square width reads
\begin{equation}
	\label{33} 
	\scalemath{0.8}{ W^2 (R,L_{T})= W^2_{\rm{NG_{(\ell o)}}}(R,L_{T})+ W^2_{\rm{NG_{(n\ell o)}}}(R,L_{T})+ W^2_{\rm{Ext}}(R,L_{T})},
\end{equation}
with
\begin{equation}
	W^2_{\rm{NG_{(n\ell o)}}}(R,L_{T})=-\left\langle \mathbf{X}^{2}(\zeta,\zeta_{0})S^{\rm{NG}}_{n\ell o} \right\rangle,
	\label{eq:W2_NG_NLO} 
\end{equation}
and
\begin{align}
	W^2_{\rm{Ext}}(R,L_{T})& = -\left\langle \mathbf{X}^{2}(\zeta,\zeta_{0})(S^{(1)}+S^{(2)}) \right\rangle\\
	& = W^2_{\rm{Ext_{(\ell o)}}}(R,L_{T}) +W^2_{\rm{Ext_{(n\ell o)}}}(R,L_{T}).
	\label{34}        
\end{align}

Let us define the Green's function $G(\mathbf{\zeta;\zeta'}) = \langle \mathbf{X}(\zeta,\zeta_{0}) \mathbf{X}(\zeta^{'},\zeta^{'}_{0})\rangle$ as the two-point Gaussian propagator. 
The correlator on a cylindrical sheet~\cite{Gliozzi:2010zt} of surface area $RL_{T}$  is given by
\begin{align}
	G(\mathbf{\zeta;\zeta'}) & \scalemath{0.9}{ =  \frac{1}{\pi  \sigma }\sum _{n=1}^{\infty } \frac{1}{n \left(1-q^n\right)}\sin \left(\frac{\pi  n \zeta}{R}\right) \sin \left(\frac{\pi  n \zeta'}{R}\right) } \nonumber \\
	&~~~\scalemath{0.9}{ \times	\left(q^n e^{\frac{\pi  n (\zeta_0-\zeta^{'}_{0})}{R}}+e^{-\frac{\pi n (\zeta_0-\zeta^{'}_{0})}{R}}\right) },
	\label{eq:GaussCorr}
\end{align}
where $q=e^{-\pi \frac{L_{T}}{R} }$ is the complementary nome. The Dirichlet boundary condition ~\cite{Gliozzi:2010zt} is encoded in the above propagator. 

After some algebraic manipulations and using the following auxiliary identities 
\begin{equation}
	\xi_{j} = 
	\begin{cases} 
		-i \frac{\pi}{2 R} (\zeta_{0}^{'}-\zeta_0+(-1)^{j} i \zeta - i \zeta'-L_T), & \mbox{for }~j=1,2\\ 
		-i \frac{\pi}{2 R} (\zeta_{0}^{'}-\zeta_0 +(-1)^{j}i \zeta +i \zeta'-L_{T}), & \mbox{for }~j=3,4\\
	\end{cases}
\end{equation}
the correlator Eq.(\ref{eq:GaussCorr}) can be recast into the following closed form:
\begin{equation}
	\label{Jacob}   
	G(\xi)=\frac{1}{4 \pi \sigma }\log\left[\frac{\vartheta_{4}(\xi _{1},q) \vartheta_{4}(\xi _{4},q)}{\vartheta_{4}(\xi _{2},q)\vartheta_{4}(\xi _{3},q)}\right],  
\end{equation}
where $\vartheta_4\left(\xi,q\right)$ is the Jacobi theta function.

The expectation value of the free NG string can be derived from the above form using Jacobi triple product formula. The mean-square width of the noninteracting string at the middle plane perpendicular to the string is 
\begin{align} 
	W^2_{NG_{(\ell o)}}(R/2)& = \lim_{\epsilon \rightarrow 0} G(\mathbf{\zeta;\zeta'}),\nonumber \\
	& =\frac{D-2}{4 \pi \sigma }\log\left[\frac{\vartheta_{3}(0,\sqrt{q})}{\vartheta_{4}(0,\sqrt{q})}\right]+R_{0}^{2},
	\label{NGLED}
\end{align}
with $R_{0}^{2}$ as the regularization scale, (see Appx.(\ref{Appendix_II}) for details). The above formula can be pinned down through the standard relations between the elliptic Jacobi and Dedekind $\eta$ functions to an equivalent to either of the forms derived in Refs.\cite{allais,Gliozzi:2010zt}.  

The first term in Eq.\eqref{34} representing the LO perturbation due to the smooth string in terms of the corresponding Green's functions is
\begin{align}
	\label{eq:Integral}	
	W^2_{\rm{Ext_{(\ell o)}}} & \scalemath{0.8}{ =-(D-2)\alpha_{r} \lim_{\substack{\epsilon' \to 0 \\ \epsilon \to 0}} \int_{0}^{R}d\zeta'\int_{0}^{L_T}d \zeta'_{0}  \Big[ G(\mathbf{\zeta;\zeta'})\partial_{\alpha}^{2} \partial_{\alpha'}^{2} G(\mathbf{\zeta';\zeta}) } \nonumber\\	
	&~~~~~~~~~~~~~~~~~~~~~~~~~~~~~~~~~~ \scalemath{0.8}{ + \partial_{\alpha}^{2} G(\mathbf{\zeta;\zeta'})  \partial_{\alpha'}^{2} G(\mathbf{\zeta';\zeta}) \Big] }.
\end{align}

The limit, $\epsilon, \epsilon' \to 0 $, is such that the integral is ultimately regularized by the point-split method. Therefore, one may recast Eq.\eqref{eq:Integral} as
\begin{equation}
	\scalemath{0.9}{ W_{\rm{Ext_{(\ell o)}}}^2= \frac{-\pi(D-2)\alpha_{r}}{4 R^2 \sigma ^2} \sum _{n=1}^{\infty }\sum _{m=1}^{\infty }  \left(q^m+1\right)^2  q^{-m}   \frac{ n \left(q^n+1\right) }{\left(q^n-1\right) } }.
	\label{eqsums}
\end{equation} 
With the use of the Riemann $\zeta$ function regularization of the divergent sums in Eq.\eqref{eqsums}, a generic form for the contribution of the extrinsic-curvature term to the mean-square width can be extracted as   
\begin{equation}
	\label{WExt}  
	W_{\rm{Ext_{(\ell o)}}}^2=\frac{-\pi(D-2)\alpha_{r}}{24 R^2 \sigma ^2 }  E_2 \left( \tau \right).
\end{equation}
Using the standard modular transform of an Eisenstein series \cite{Zagier2008,Freitag2009}
\begin{equation}
	E_2\left( \tau \right) = \frac{-4R^2}{L_{T}^{2}} E_2 \left( \frac{1} {\tau}\right)+\frac{6}{i \pi \tau}.
\end{equation}
The contribution of the rigidity term to the mean-square width of the string at the limit $R\gg L_{T}$ can be read off from Eq.\eqref{WExt}, 
\begin{equation}
	W_{\rm{Ext_{(\ell o)}}}^2 =-\left(\frac{ (D-2)\alpha_{r}} {2 \sigma^2 R L_{T}} \right)+C(L_{T}),
	\label{WExtlT}  
\end{equation}
with $C(L_{T})$ being only a function in the thermodynamic scale, see Appx.(\ref{Appendix_III}) for details.

%%%%%%%%%%%%%%%%%%%%%%%%%%%%%%%%%%%%%%%%%%%%%%%%%%%%%%%%%%%%%
%%%%%%%%%%%%%%%%%%%%%%%%%%%%%%%%%%%%%%%%%%%%%%%%%%%%%%%%%%%%%
\section{The Interaction-term Contribution to the Width}
\label{sec:IntTerm_contribution}

In the following, we evaluate the NLO term in the perturbation expansion~\cite{German:1989vk} of the extrinsic-curvature of rigid string $S^{(2)}$ of the PK action Eq.\eqref{Pert}, which is given by
\begin{eqnarray}
	S^{(2)} & = \int_{0}^{L_T} d\zeta_0 \int_{0}^{R} d\zeta \Bigg[  \frac{\sigma}{8}\left( \frac{\partial \mathbf{X}} {\partial \zeta_{\alpha}} \right)^4 -\frac{\sigma}{4} \left( \frac{\partial \mathbf{X}} {\partial \zeta_{\alpha}} \cdot \frac{\partial \mathbf{X}} {\partial \zeta_{\beta}}  \right)^2 \label{eq:S_ext} \\
	&~~~~~~~~  + 2 \alpha_{r} \left(\frac{\partial^{2} \mathbf{X}} {\partial \zeta_{\alpha}^2 }\right)^2 \left(\frac{\partial \mathbf{X}}{\partial \zeta_{\beta}}\right)^{2}  - \frac{\alpha_{r}}{2} \left(\frac{\partial \mathbf{X}}{\partial \zeta_{\alpha}} \cdot \frac{\partial^{2} \mathbf{X}} {\partial \zeta_{\beta}^{2}} \right)^2 \nonumber \\
	&~~~~~~~~ -\alpha_{r} \left( \frac{\partial \mathbf{X}}{\partial \zeta_{\alpha}} \cdot \frac{\partial \mathbf{X}}{\partial \zeta_{\beta}} \right) \left( \frac{\partial^{2} \mathbf{X}} {\partial \zeta_{\alpha} \partial \zeta_{\beta}} \cdot \frac{\partial^{2} \mathbf{X}}{\partial \zeta_{\gamma}^{2}} \right)\Bigg] \nonumber. 
\end{eqnarray}

\begin{figure}[t]
	\begin{center}			
		\includegraphics[scale=0.5]{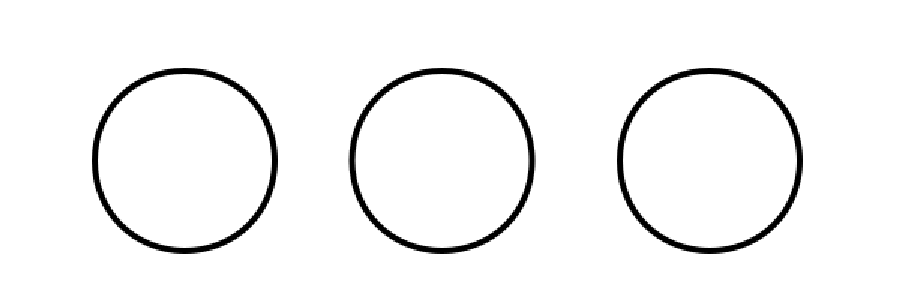}
		\caption{Feynman diagrams corresponding to disconnected loops~\cite{German:1989vk} of the first two terms of NLO expansion, Eq.\eqref{tbyt}.}
	\label{fig:FeynmanDiag_discLoops}						
	\end{center}
\end{figure}

Using notation from Ref.\cite{German:1989vk} one can write the contribution to a width of the stiff string due to second loop interactions in the following series:
\begin{align}
	W^{2}_{\rm{Ext_{(n\ell o)}}}&=\left\langle  \mathbf{X}^2 S^{(2)} \right\rangle \nonumber \\
	& = \left\langle  \mathbf{X}^2 \left( S^{(2)}_{(a)}+S^{(2)}_{(b)}+S^{(2)}_{(c)}+S^{(2)}_{(d)}+S^{(2)}_{(e)} \right) \right\rangle.
	\label{tbyt} 
\end{align}
Here each expectation value is written with respect to the free NG string and corresponds to the respective term in Eq.(\ref{eq:S_ext}). The first two interaction terms have the dimensional coupling $\sigma$ with the Feynman diagrams corresponding to disconnected loops shown in Fig.\ref{fig:FeynmanDiag_discLoops}.
%---------------------------------------------  
%--------------------------------------------- 

After Wick contraction~\cite{German:1989vk}, the first interaction term in Eq.\eqref{tbyt} reads
\begin{align}
	\scalemath{0.8}{ \left\langle \mathbf{X}^2 S^{(2)}_{(a)} \right\rangle_0}  & \scalemath{0.8}{ =\frac{\sigma} {8} \left\langle \mathbf{X}\cdot \mathbf{X} \left(\frac{\partial \mathbf{X}}{ \partial \zeta_{\alpha}} \right)^{4} \right\rangle_0 } \\
	&\scalemath{0.8}{ = \frac{\sigma} {8}\Bigg[(D-2)^2 \Bigg( \left\langle \mathbf{X} \frac{\partial \mathbf{X}}{\partial \zeta_{\alpha}} \right\rangle_{0} \left\langle \mathbf{X} \frac{\partial\mathbf{X}}{\partial \zeta_{\alpha}} \right\rangle_{0} \left\langle \frac{\partial \mathbf{X}}{\partial \zeta_{\beta}} \frac{\partial \mathbf{X}}{\partial \zeta_{\beta}}  \right\rangle_{0} }\nonumber \\
	&~~~~~~~~~~~~~~\scalemath{0.8}{ + \left\langle \mathbf{X}^2 \right\rangle \left\langle \frac{\partial \mathbf{X}}{\partial \zeta_{\beta}} \frac{\partial \mathbf{X}}{\partial \zeta_{\beta}} \right\rangle \left\langle \frac{\partial \mathbf{X}}{\partial \zeta_{\alpha}} \frac{\partial \mathbf{X}}{\partial \zeta_{\alpha}}\right\rangle \Bigg) } \nonumber \\ 
	&\scalemath{0.8}{ + (2D-2) \Bigg(\left\langle \mathbf{X} \frac{\partial\mathbf{X}}{\partial \zeta_{\alpha}} \right\rangle_{0} \left\langle \mathbf{X} \frac{\partial \mathbf{X}}{\partial\zeta_{\beta}} \right\rangle_{0} \left\langle \frac{\partial \mathbf{X}}{\partial\zeta_{\alpha}}  \frac{\partial \mathbf{X}}{\partial \zeta_{\beta}} \right\rangle_{0} } \nonumber \\ 
	&~~~~~~~~~~~~~\scalemath{0.8}{ + \left\langle \mathbf{X}^2 \right\rangle_{0} \left\langle \frac{\partial \mathbf{X}}{\partial \zeta_{\alpha}} \frac{\partial\mathbf{X}}{\partial \zeta_{\beta}} \right\rangle_{0}  \left\langle \frac{\partial\mathbf{X}}{\partial\zeta_{\alpha}} \frac{\partial\mathbf{X}}{\partial \zeta_{\beta}} \right\rangle_{0} \Bigg) \Bigg] } \nonumber.
\end{align}
The above correlator upon point split becomes an integral over the sum of the triple product of derivatives of Green's functions 
\begin{align}
	\label{AG}	
	&\scalemath{0.75}{ \left\langle \mathbf{X}^2 S^{(2)}_{(a)} \right\rangle_{0} } \\
	&\scalemath{0.75}{  = \lim_{\substack{\epsilon' \to 0 \\  \epsilon \to 0}} \int_{0}^{R} \int_{0}^{L_{T}}  \frac{\sigma}{8} \Bigg[ (D-2)^2  \bigg( \partial _{\alpha '} G\left(\mathbf{\zeta;\zeta'} \right)\partial _{\beta '}\partial _{\beta ''}G\left(\mathbf{\zeta';\zeta''}\right)\partial _{\alpha ''}G\left(\mathbf{\zeta'';\zeta} \right) } \nonumber \\ 
	&~~~~~~~~~~~~\scalemath{0.75}{ +G\left(\mathbf{\zeta;\zeta'}   \right)  \partial _{\alpha '} \partial _{\alpha ''}G\left(\mathbf{\zeta';\zeta''}\right) \partial _{\beta''}\partial _{\beta } G\left(\mathbf{\zeta'';\zeta}  \right) \bigg) } \nonumber \\
	&~~~~~~~~~~~~\scalemath{0.75}{ +(2D-2) \bigg( \partial _{\alpha '}G\left( \mathbf{\zeta;\zeta'} \right)\partial _{\alpha '}\partial _{\beta ''}G\left(\mathbf{\zeta';\zeta''}   \right) \partial _{\beta ''} G\left(\mathbf{\zeta'';\zeta} \right) } \nonumber \\
	&~~~~~~~~~~~~\scalemath{0.75}{ +G\left( \mathbf{\zeta;\zeta'} \right)\partial _{\alpha '}\partial _{\beta ''} G\left( \mathbf{\zeta';\zeta''} \right) \partial _{\alpha ''} \partial _{\beta }G\left( \mathbf{\zeta'';\zeta} \right) \bigg) \Bigg]  d\zeta d\zeta_{0}}.
\end{align}

The second Wick-contracted interaction term in Eq.\eqref{tbyt} reads
\begin{align}
	\label{eq:2ndWickContractedTerm}
	\left\langle \mathbf{X}^2 S_{(b)}\right\rangle_0 & \scalemath{0.75}{ = \frac{-\sigma}{4} \left\langle \mathbf{X}\cdot \mathbf{X}  \left(\frac{\partial \mathbf{X}}{\partial \zeta_{\alpha}}\cdot \frac{\partial \mathbf{X}}{\partial \zeta_{\beta}} \right)^2 \right\rangle_0 } \\
	&\scalemath{0.75}{ = \frac{-\sigma(D-2)}{4}\Bigg[\bigg(\left\langle \mathbf{X} \frac{\partial \mathbf{X}}{\partial \zeta_{\alpha}} \right\rangle_{0} \left\langle \mathbf{X} \frac{\partial \mathbf{X}}{\partial \zeta_{\alpha}} \right\rangle_{0} \left\langle \frac{\partial \mathbf{X}}{\partial \zeta_{\beta}}\frac{\partial \mathbf{X}}{\partial \zeta_{\beta}} \right\rangle_{0} } \nonumber \\
	&~~~~~~~~~~~~~~~~~ \scalemath{0.75}{ +\left\langle \mathbf{X}^{2} \right\rangle_{0} \left\langle \frac{\partial \mathbf{X}}{\partial \zeta_{\beta}}\frac{\partial \mathbf{X}}{\partial \zeta_{\beta}} \right\rangle_{0} \left\langle \frac{\partial \mathbf{X}}{\partial \zeta_{\alpha}}\frac{\partial\mathbf{X}}{\partial \zeta_{\alpha}} \right\rangle_{0}\bigg) }\nonumber \\
	&\scalemath{0.75}{  +(D-1) \bigg( \left\langle \mathbf{X}^{2} \right\rangle_{0} \left\langle \frac{\partial \mathbf{X}}{\partial \zeta_{\alpha}}\frac{\partial \mathbf{X}}{\partial \zeta_{\beta}}  \right\rangle_{0} \left\langle \frac{\partial \mathbf{X}}{\partial \zeta_{\alpha}}\frac{\partial \mathbf{X}}{\partial \zeta_{\beta}}  \right\rangle_{0} } \nonumber \\
	&~~~~~~~~~~~~~~~~~\scalemath{0.75}{ +\left\langle \frac{\partial \mathbf{X}}{\partial \zeta_{\alpha}}\frac{\partial X}{\partial \zeta_{\beta}} \right\rangle_{0} \left\langle \mathbf{X}  \frac{\partial\mathbf{X}}{\partial \zeta_{\alpha}} \right\rangle_{0}  \left\langle \mathbf{X} \frac{\partial \mathbf{X}}{\partial\zeta_{\beta}} \right\rangle_{0}\bigg)\Bigg] } \nonumber.
\end{align}
%---------------------------------------------  
\begin{figure}%[t]
	\begin{center}					
		\includegraphics[scale=0.5]{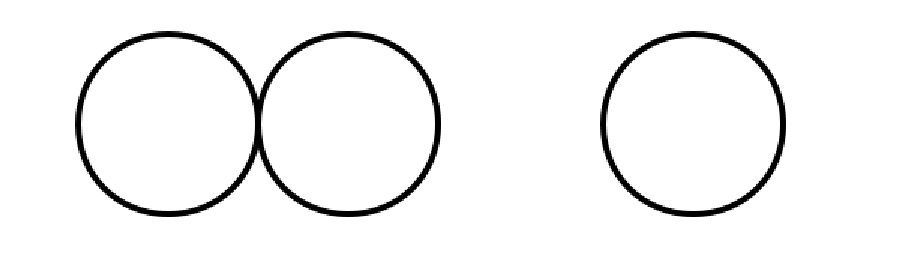}
		\caption{Connected and disconnected loops corresponding to the interaction terms $\left\langle \mathbf{X}^{2}S^{(2)}_{(c)}\right\rangle$, $\left\langle \mathbf{X}^{2}S^{(2)}_{(d)}\right\rangle$ and $\left\langle \mathbf{X}^{2}S^{(2)}_{(e)}\right\rangle$ in Eq.\eqref{tbyt} with dimensionless couplings $\alpha_{r}$.}
	\label{FeymannDiagConnDisc}			
	\end{center}			
\end{figure}
%---------------------------------------------  

The sum of the corresponding Green's correlators is given by
\begin{align}
	& \scalemath{0.8}{\left\langle \mathbf{X}^2 S^{(2)}_{(b)} \right\rangle_0 } \\
	& \scalemath{0.8}{= \lim_{\substack{\epsilon' \to 0\\ \epsilon \to 0} }
		\int_{0}^{R} \int_{0}^{L_{T}}  \frac{-\sigma}{4} (D-2)\Bigg[\bigg( \partial _{\alpha'} G\left(\mathbf{\zeta;\zeta'} \right)  \partial _{\beta'} \partial _{\beta''} G\left(\mathbf{\zeta';\zeta''} \right) \partial _{\alpha ''}G\left(\mathbf{\zeta'';\zeta} \right) }\nonumber \\
	&~~~~~~~~~~~~~~~~~\scalemath{0.85}{ + G\left(\mathbf{\zeta;\zeta'} \right)\partial _{\beta'}\partial _{\beta''} G\left(\mathbf{\zeta';\zeta''} \right) \partial _{\alpha''}\partial _{\alpha } G\left(\mathbf{\zeta'';\zeta} \right) \bigg) } \nonumber\\
	&~~~~\scalemath{0.85}{ +(D-1) \bigg( \partial _{\alpha '} G\left(\mathbf{\zeta;\zeta'}\right)\partial _{\alpha '}\partial _{\beta ''} G\left(\mathbf{\zeta';\zeta''} \right)\partial _{\beta ''} G\left(\mathbf{\zeta'';\zeta} \right) } \nonumber \\
	&~~~~~~~~~\scalemath{0.85}{ +G\left(\mathbf{\zeta;\zeta'}  \right) \partial _{\alpha '}\partial _{\beta ''} G\left( \mathbf{\zeta';\zeta''} \right) \partial _{\alpha ''}\partial _{\beta }G\left( \mathbf{\zeta'';\zeta} \right)\bigg)\Bigg] d\zeta d\zeta_0 }
	\label{BG}.
\end{align}
The above two correlators are given by the sum of the Green's triple product and the remaining integrals vanish (see Appx.(\ref{Appendix_IV}) for detailed calculations),
%---------------------------------------------  
\begin{align}
	&I^{(2)}_{1}(R,L_{T}) \label{eq:GreenString11_01}\\
	&\scalemath{0.7}{ =\lim_{\substack{\epsilon' \to 0\\ \epsilon \to 0}}
		\int _0^R\int _0^{L_{T}}  d\zeta d\zeta_{0} \Big[ \partial_{\alpha'} G\left(\zeta,\zeta_{0};\zeta',\zeta'_{0}\right)\partial_{\beta'}\partial_{\beta''} G\left(\zeta',\zeta'_{0};\zeta'',\zeta''_{0}\right)\partial_{\alpha''} G\left(\zeta'',\zeta''_{0};\zeta,\zeta_{0}\right) \Big] }\nonumber
\end{align}
and
\begin{align}
	&I^{(2)}_{2}(R,L_{T}) \label{eq:GreenString12_02}\\
	&\scalemath{0.7}{=\lim_{\substack{\epsilon' \to 0\\ \epsilon \to 0}}
		\int _0^R\int _0^{L_{T}}  d\zeta d\zeta_{0} \Big[\partial_{\alpha'}G\left(\zeta,\zeta_{0};\zeta',\zeta^{\prime}_{0}\right) \partial_{\alpha'} \partial_{\beta''} G\left(\zeta',\zeta^{\prime}_{0};\zeta'',\zeta_{0}''\right) \partial_{\beta''} G\left(\zeta'',\zeta''_{0};\zeta,\zeta_{0} \right) \Big]} \nonumber
\end{align}

%--------------------------------------------- 
\begin{figure*}[!ht]	
	\begin{center}
		\subfigure[]{\includegraphics[scale=0.37]{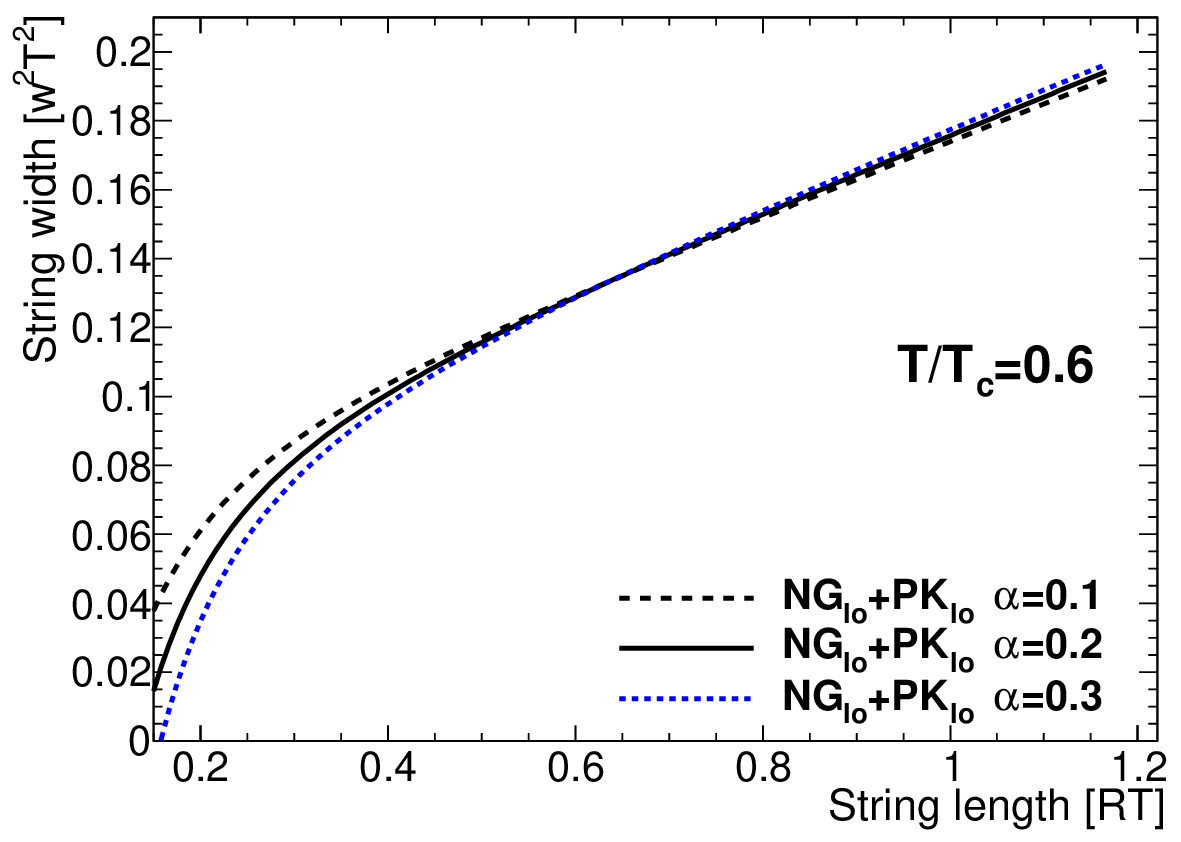}}
		\subfigure[]{\includegraphics[scale=0.37]{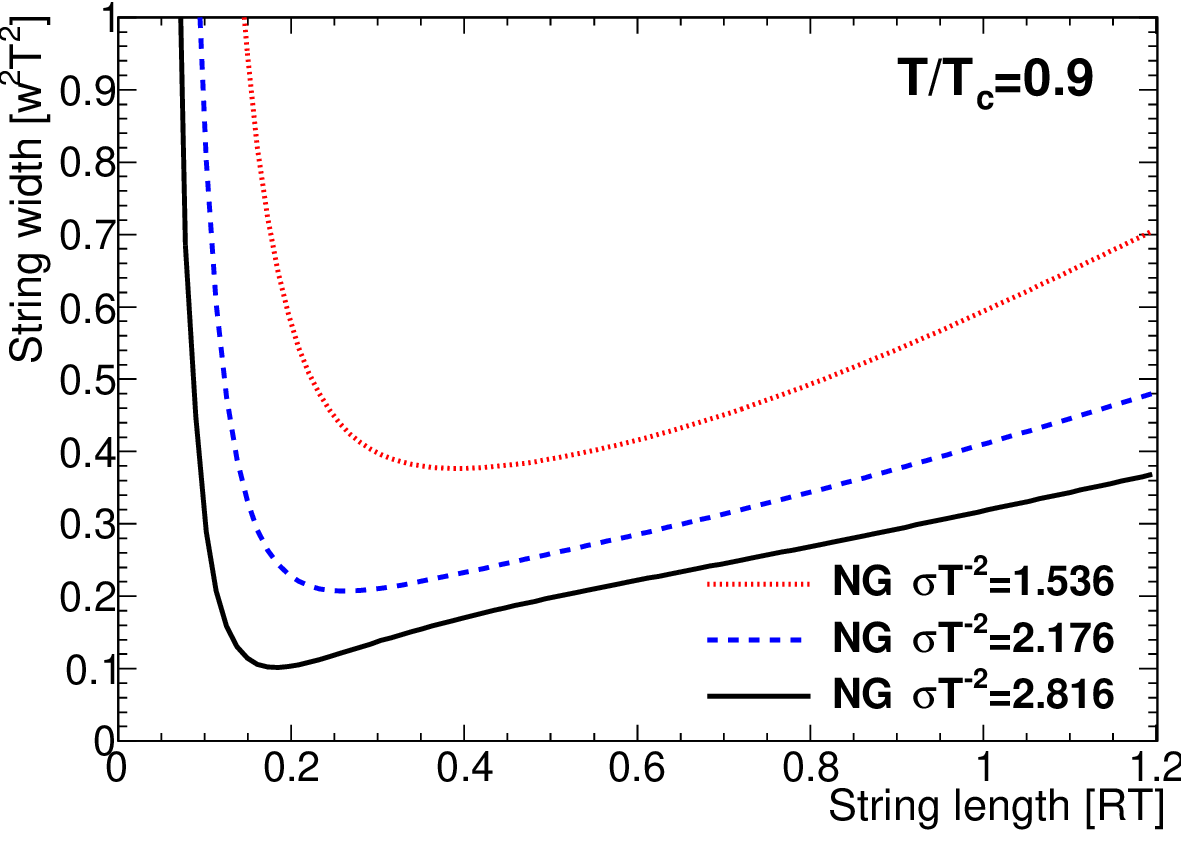}}
		\caption{
			The mean-square width of the string, $W^{2}(\frac{R}{2})$, at the middle plane versus the string length $R$. The temperature scales $T/T_{c}$ mimics a Yang-Mills theory at $\beta=6$.0~\cite{PhysRevD.85.077501} . (a) The plot compares  $W^{2}=W^2_{\rm{NG_{(\ell o)}}}$ of the free NG string Eq.(\ref{NGLED}) and the corresponding square width of PK string $W^2=W^2_{\rm{NG_{(\ell o)}}}+W^2_{\rm{NG_{(n\ell o)}}}+W^2_{\rm{Ext_{(\ell o)}}}$ given by Eq.\eqref{WExt} at one loop in extrinsic-curvature, at the standard value~\cite{Schilling93} of string tension $\sqrt{\sigma} =440$ MeV. (b) Two-loop mean-square width of pure NG string Eqs.\eqref{NGLED} and \eqref{NGNLO}, plotted versus the string length for each depicted value of $\sigma T^{-2}$.			
		}   
	\label{LOPKNG}			
	\end{center}		
\end{figure*}
%---------------------------------------------

The sum of the first two interaction terms with the dimensional weights yields the NLO term \eqref{eq:W2_NG_NLO} of the pure NG action  
\begin{equation}
	W^{2}_{NG_{(n\ell o)}}=  \left( (D-2)^2-(D-2)\right) I^{(2)}_{1}+ (D-2) I^{(2)}_{2}.
\end{equation}

The integrals of Eqs.\eqref{eq:GreenString11_01} and \eqref{eq:GreenString12_02} have been worked out in detail in Ref.~\cite{Gliozzi:2010zt}. The NLO correction to the width of the NG string then reads 
\begin{align}
	\label{NGNLO}	
	W^{2}_{NG_{(n\ell o)}}& \scalemath{0.7}{ =\frac{\pi}{12 \sigma R^2} \Big[E_2(\tau)-4E_2(2\tau)\Big] \left(W_{lo}^2-\frac{D-2}{4\pi \sigma}\right) } \\
	&\scalemath{0.7}{ +\frac{(D-2)\pi}{12\sigma^2 R^2}\Bigg\{\tau \Big(q \frac{d}{dq}-\frac{D-2}{12}E_2(\tau)\Big) \Big[ E_2(2\tau)-E_2(\tau) \Big]-\frac{D-2}{8 \pi} E_2(\tau)\Bigg\} } \nonumber,
\end{align}
with $\tau=\frac{i L_{T}}{2 R}$ as the modular parameter of the cylindrical sheet.

The next three interaction terms in Eq.\eqref{tbyt}, namely $\left\langle \mathbf{X}^{2}S^{(2)}_{(c)}\right\rangle$, $\left\langle \mathbf{X}^{2}S^{(2)}_{(d)}\right\rangle$ and $\left\langle \mathbf{X}^{2}S^{(2)}_{(e)}\right\rangle$  have dimensionless couplings $\alpha_{r}$ and correspond to the loops ~\cite{German:1989vk} shown in Fig.~\ref{FeymannDiagConnDisc}.

The contraction of the third interaction term in Eq.\eqref{tbyt} reads
\begin{align}
	\scalemath{0.9}{ \left\langle \mathbf{X}^2 S_{(c)} \right\rangle_0} & \scalemath{0.9}{ = 2 \alpha_{r} \left\langle (\mathbf{X}\cdot\mathbf{X}) \left(\frac{\partial^{2} \mathbf{X}}{\partial \zeta_{\alpha}^{2}} \right)^2 \left(\frac{\partial \mathbf{X}} {\partial \zeta_{\beta}} \right)^2\right\rangle_0 } \\
	&\scalemath{0.9}{=2 \alpha_{r}\Bigg[(D-2)^2 \Bigg(\left\langle \mathbf{X}^{2} \right\rangle_0 \left\langle \frac{\partial^{2} \mathbf{X}}{\partial \zeta_{\alpha}^{2}} \frac{\partial^2 \mathbf{X} }{\partial \zeta_{\beta}^2 } \right\rangle_0 \left\langle \frac{\partial^{2} \mathbf{X}}{\partial \zeta_{\gamma}^{2}} \right\rangle_0 }\nonumber \\
	&~~~~ \scalemath{0.9}{ + \left\langle \frac{\partial^{2}\mathbf{X}}{\partial \zeta_{\gamma}^{2}} \right\rangle_0 \left\langle \mathbf{X} \frac{\partial^{2}\mathbf{X}}{\partial \zeta_{\alpha}^{2}} \right\rangle_0 \left\langle \mathbf{X} \frac{\partial^{2}\mathbf{X}}{\partial \zeta_{\beta}^{2}} \right\rangle_0 } \nonumber \\
	&~~~~ \scalemath{0.9}{ +\left\langle \mathbf{X} \frac{\partial \mathbf{X}}{\partial \zeta_{\gamma}} \right\rangle_0 \left\langle \mathbf{X} \frac{\partial \mathbf{X}}{\partial \zeta_{\gamma}} \right\rangle_0 \left\langle \frac{\partial^{2} \mathbf{X}}{\partial \zeta_{\alpha}^{2}} \frac{\partial^2 \mathbf{X}}{\partial \zeta_{\beta}^2 } \right\rangle_0 \Bigg) } \nonumber \\
	&~~~~ \scalemath{0.9}{ +2 (D-2) \Bigg(\left\langle \mathbf{X}^2 \right\rangle_0 \left\langle \frac{\partial \mathbf{X}}{\partial \zeta_{\alpha}} \frac{\partial^2 \mathbf{X}}{\partial \zeta_{\gamma}^2} \right\rangle_0 \left\langle \frac{\partial \mathbf{X}}{\partial \zeta_{\alpha}} \frac{\partial^{2} \mathbf{X}}{\partial \zeta_{\gamma}^2} \right\rangle_0 } \nonumber \\
	&~~~~ \scalemath{0.9}{ +\left\langle \frac{\partial \mathbf{X}}{\partial \zeta_{\alpha}} \frac{\partial^2 \mathbf{X}}{\partial \zeta_{\gamma}^2} \right\rangle_0\left\langle \mathbf{X} \frac{\partial \mathbf{X}}{\partial \zeta_{\alpha}} \right\rangle_0 \left\langle \mathbf{X} \frac{\partial^{2} \mathbf{X}}{\partial \zeta_{\gamma}^{2}} \right\rangle_0 \Bigg)\Bigg] } \nonumber. 
\end{align}

The corresponding sum of the point-split Green's correlators reads
\begin{align}
	\label{CG}	
	\scalemath{0.7}{ \left\langle \mathbf{X}^2 S^{(2)}_{(c)} \right\rangle_0 } & \scalemath{0.7}{ = \lim_{\substack{\epsilon \to 0\\ \epsilon' \to 0} } \int_{0}^{R}\int_{0}^{L_{T}}2 \alpha_{r} \bigg[ (D-2)^2 \bigg(
		G\left(\mathbf{\zeta;\zeta'} \right) \partial^2_{\alpha'}\partial _{\beta''}^{2} G\left(\mathbf{\zeta';\zeta''} \right) \partial_{\alpha''} \partial _{\alpha } G\left( \mathbf{\zeta'';\zeta}\right)I_{6} } \nonumber\\  
	&~~~ \scalemath{0.8}{ +\partial _{\alpha '}^{2} G\left(\mathbf{\zeta;\zeta'} \right)\partial _{\alpha '}\partial _{\alpha ''}G\left(\mathbf{\zeta';\zeta''} \right)\partial _{\beta }^{2} G\left(\mathbf{\zeta'';\zeta} \right) I_{5}\bigg) } \nonumber\\ 
	&~~~ \scalemath{0.8}{  +\partial _{\beta } G\left( \mathbf{\zeta;\zeta'} \right) \partial_{\alpha'}^{2}\partial_{\alpha''}^{2}G\left(\mathbf{\zeta';\zeta''}\right) \partial _{\beta ''} G\left(\mathbf{\zeta'';\zeta}\right)I_{7} } \nonumber\\ 
	&~~~ \scalemath{0.8}{  +2(D-2) \bigg(\partial_{\alpha' } G\left( \mathbf{\zeta;\zeta'} \right) \partial_{\alpha'}\partial_{\beta''}^{2}G\left(\mathbf{\zeta';\zeta''}\right) \partial_{\beta}^{2} G\left(\mathbf{\zeta'';\zeta}\right)I_{10} } \nonumber\\ 
	&~~~ \scalemath{0.8}{  +G\left(\mathbf{\zeta;\zeta'} \right)\partial _{\alpha '}^{2}\partial _{\beta ''} G\left(\mathbf{\zeta';\zeta''} \right) \partial _{\alpha ''}^{2}\partial _{\beta } G\left(\mathbf{\zeta'';\zeta} \right)I_{8}\bigg)\bigg]d\zeta d\zeta_0 }.		
\end{align}
The symmetries of the diagonal and off-diagonal components can be elaborated from the equation of motion
\begin{eqnarray}
	&\partial_{\alpha} \partial_{\alpha} G\left(\mathbf{\zeta;\zeta'} \right) =-\partial_{\beta} \partial_{\beta} G\left(\mathbf{\zeta;\zeta'} \right),\label{EOM} \\
	&\partial_{\alpha} \partial_{\alpha'} \partial_{\alpha} \partial_{\alpha'} G\left(\mathbf{\zeta;\zeta'} \right) =\partial_{\beta} \partial_{\beta'} \partial_{\beta} \partial_{\beta'}  G\left(\mathbf{\zeta;\zeta'} \right). \nonumber
\end{eqnarray}
The only nonvanishing Green's integral is given by twice the product Eq.~\eqref{eq:GreenString_07}, 
\begin{equation}
	\left\langle \mathbf{X}^{2} S^{(2)}_{(c)} \right\rangle_0=2 \alpha_{r} \Bigg[(D-2)^2 \left(2 \,I^{(2)}_{7}\right)  \Bigg].  
\end{equation}
The expectation value of the correlator is explicitly 
\begin{align}
	\label{cExact}	
	\scalemath{0.85}{ \left\langle \mathbf{X}^2 S^{(2)}_{(c)} \right\rangle_0} & \scalemath{0.8}{ =\frac{\pi ^3 \alpha_r (D-2)^2}{480 R^5 \sigma ^3} (E_4(\tau )+1) } \\
	&~~~~\scalemath{0.85}{ \times \Bigg[\frac{4 R}{\pi} \left(\log \left( \eta \left(\frac{\tau}{2} \right)\right)+  \gamma \right) +\frac{L_{T}}{12} \left(2-E_{2}\left( \frac{\tau}{2} \right) \right)\Bigg]}. \nonumber
\end{align}

The fourth interaction term in Eq.\eqref{tbyt} can be Wick contracted as
\begin{align}
	&\scalemath{0.85}{ \left\langle \mathbf{X}^2  S_{(d)} \right\rangle_0 =-\frac{\alpha_{r}}{2} \left\langle \mathbf{X}\cdot\mathbf{X} \left(\frac{\partial \mathbf{X}}{\partial \zeta_{\alpha}} \cdot \frac{\partial^{2} \mathbf{X}} {\partial \zeta_{\beta}^{2}} \right)^2 \right\rangle_0 \label{d} } \\
	&\scalemath{0.85}{ = -\frac{\alpha_{r}}{2} \Bigg[(D-2) (D-1) \Bigg( \left\langle \mathbf{X} \frac{\partial \mathbf{X}}{\partial \zeta_{\alpha}} \right\rangle_0 \left\langle \mathbf{X} \frac{\partial^{2} \mathbf{X}}{\partial \zeta_{\gamma}^2} \right\rangle_0 \left\langle \frac{\partial \mathbf{X}}{\partial \zeta_{\alpha} }  \frac{\partial^{2}\mathbf{X}}{\partial \zeta_{\gamma}^{2}} \right\rangle_0 } \nonumber \\
	&~~~~~~~~~~~~~~~~~~~~~~~ \scalemath{0.85}{ + \left\langle \mathbf{X}^2 \right\rangle_0 \left\langle \frac{\partial\mathbf{X}}{\partial \zeta_{\alpha}}\frac{\partial^{2} \mathbf{X}}{\partial \zeta_{\gamma}^{2}}\right\rangle_0
		\left\langle \frac{\partial \mathbf{X}}{\partial \zeta_{\alpha}} \frac{\partial^{2} \mathbf{X}}{\partial \zeta_{\gamma}^{2}} \right\rangle_0 \Bigg) } \nonumber \\
	&~~~~~~~~~~ \scalemath{0.85}{ + (D-2) \Bigg(
		\left\langle \frac{\partial^{2} \mathbf{X}}{\partial \zeta_{\alpha}^{2}} \right\rangle_0 \left\langle \mathbf{X} \frac{\partial^{2}\mathbf{X}}{\partial\zeta_{\beta}^{2}} \right\rangle_0 \left\langle \mathbf{X} \frac{\partial^{2}\mathbf{X}}{\partial\zeta_{\gamma}^{2}} \right\rangle_0 } \nonumber \\
	&~~~~~~~~~~~~~~~~~~~~~~~~~	\scalemath{0.85}{ + \left\langle \mathbf{X}^{2} \right\rangle_0 \left\langle  \frac{\partial^{2} \mathbf{X}}{\partial \zeta_{\alpha}^{2}} \right\rangle_0 \left\langle \frac{\partial^2 \mathbf{X}}{\partial \zeta_{\beta}^2} \frac{\partial^{2} \mathbf{X}}{\partial \zeta_{\gamma}^{2} } \right\rangle_0 } \nonumber \\
	&~~~~~~~~~~~~~~~~~~~~~~~~~ \scalemath{0.85}{ +\left\langle \mathbf{X} \frac{\partial \mathbf{X}}{\partial \zeta_{\alpha}} \right\rangle_0 \left\langle \mathbf{X} \frac{\partial \mathbf{X}}{\partial \zeta_{\alpha}}  \right\rangle_0 \left\langle \frac{\partial^{2} \mathbf{X}}{\partial \zeta_{\beta}^{2} }  \frac{\partial^{2} \mathbf{X}}{\partial \zeta_{\gamma}^{2} }  \right\rangle_0 \Bigg)\Bigg] }.\nonumber 
\end{align}
The point-split form of the above correlator in terms of Green's functions reads 
\begin{align}
	\label{DG}	
	\scalemath{0.85}{ \left\langle \mathbf{X}^2 S^{(2)}_{(d)} \right\rangle_0 } & \scalemath{0.85}{= -\frac{\alpha_{r}}{2} \lim_{\substack{\epsilon' \to 0\\ \epsilon \to 0} }\int_{0}^{R} \int_{0}^{L_{T}}  \bigg[ (D-2) (D-1) }\nonumber \\		
	&~~~~~~~~~~ \scalemath{0.85}{	\times \bigg(\partial _{\alpha' }G\left(\mathbf{\zeta;\zeta'} \right) \partial_{\alpha'} \partial_{\beta''}^{2}  G\left( \mathbf{\zeta';\zeta''} \right) \partial _{\beta''}^{2} G\left(\mathbf{\zeta'';\zeta} \right) } \nonumber \\
	&~~~~~~~~~~~\scalemath{0.85}{ +G\left(\mathbf{\zeta;\zeta'} \right)  \partial_{\beta''}^{2} \partial_{\alpha } G\left(\mathbf{\zeta'';\zeta'}\right) \partial_{\beta'}^{2} \partial_{\alpha } G\left(\mathbf{\zeta';\zeta}  \right) \bigg) }\nonumber \\
	&~~~\scalemath{0.85}{ +(D-2)\bigg(\partial _{\alpha '}^{2} G\left(\mathbf{\zeta;\zeta'}\right) \partial_{\beta}^{2} G\left( \mathbf{\zeta'';\zeta} \right)\partial _{\alpha '}\partial _{\alpha ''}G\left( \mathbf{\zeta';\zeta''} \right) }\nonumber \\
	&~~~~~~~~~~~\scalemath{0.85}{  +G\left(\mathbf{\zeta;\zeta'} \right)\partial_{\alpha'} \partial_{\alpha ''} G\left(\mathbf{\zeta';\zeta''}\right) \partial_{\alpha ''}^{2} \partial_{\beta }^2 G\left(\mathbf{\zeta'';\zeta} \right) } \nonumber \\
	&~~~ \scalemath{0.85}{ +\partial _{\alpha '}G\left(\mathbf{\zeta;\zeta'}\right) \partial_{\beta'}^{2}\partial_{\beta''}^{2} G\left( \mathbf{\zeta';\zeta''} \right)  \partial_{\alpha ''}G\left(\mathbf{\zeta'';\zeta}  \right) \bigg) \bigg] d\zeta d\zeta^{0} }.  
\end{align}

The terms involving the sum over $\gamma$ are either related via the equation of motion \eqref{EOM} or identically vanish. The surviving terms in the above sum are again the integrals over the triple Green's product Eq.~\eqref{eq:GreenString_07} -Appx.(\ref{Appendix_IV}),
\begin{equation}
	\left\langle \mathbf{X}^{2} S^{(2)}_{(d)} \right\rangle_0 =\frac{-\alpha_{r}}{2} \Bigg[(D-2) \left(2\,I^{(2)}_{7} \right)  \Bigg].  
\end{equation}

The Wick contraction of the last interaction term in Eq.\eqref{tbyt} is 
\begin{align}
	\label{e}	
	\scalemath{0.85}{ \left\langle \mathbf{X}^2 S^{(2)}_{(e)} \right\rangle_0 } & \scalemath{0.85}{=-\alpha_{r} \left\langle \mathbf{X}\cdot\mathbf{X}\left( \frac{\partial \mathbf{X}}{\partial \zeta_{\alpha}}\cdot \frac{\partial \mathbf{X}}{\partial \zeta_{\beta}} \right) \left( \frac{\partial^{2} \mathbf{X}} {\partial \zeta_{\alpha} \partial \zeta_{\beta}} \cdot \frac{\partial^{2} \mathbf{X}}{\partial \zeta_{\gamma}^{2}} \right)\right\rangle_0  } \\
	&\scalemath{0.85}{ = -\alpha_{r}\bigg[(D-2)^2 \bigg( \left\langle \frac{\partial^{2} \mathbf{X}}{\partial \zeta_{\alpha}\partial \zeta_{\beta}} \right\rangle \left\langle \mathbf{X} \frac{\partial^{2} \mathbf{X}}{\partial \zeta_{\alpha} \partial \zeta_{\beta} }\right\rangle \left\langle \mathbf{X} \frac{\partial^{2} \mathbf{X}}{\partial \zeta_{\gamma}^{2} } \right\rangle_0 } \nonumber \\
	&~~~~~~~~~ \scalemath{0.85}{ + \left\langle \mathbf{X}^{2}  \right\rangle_0 \left\langle \frac{\partial^{2}\mathbf{X}}{\partial \zeta_{\alpha}\partial \zeta_{\beta}} \right\rangle_0 \left\langle \frac{\partial^{2}\mathbf{X}}{\partial \zeta_{\alpha}\zeta_{\beta}} \frac{\partial^{2}\mathbf{X}}{\partial \zeta_{\gamma}^{2}}\right\rangle_0 } \nonumber \\
	&~~~~~~~~~ \scalemath{0.85}{ + \left\langle \mathbf{X} \frac{\partial \mathbf{X}}{\partial \zeta_{\alpha}} \right\rangle_0 \left\langle \mathbf{X} \frac{\partial \mathbf{X}}{\partial \zeta_{\beta}} \right\rangle_0 \left\langle \frac{\partial^{2} \mathbf{X}}{\partial \zeta_{\alpha}\zeta_{\beta}} \frac{\partial^{2} \mathbf{X}}{\partial \zeta_{\gamma}^{2} } \right\rangle_0 \bigg) } \nonumber \\
	&\scalemath{0.85}{ +2(D-2)\bigg(\left\langle \frac{\partial^2 \mathbf{X}}{\partial\zeta_{\gamma}^2}\frac{\partial X}{\partial\zeta_{\beta}} \right\rangle_0 \left\langle \mathbf{X} \frac{\partial\mathbf{X}}{\partial \zeta_{\alpha}} \right\rangle_0 \left\langle \mathbf{X} \frac{\partial^{2}\mathbf{X}}{\partial \zeta_{\alpha}\zeta_{\beta}} \right\rangle } \nonumber \\
	&~~~~~~~~~ \scalemath{0.85}{ +\left\langle \mathbf{X}^{2} \right\rangle \left\langle \frac{\partial \mathbf{X}}{\partial \zeta_{\alpha}} \frac{\partial^{2}\mathbf{X}}{\partial \zeta_{\alpha}\zeta_{\beta}} \right\rangle_0 \left\langle \frac{\partial\mathbf{X}}{\partial \zeta_{\beta}} \frac{\partial^{2} \mathbf{X}}{\partial \zeta_{\gamma}^{2} } \right\rangle_0 } \nonumber \\
	&~~~~~~~~~ \scalemath{0.85}{ + \left\langle \mathbf{X} \frac{\partial \mathbf{X}}{\partial \zeta_{\beta}} \right\rangle_0 \left\langle \mathbf{X} \frac{\partial^{2} \mathbf{X}}{\partial \zeta_{\gamma}^2} \right\rangle \left\langle \frac{\partial \mathbf{X}}{\partial\zeta_{\alpha}}\frac{\partial^2 \mathbf{X}}{\partial\zeta_{\alpha}\zeta_{\beta}} \right\rangle_0 \bigg)\bigg] },\nonumber
\end{align}
for which the integral over the sum of the point-splitting Green's triple product is
\begin{align}
	\label{EG}	
	& \scalemath{0.75}{	\left\langle \mathbf{X}^2 S^{(2)}_{(e)} \right\rangle_0 } \\
	& \scalemath{0.75}{= -\alpha_{r}\lim_{\substack{\epsilon' \to 0\\ \epsilon \to 0} } \int_{0}^{R}\int_{0}^{L_{T}} \Bigg[ (D-2)^2 \bigg( \partial_{\alpha}\partial_{\beta'} G\left(\mathbf{\zeta;\zeta'} \right) \partial _{\alpha''}\partial _{\beta'' }G\left(\mathbf{\zeta';\zeta''} \right) \partial _{\beta}^{2} G\left( \mathbf{\zeta'';\zeta} \right) }  \nonumber\\
	&~~~~~~~~~~~~~~~~~~~~~ \scalemath{0.75}{ +G\left(\mathbf{\zeta;\zeta'} \right) \partial _{\alpha' }\partial _{\beta'' } G\left(\mathbf{\zeta';\zeta''} \right) \partial_{\alpha'' } \partial _{\beta''} \partial_{\alpha}^{2} G\left(\mathbf{\zeta'';\zeta} \right) } \nonumber\\
	&~~~~~~~~~~~~~~~~~~~~~ \scalemath{0.75}{ +\partial _{\alpha }G\left(\mathbf{\zeta;\zeta'} \right)\partial_{\alpha''} \partial_{\beta''} \partial_{\alpha'}^{2} G\left(\mathbf{\zeta';\zeta''}\right)\partial _{\beta ''}G\left(\mathbf{\zeta'';\zeta} \right) \bigg )  } \nonumber\\
	&~~ \scalemath{0.75}{ +2(D-2)\bigg( \partial _{\alpha } G\left(\mathbf{\zeta;\zeta'} \right) \partial _{\alpha ''}\partial _{\beta'' }G\left(\mathbf{\zeta';\zeta''} \right)\partial_{\beta''} \partial_{\alpha}^{2}  G\left(\mathbf{\zeta'';\zeta} \right) }\nonumber\\
	&~~~~~~~~~~~~~~~~ \scalemath{0.75}{ +G\left(\mathbf{\zeta;\zeta'} \right)\partial_{\alpha' }\partial _{\alpha'' }\partial _{\beta'' } G\left(\mathbf{\zeta';\zeta''} \right)\partial _{\beta' } \partial_{\alpha }^{2} G\left(\mathbf{\zeta'';\zeta} \right) } \nonumber\\
	&~~~~~~~~~~~~~~~~ \scalemath{0.75}{ +\partial_{\alpha'}^2 G\left(\mathbf{\zeta;\zeta'} \right)\partial_{\alpha' }\partial _{\alpha'' }\partial _{\beta'' } G\left(\mathbf{\zeta';\zeta''} \right)\partial _{\beta } G\left(\mathbf{\zeta'';\zeta} \right)  \bigg) \Bigg]  d\zeta d\zeta_0 }. 
\end{align}
The nontrivial terms are given by the integrals  
\begin{equation}
	\left\langle \mathbf{X}^{2} S^{(2)}_{(e)} \right\rangle_0 =-\alpha_{r} \Bigg[(D-2)^{2} I^{(2)}_{16} +2(D-2)I^{(2)}_{14}  \Bigg],
	\label{eExact}
\end{equation}
which are accordingly given by Eqs.~\eqref{eq:GreenString_14} and \eqref{eq:GreenString_16} of Appx.(\ref{Appendix_IV}). The resultant interaction term is
%---------------------------------------------  
\begin{align}
	\label{dExact} 	
	& \scalemath{0.8}{ \left\langle \mathbf{X}^{2} S^{(2)}_{(e)} \right\rangle_0 = \frac{\pi^2 \alpha_{r}(D-2)}{8 R^4 \sigma^3} -\frac{\pi ^3 \alpha_r (D-2)^2}{8 R^5  \sigma ^3} \Bigg(\frac{E_4(\tau )+1}{240}\Bigg)   } \\ 
	&~~~~~~~~~~~~~~~~~~~~~\scalemath{0.8}{ \times \Bigg[ \frac{4 R}{\pi} \left(\log \left( \eta \left(\frac{\tau}{2} \right)\right)+  \gamma \right) +\frac{L_{T}}{12} \left(2-E_{2}\left(\frac{\tau}{2} \right) \right) \Bigg] }. \nonumber    
\end{align}

\begin{figure*}[!ht]
	\begin{center}						
		\subfigure[]{\includegraphics[scale=0.37]{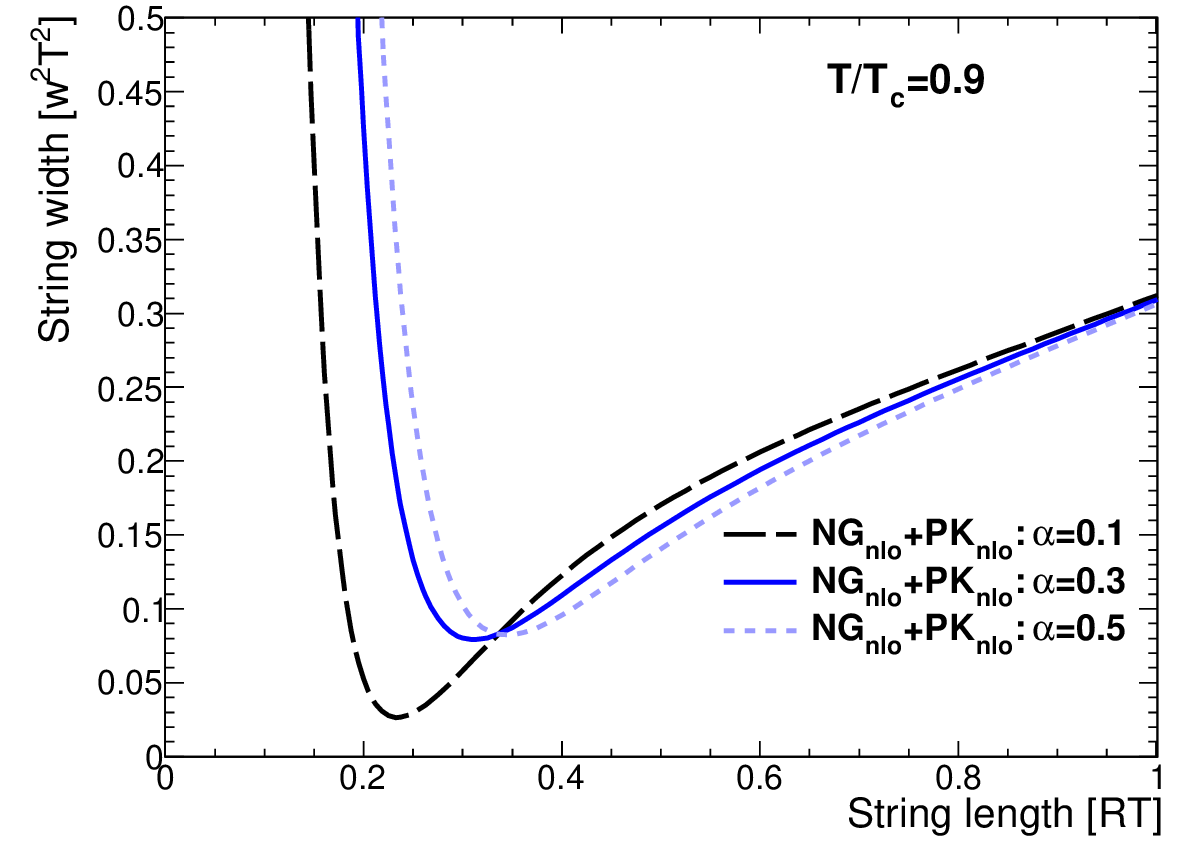}}
		\subfigure[]{\includegraphics[scale=0.37]{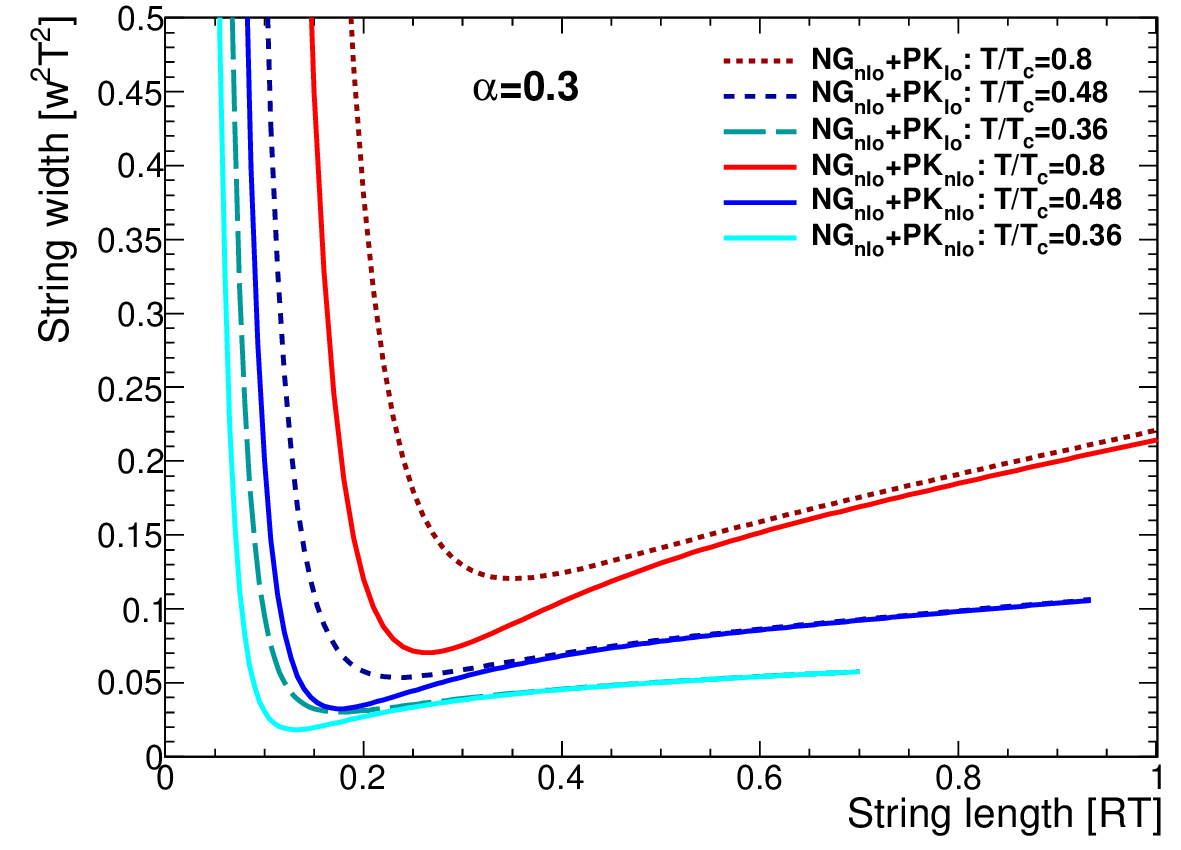}}
		\caption{
			(a) The plot compares the profile of the rigid string for the depicted values of the rigidity parameter $\alpha_r$ at next-to-leading-order ($W^2=W^2_{\rm{NG_{(\ell o)}}}+W^2_{\rm{NG_{(n\ell o)}}}+W^2_{\rm{Ext_{(\ell o)}}}+W^2_{\rm{Ext_{(n\ell o)}}}$) and high temperature $T/T_c=0.9$ in the middle plane. 
			(b) The plot shows the profile of the rigid string with respect to the different temperatures. The dashed lines are for the middle plane's width ($W^2=W^2_{\rm{NG_{(n\ell o)}}}+W^2_{\rm{Ext_{(\ell o)}}}$) at one loop in the rigidity Eqs.~\eqref{NGLED} and \eqref{WExt}; whereas the solid lines are the same as (a) and represent the width at NLO of the rigid string given by Eq.\eqref{Ext2loopT}.			
		}
	\label{CPKNG}			
	\end{center}
\end{figure*}

Compiling all the outcomes of the contribution of the interaction terms to the string width at finite temperature given in Eqs.~\eqref{cExact},\eqref{dExact}, and \eqref{eExact}, the mean-square width is then
\begin{equation}
	\scalemath{0.7}{ W^{2}_{\rm{Ext_{(n\ell o)}}}=\alpha_{r}\Bigg[\left(4(D-2)^2-(D-2)\right)\,I^{(2)}_7+2(D-2)I^{(2)}_{14}-(D-2)^2\,I_{16}^{(2)}\Bigg] }.
\end{equation}
Explicitly, the NLO correction, the second term in Eq.~\eqref{34}, of the smooth string is   
\begin{align}
	\label{Ext2loopT}
	&\scalemath{0.7}{ W^{2}_{\rm{Ext_{(n\ell o)}}}= \frac{\pi^2 \alpha_{r}(D-2)}{8 R^4 \sigma^3} }\\
	&\scalemath{0.7}{ +\frac{ \pi^3  \alpha_r (3(D-2)^2-(D-2))}{8 R^5 \sigma^3} \Bigg[ \frac{4 R}{\pi} \left(\log \left( \eta \left(\frac{\tau}{2}  \right)\right)+\gamma \right) +\frac{L_{T}}{12} \left(2-E_{2}\left(\frac{\tau}{2} \right) \right) \Bigg] } \nonumber \\
	&\scalemath{0.7}{ \times\left(\frac{1}{240}E_4\left(\tau\right)+\frac{1}{240}\right)} .\nonumber
\end{align}

%%%%%%%%%%%%%%%%%%%%%%%%%%%%%%%%%%%%%%%%%%%%%%%%%%%%%%%
%%%%%%%%%%%%%%%%%%%%%%%%%%%%%%%%%%%%%%%%%%%%%%%%%%%%%%%
\section{Numerical Discussion}
\label{sec:NumericalDiscussion}

The broadening aspects of the color tube taking into account the rigidity property, can be read off from the LO contribution of the stiff string solution \eqref{WExt}. The rigidity impacts on the mean-square width inversely decrease with the square of the length scale $R$. As a result, the influence of stiffness is likely to be more prominent at short and intermediate distances.

The mean-square width $W^{2}$ of the string of the free NG string Eq.(\ref{NGLED}) in the center plane is displayed versus string length in Fig.~\ref{LOPKNG}-(a). In the same figure, we plot the corresponding string width with the received LO geometrical corrections \eqref{WExt} for the depicted values of the rigidity parameter $\alpha_{r}$. One can read off a decrease in the mean-square width for the positive values of the rigidity parameter. The geometrical component in the string action filters out the highly curved configurations and the width profile is finely reduced over a long string length scale. The stiffness, however, has more influence on intermediate and short string length scales, where highly wrinkled surfaces appear more prone to emerge in conventional free effective bosonic theory at high energy.

The mean-square width in Fig.~\ref{LOPKNG}-(a) appears to assume negative nonphysical values at finite and positive values of $R$ when $\alpha_r = 0.3$ or larger. The origin of this effect is readily delineated from the simplest instance when taking the free massless string limit, where the width square at temperature $T=0$ shows logarithmic broadening as~\cite{Luscher:1980iy}   
\begin{equation}
	W^{2}=\frac{D-2}{4\pi \sigma}\log{\frac{R}{\mu_0}}.
\end{equation}

The mean-square width returns a negative sign wherever the separation distance between the string end points $R$ is smaller than the UV cutoff $\mu_0$. This is a common property among all effective string model solutions~\cite{Luscher:1980iy,allais,Gliozzi:2010zt}. The unphysical features at short distance signify the breakdown of the long string approximation over which the effective string theory resides \cite{PhysRevLett.67.1681}. The cutoff appears from the standard point-split regularization. The UV-cutoff scale determines $R^2_{0}$ of Eq.(\ref{NGLED}), which is typically regarded as the intrinsic thickness of the string~\cite{Caselle:2012rp} and is intertwined with the long string limit, where $R$ is sufficiently greater than the intrinsic width of the string.

When the rigidity LO correction terms $W^2_{{\rm{Ext}}_{(\ell o)}}$  of the PK string \eqref{WExt} are involved, a more stringent form of this effect does occur [Fig.\ref{LOPKNG}-(a)]. This also appears to be the case when solutions involving perturbative expansions of boundary terms are considered~\cite{Aharony:2010cx,Billo:2012da}. It is worth noting that the same observation holds even when the low-energy/long string expansion is truncated at the NLO for the pure NG case, Eq.\eqref{NGNLO}. In Fig.~\ref{LOPKNG}-(b), for example, a specific value of the UV-cutoff $\mu_0$ parameter is adopted to eliminate nonphysical $W^2$ values at the displayed string tensions.

To renormalize to the positive physical values of the width square, the UV cutoff $\mu_0 $ must subsequently be reset. The renormalization of the string model parameters is not arbitrary and depends on the theory cutoff. For example, the exact dependence of the parameters $\sigma$, $\alpha_r$ on the cutoff has been derived in detail in Refs.~\cite{Nesterenko:1997ku,PhysRevD.27.2944,German:1989vk} with the use of two different regularization schemes. In these expressions, the renormalized parameters appear as a function of an UV scale. When comparing effective string theory predictions with lattice gauge theory, setting an energy scale inverse to lattice spacing implies that cutoff and renormalization effects are coupled. In other words, the renormalization effects have an impact on the model's adjustable parameters via the lattice spacing, which is the characteristic cutoff of the underlying quantum field theory (QFT).

Using the numerical lattice data, the values of the renormalized rigid-string model parameters are determined when fitting the string solution to the lattice, as shown, for example in Refs.~\cite{Caselle:2014eka,Brandt:2017yzw} for the parameter $\alpha_r$. Also, we shall see in the foregoing numerical discussions of the fit of the string model to lattice data, in  Table.~\ref{Fits}, the comparison between the string models $model_1$ and $model_2$ of the NG string and $model_3, model_4$, and $model_5$ with rigidity corrections. This reveals how the parameter $R_{0}^2$ resets its value when including the renormalized $\alpha_r$. 

In Fig.~\ref{CPKNG}-(a) the mean-square width corresponding to the NLO corrections of the rigidity \eqref{Ext2loopT} is plotted  at high temperature $T/T_c=0.9$. We see that the extrinsic-curvature term Eq.\eqref{Ext2loopT} brings on essential corrections on the extracted width of the energy profile at short distances.

A similar plot in Fig.~\ref{CPKNG}-(b) illustrates, however, the perturbative mean-square width with the decrease of the temperature, Eq.~\eqref{Ext2loopT}. The rigidity corrections to the string profile appear to be less compared to the other limit at high temperature/small distances. The curves depicting the mean-square width profile exhibit smaller corrections at low temperatures. This is consistent with the experience that the sharp fluctuations are less probable to occur near the system's ground state, but rather induced by the thermal effects.

In the limit, $R m \rightarrow \infty $ with the mass equal to $m=1/\alpha_{r}$, the rigidity term vanishes and the theory approaches the free massless NG string behavior. On the other hand, we identify a  divergence at short distances $R$ in the mean-square width of a PK string, as depicted in Fig.\ref{CPKNG}-(b). The PK string model envisaged in the figure is composed from the NG action truncated at NLO, along with the LO and NLO rigidity corrections. The divergence taking place at a relatively short length is not surprising given that the two-loop perturbative expansion converges upon the long string limit.

On the other hand, the extended solutions $W^2_{{\rm{Ext}}_{(lo)}}$ and $W^2_{{\rm{Ext}}_{(nlo)}}$ of the PK string have been regularized using the $\zeta$ function scheme. Apart from the NG string, examination of the corresponding formulas \eqref{WExt} and \eqref{Ext2loopT} demonstrates that they do converge at all separation distances of $R>0$. The NLO terms of the NG action \eqref{NGNLO}, as illustrated in Fig.\ref{LOPKNG}-b, are the source of the divergence observed at a comparatively small distance. The divergence arises as a result of terms proportional to the cutoff in Eq.\eqref{NGNLO}.

The long string limit, which sets in the radius of convergence of the low-energy expansion, is intimately related to the intrinsic width of the string and the corresponding cutoff $\mu_0$. As previously discussed, applying the LO correction term of the PK string $W^2_{{\rm{Ext}}_{(\ell o)}}$ without the corresponding renormalization of the model's parameter~\cite{Nesterenko:1997ku,PhysRevD.27.2944,German:1989vk} shifts the radius of convergence of the long string limit. It is why the well-behaved region of the width square appears in Fig.~\ref{CPKNG}-(b) to commence at larger $R$. Nevertheless, the plot in Fig.~\ref{CPKNG}-(b) indicates that including the NLO corrections $W^2_{{\rm{Ext}}_{(nlo)}}$ has offset this shift to some extent, with the divergence occurring at a smaller length scale.

The NLO rigidity corrections $W^2_{{\rm{Ext}}_{(n \ell o)}}$ rapidly scales down the width as one draws closer to the relatively short length of $R$. This is anticipated in the proximity of the radius of convergence of the low-energy perturbative expansion, where higher-order terms ought to provide greater corrections to the mean-square width. Nonetheless, with a reasonable choice of $R$ and $\alpha_r$, the precision of the width of the PK string with the NLO rigidity correction terms should be sufficient. For example, if one considers the changes in mean-square width among the curves in Fig.~\ref{CPKNG}-(a) at different values of $\alpha_r$, we see that the corrections triggered by the increase in $\alpha_r$ exhibit modest variances all the way across the manifested convergence zone.

The effects of the smooth string, though not dominant~\cite{Caselle:2014eka,Brandt:2017yzw}, are expected to be very relevant to non-Abelian gauge models and in the continuum limit~\cite{Caselle:2014eka}. Nevertheless, the truncation of the perturbative expansion keeping the rigidity corrections at NLO would be favorable when scrutinizing the energy profile on high-resolution lattice data and high temperatures.

In this section, we briefly discuss the rigidity property of QCD flux tube considering the lattice data generated from the simulation setup detailed in Refs.~\cite{Bakry:2017utr,Bakry:2018kpn,Bakry:2020flt}. The Monte Carlo data corresponds to pure gluonic configurations of $SU(3)$ Yang-Mills gauge in four dimensions. The space-time dimension of the 4D torus corresponds to $36^3 \times 8$ at a coupling value of $\beta=6.0$, lattice spacing $a=0.1$ fm and temperature $T/T_{c}=0.9$.

The gauge configurations are generated using the standard Wilson gauge action employing a pseudo-heat bath ~\cite{FHKP} updating to the corresponding three $SU(2)$ subgroup elements ~\cite{Marinari}. Each update step/sweep consists of one heat bath and five microcanonical reflections. The gauge configurations are thermalized following 2000 sweeps.

The multiplication of two Polyakov loops 
\begin{equation}
	\mathcal{P}_{2Q}(\vec{r}_{1},\vec{r}_{2}) =  P(\vec{r}_{1})P^{\dagger}(\vec{r}_{2}),
\end{equation}
constructs two infinitely heavy static quark-antiquark $Q\bar{Q}$ states on the lattice. After constructing the static meson at a finite temperature, measurement of the Euclidean action density is taken at each point of $SU(3)$ gauge configurations.
\begin{table*}[!ht]
	\caption{The returned $\chi^2$ from the fits of the theoretical models of both NG and PK string, Eqs.~\eqref{fitmod1}-\eqref{fitmod_3}, to the Monte Carlo lattice data of the mean-square width of the Euclidean action density.}
	\begin{center}
		%		\begin{ruledtabular}
			\begin{tabular}{cc|ccc|ccc|ccc}\hline
				\multirow{2}{*}{} &{Range} 
				&\multicolumn{3}{c}{$R \in[0.5,0.7]$ fm} &\multicolumn{3}{c}{$R \in[0.5,0.8]$ fm}&\multicolumn{3}{c}{$ R \in [0.5,1.2]$ fm}\\
				&{Model} 
				&\multicolumn{1}{c}{$\chi^{2}$} &\multicolumn{1}{c}{$R_{0}^{2}$} 
				&\multicolumn{1}{c}{$\alpha$} &\multicolumn{1}{c}{$\chi^2$}&\multicolumn{1}{c}{$R_{0}^{2}$}&\multicolumn{1}{c}{$\alpha$}&\multicolumn{1}{c}{$\chi^2$}&\multicolumn{1}{c}{$R_{0}^{2}$}&\multicolumn{1}{c}{$\alpha$}\\
				\hline				
				&\small{$\rm{Model_1}$} &40.61   & 2.78(7)   &   0.0  & 82.71 &2.84(7) & 0.0 &128.15     & 2.90 (7)   &0.0         \\
				\multicolumn{1}{c}{} &\small{$\rm{Model_2}$} &9.79   &3.2(2)    &  0.0  &22.04    &3.3(2) & 0.0 &31.90     &3.4 (4)   & 0.0          \\
				\multicolumn{1}{c}{} &\small{$\rm{Model_3}$} &5.35    &3.8(2)  &0.36(1)   &13.82 &4.09(2) &0.45(5) &23.00  & 4.3 (2) &0.51(5)       \\
				\multicolumn{1}{c}{} &\small{$\rm{Model_4}$} & 3.25  &3.7(3) & 0.1(4) & 7.96 &4.0(2) &0.13(3) & 11.095  &4.2(2)  &0.15(3)     \\
				\multicolumn{1}{c}{} &\small{$\rm{Model_5}$} & 3.18   &4.6(6) &0.12(5)&7.815  &5.1(5) &0.16(4) &10.91  & 5.4(5) & 0.18(4)     \\\hline
			\end{tabular}		
			%		\end{ruledtabular}
	\end{center}
	\label{Fits}	
\end{table*}

To characterize the Euclidean action density on the lattice we utilize a plaquette operator defined by 
\begin{equation}
	\square_{\mu\nu}(\vec{\rho})=1-\frac{1}{3} Re Tr \left[U_{\mu}(\vec{\rho})U_{\nu}(\vec{\rho}+\vec{\mu})U_{\mu}^{\dagger}(\vec{\rho}+\vec{\nu})U_{\nu}^{\dagger}(\vec{\rho})\right],   
\end{equation}
which corresponds to the minimal loop structure on the lattice with the indices $\mu$ and $\nu$ corresponding to Lorentz indices. The plaquette operator can be expanded in a power series \cite{Gattringer} of the symmetric tensor $F_{\mu\nu}^{c}$ such that
\begin{equation}
	\square_{\mu\nu}(\vec{\rho}) =1-\frac{1}{3} Re \, Tr \exp\left[ \,i\, g \,a^{2} \sum_{c} F^{c}_{\mu\nu}(\vec{\rho}) T^{c}\right],  
\end{equation}
where $g$ is the coupling of Yang-Mills theory, the index $c$ is color indices and $T^{c}$ are the generators of Lie algebra of $SU(3)_{c}$, 
\begin{equation}
	S(\vec{\rho})  = \frac{1}{2}  (E^2(\vec{\rho})-B^2(\vec{\rho})).                         
\end{equation}
The chromoelectric and chromomagnetic components fields are related to the plaquette components at position $\vec{\rho}$ as
\begin{eqnarray}
	E^2(\vec{\rho})= \sum_{i}^{} E^2_{i}(\vec{\rho}) &\rightarrow \sum_{i}^{} \square_{0i}(\vec{\rho})\\
	B^2(\vec{\rho})= \sum_{i}^{} B_{i}^2(\vec{\rho}) &\rightarrow \sum_{i}^{} \square_{kj}(\vec{\rho}),
	\label{square1}
\end{eqnarray}
where the index $i$ of the magnetic field is the complement of the $j\,k$ plaquette component.

A dimensionless scalar field characterizing the Euclidean action density distribution in the Polyakov vacuum, i.e., in the presence of color sources \cite{Bissey:2006bz} can be defined as
\begin{equation}
	\mathcal{C}(\vec{\rho};\vec{r}_{1},\vec{r}_{2} )= \frac{\langle\, \mathcal{P}_{2Q}(\vec{r}_{1},\vec{r}_{2})\,\rangle\, \,\langle S(\vec{\rho})\, \rangle-\langle\mathcal{P}_{2Q}(\vec{r}_{1},\vec{r}_{2}) \, S(\vec{\rho})\,\rangle } {\langle\, \mathcal{P}_{2Q}(\vec{r}_{1},\vec{r}_{2})\,\rangle\, \,\langle S(\vec{\rho})\, \rangle},
	\label{Actiondensity}
\end{equation}
with the vector $\vec{\rho}$ referring to the spatial position of the energy probe with respect to some origin, and the bracket $\langle ...\rangle$ stands for averaging over gauge configurations and lattice symmetries. Other dimensionful definitions of the correlator \eqref{Actiondensity} yield equivalent representation of the width (see, for example Ref.~\cite{Okiharu:2003vt}).

The above equation is dimensionless. However, the field densities in physical units~\cite{Bicudo} would read
\begin{equation}
	\sum_{c}^{} F^{c}_{\mu \nu}F^{c}_{\mu \nu} =\frac{2\beta}{a^{4}}\left[ 1 - \frac{1}{3} Tr(\square_{\mu\nu}) \right]+\mathcal{O}(a).
	\label{square2}
\end{equation}
The sum over the color index $c$ ensures the gauge invariance of the square of the field densities.

The measurements are taken on 20 bins separated by 70 sweeps of updates each bin consists of 500 configurations. To avoid artificial reduction of the error bars, each set of 70 update configurations has been included in the same jackknife subensemble, such that the variances are calculated with respect to decorrelated bins.

To eliminate statistical fluctuations with the physical observable left intact, only 20 sweeps of UV filtering of stout-link smearing \cite{Morningstar:2003gk} have been applied on each considered gauge configuration. We make use of the symmetry of the 4D torus, that is, the measurements that are taken at a fixed color source's separations $R$ are repeated and averaged. The characterization \eqref{Actiondensity} yields $0<C<1$ with $C \rightarrow  0$ away from the quarks by virtue of the cluster decomposition of the operators.

To estimate the mean-square width of the gluonic action density along each transverse plane to the quark-antiquark axis by taking into consideration the axial cylindrical symmetry of the tube, we choose a double Gaussian function of amplitude $A$ and fit parameters $\sigma_1,\sigma_2$ and $\kappa$, such that
\begin{equation}  
	G(r,\theta;z)= A (e^{-r^2/\sigma_1^2}+e^{-r^2/\sigma_2^2})+\kappa.
	\label{conGE}
\end{equation}

A measurement of the width of the Euclidean action density may be taken by fitting the density distribution $\mathcal{C}(\vec{\rho};z)$ to Eq.\eqref{conGE} through each transverse plane to the cylinder's axis $z$ to Eq.\eqref{conGE}, with $r^2=x^2+y^2$ in each selected transverse plane $\vec{\rho}(r,\theta;z)$, which returns acceptable values of $\chi^{2}$ at the intermediate and long distances~\cite{Bakry:2020ebo}.

There is another justification behind the selection of this ansatz other than the good parametrization behavior over the targeted distance scale~\cite{Bakry:2020ebo}. The reason comes from the observation of a large difference in the value of the returned fit parameters satisfying the width $\sigma_1 \gg \sigma_2$. That is, when parametrizing the transverse profile with this particular form, the usual Gaussian function receives corrections from a narrower Gaussian with the same amplitude. The overlap of a two Gaussian of the same amplitude and different width reweighs the fit function at the center of the profile with respect to the tail of the bell-shaped region.

The form is inspired by the fit function as shown in Ref.~\cite{Bicudo} where a classical flux-tube profile is convoluted with a Gaussian distribution relevant to the quantum oscillations. The form in Ref.~\cite{Gliozzi:2010zv} also corrects the tube center near the $Q\bar{Q}$ axis and asymptotically appears to reweigh the tail region. It should be noted that, with the use of these fit ansatz introduced in the aforementioned reference, we obtained good fits at small separation distances; however, we received large error bars at large distances. For convenience, we have recoursed to the two Gaussian forms, which describe one way to mimic the same function.

The second moment of the action density distribution with respect to the cylinder's axis $z$ joining the two quarks is
\begin{equation} 
	W^{2}(z_i)=\quad \frac{\int \, dr\,r^{3}\,G(r,\theta;z_i)} {\int \,dr \,r \: G(r,\theta;z_i) },
	\label{widthG}
\end{equation}
which defines the $i$-th data entry of the mean-square width of the tube at the plane $z_i$ on the lattice.

The measurements of the mean-square width of the Euclidean action density have been achieved with a good signal-to-noise ratio and are obtained by employing only 20 sweeps of stout-link smearing~\cite{Bakry:2017utr,Bakry:2018kpn,Bakry:2020ebo}. The small number of stout-link smearing sweeps should increase the accuracy of the measurement and keep the relevant physics intact, for the targeted color separation distance scale~\cite{Bakry:2014ina}. All the most recent calculated data relevant to the Euclidean action density with the fit analysis and width measurements have been included in Refs.~\cite{Bakry:2017utr,Bakry:2018kpn,Bakry:2020ebo} in detail.

In general, renormalization effects~\cite{Peter} impact lattice measurements via lattice spacing, which establishes a natural cutoff for the corresponding QFT. Setting an energy scale inverse to lattice spacing entails that cutoff and renormalization effects are entangled; any change in the scale is also a change in the cutoff. When measuring the width, a change in the cutoff offsets the width by an overall constant. This induction is based on the observation of an identical impact on the width~\cite{Bakry:2010sp} due to the application of a different number of cooling sweeps, which effectively increases lattice spacing or changes the UV scale. In the present setup, we study the behavior of the quantum broadening of the width, which would be unaffected by the same overall shifts at a given $R$.

Another subtlety concerning the width measurement comes from the issue of possible cross-correlations among the correlators at various distances. The computations of the field densities involve a three-point correlation function consisting of two Polyakov loops in opposite directions and the flux-tube operator, Eq.~\eqref{Actiondensity}. In this setup, when the size of the flux probe operator is substantially large, cross-correlations among measurements at a different distances could be prone to occur on the same gauge configuration. Large size/extended operators involve nested loops with different sizes, which is a well-known strategy in LGTs to minimize the discretization errors owing to the lattice spacing. In the present calculations, we applied the simplest ``one-loop field strength tensor operator'', which is proportional to one plaquette and is the minimal-sized loop structure on hypercubic lattices, to compute chromoelectric and chromomagnetic field densities Eqs.\eqref{square1} and \eqref{square2}. For the same reason, we avoided using algorithms utilizing extended operators in the stout-link smearing besides the small number of smearing sweeps to minimize blurring (Brownian motion\cite{Takahashi:2002bw}) of gauge links.

We may add to the above one more point that comes from the lattice setup where the Monte Carlo simulations with the standard Wilson gauge action ensure locality~\cite{Luscher:2001up}. In addition, the heat bath combined with overrelaxation / microcanonical steps would assist to diverge the autocorrelation time \cite{QCD-TARO:1992iba}.

As a consistency check of cross-correlation between adjacent Polyakov loop correlators, we calculate the local string tension and convert to lattice spacing $a$ in Fermi units using the transfer matrix interpretation \cite{Luscher:2001up}, such that
\begin{align}
& \alpha(\Lambda) \, \exp(- \sigma R L_T)  = \langle P(0) P^{\dagger}(R) \rangle \\			
& \sigma a^2   = - \frac{1}{L_{T}}\log \left(\frac{\langle P(0) P^{\dagger}(R+1)\rangle}{\langle P(0) P^{\dagger}(R) \rangle}    \right)
\end{align} 
The lattice spacing in Fermi units is then
\begin{equation}
a = \frac{0.1973 \rm{GeV\, fm} }{\sqrt{0.440} \rm{GeV}}  \left[ -\frac{1}{ L_{T}}\log \left( \frac{\langle P(0) P^{\dagger}(R+1)\rangle}{\langle P(0) P^{\dagger}(R)\rangle }\right)  \right]^{\frac{1}{2}}
\end{equation}
on each gauge configuration. The above expectation values are calculated as average on all lattice points utilizing the lattice space-time symmetries to improve the statistics in a gauge-independent manner. The purpose of these measurements is to test if we could extract the correct value of the lattice spacing $a \simeq 0.098$ from two Polyakov loops correlators at a different distance. We report the correct value of the lattice spacing which indicates the minor effect of cross-correlations among adjacent Polyakov loop correlators at various distances.
%------------------------------------------------------------------------------------------------
\begin{figure*}[t]
\begin{center}						
	\subfigure[]{\includegraphics[scale=0.37]{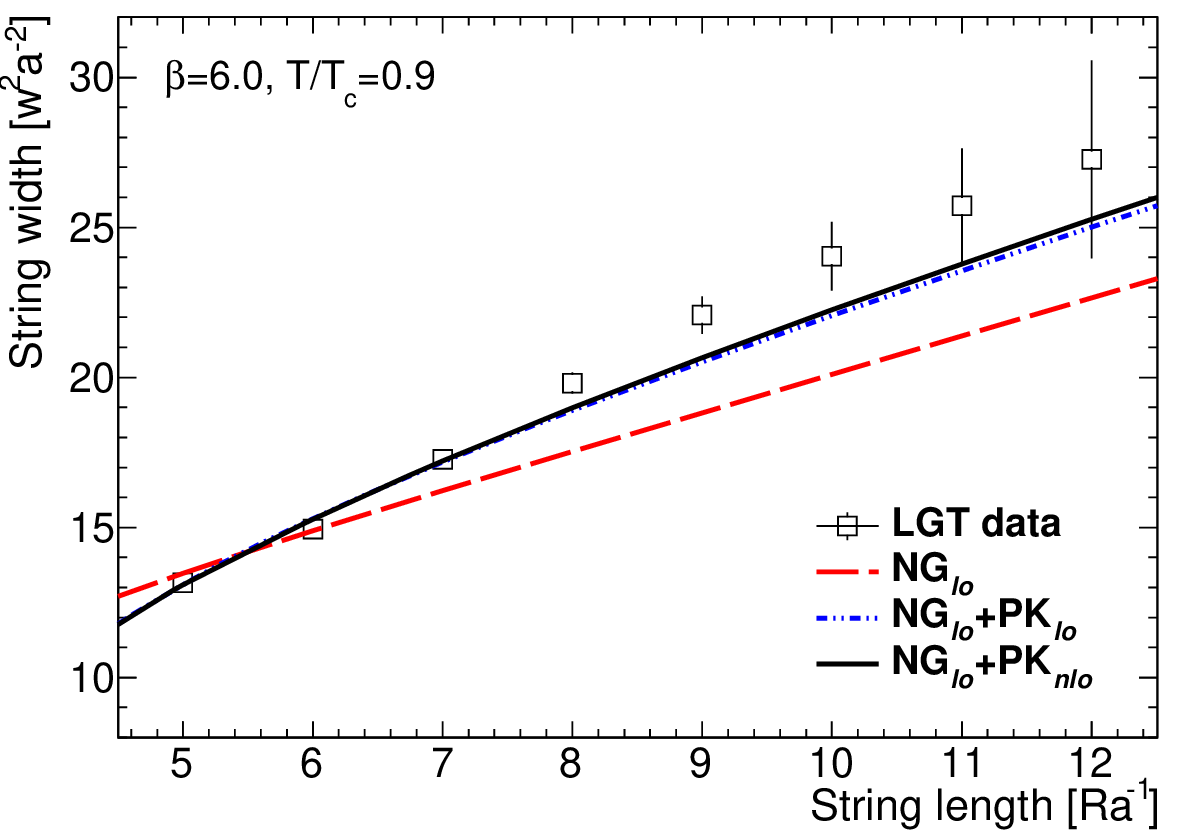}}
	\subfigure[]{\includegraphics[scale=0.37]{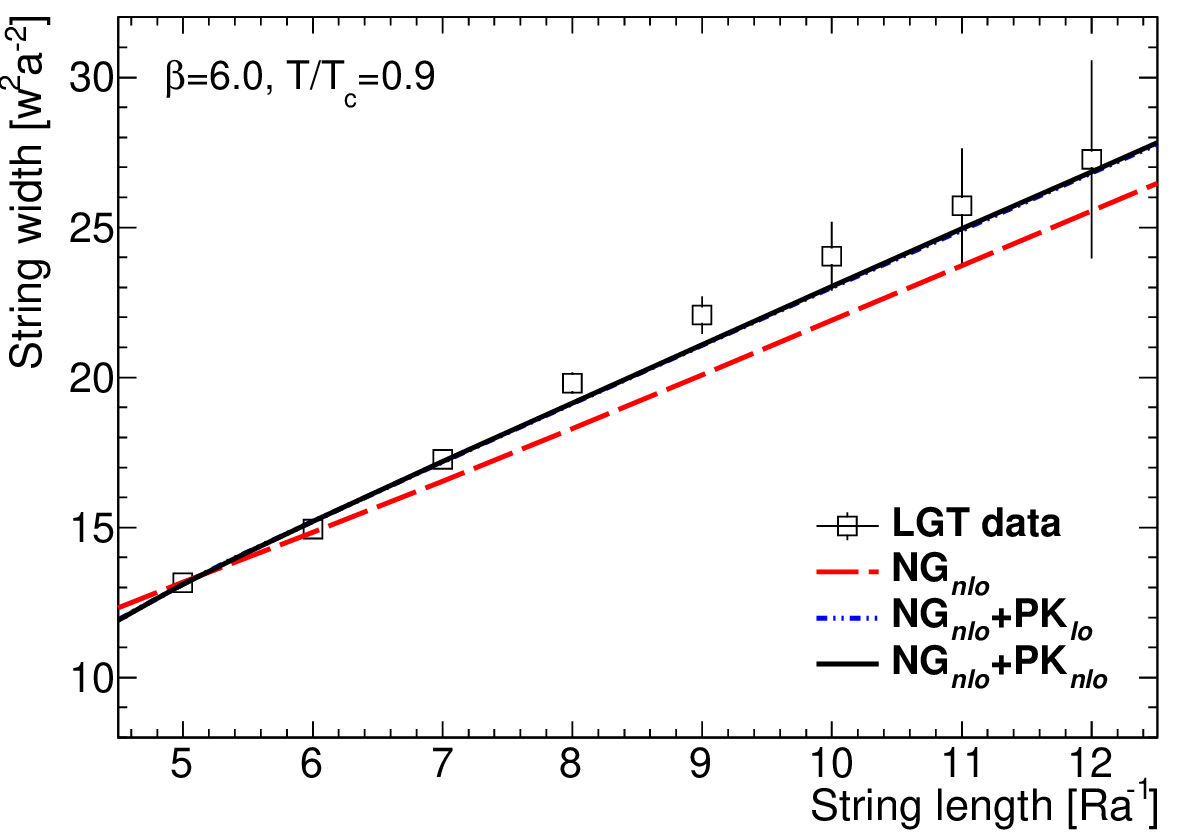}}   
	\caption{The mean-square width of the string $W^2(R/2)$ versus $Q\bar{Q}$ separations measured in the middle plane $R/2$ at $T/T_c=0.9$.(a) The fits of the LO width square of NG string, Eq.~(\ref{NGLED}) and combination of rigidity contributions given by PK string models Eqs.~\eqref{fitmod_1}-\eqref{fitmod_3}. (b) The fits of the NLO width of NG string, Eq.~\eqref{NGNLO}, and combination of rigidity contributions given by PK string models Eqs.~\eqref{fitmod_1}-\eqref{fitmod_3}. The Monte Carlo lattice data of the square width of the Euclidean action density width are shown with boxes. The lattice data have been fitted in the interval $R \in [0.5, 0.7]$ fm.}
\label{LatticeData}	
\end{center}						
\end{figure*}
%-----------------------------------------------------------------------------------------------------

The numerical values~\cite{Bakry:2017utr,Bakry:2018kpn,Bakry:2020ebo} of the mean-square width indicate a broadening in mean-square width of the string at all transverse planes $z_{i}$ as the color sources are pulled apart a distance $R_i$. The maximum color source separation distance has been taken up to $R = 1.2$ fm, which is one-third the total spatial extent of the lattice 3.6 fm to avoid cross-correlations through the toroidal lattice owing to the periodic boundary condition. At the central transverse plane $z_i=R/2$, the pattern of the width broadening \eqref{widthG} can be compared to prescribed width models that combine an assortment of potentially interesting quantum string corrections.

In addition to the pure NG string models at two orders,
\begin{equation}
W^2_{model_1}=W^2_{\rm{NG_{(\ell o)}}},
\label{fitmod1}  
\end{equation}
\begin{equation}
W^2_{model_2}=W^2_{\rm{NG_{(\ell o)}}}+W^2_{\rm{NG_{(n\ell o)}}},
\label{fitmod2}  
\end{equation}  
which are given in accordance with Eqs.(\ref{NGLED}) and \eqref{NGNLO}, respectively. The following three models define selected combinations with the rigidity corrections:
\begin{equation}
W^2_{model_3}=W^2_{\rm{NG_{(\ell o)}}}+W^2_{\rm{Ext_{(\ell o)}}},
\label{fitmod_1}   
\end{equation}
\begin{equation}
W^2_{model_4}=W^2_{\rm{NG_{(\ell o)}}}+W^2_{\rm{NG_{(n\ell o)}}}+W^2_{\rm{Ext_{(\ell o)}}},
\label{fitmod_2}   
\end{equation}
\begin{equation}
W^2_{model_5}=W^2_{\rm{NG_{(\ell o)}}}+W^2_{\rm{NG_{(n\ell o)}}}+W^2_{\rm{Ext_{(\ell o)}}}+W^2_{\rm{Ext_{(n\ell o)}}},
\label{fitmod_3}   
\end{equation}
which are provided by Eqs.\eqref{NGLED}, \eqref{NGNLO}, \eqref{WExt} and \eqref{Ext2loopT}.

The fits are obtained by optimizing the sets of the parameter space ${R_0, \alpha}$ in the above-mentioned models such that the least square residuals
\begin{equation}
\chi^2(R_0,\alpha)=\frac{ \sum_{i} (W^2_{i}(R_i/2)-W^2_{model}(R_i/2;R_0,\alpha))^2 }{ \sum_i e_i^2(R_i/2)},
\label{chisquared}  
\end{equation}  
are minimized. In the above equation \eqref{chisquared}, $e_i^2(R_i/2)$ is the square of the error in the measured mean-square width $W^2_{i}(R_i/2)$ from the lattice simulation \eqref{widthG}.

Here, the analysis of the fit behavior of the mean-square width data is discussed keeping the string tension fixed to the standard value~\cite{Schilling93} of $\sqrt{\sigma} =440$ MeV. Our interest is to divulge the rigidity parametrization behavior at each order of the perturbative expansion of the PK string with respect to numerical data. We consider the lattice data of the mean-square width of the QCD flux tube at the perpendicular plane in the middle $R/2$ between the color sources~\cite{Bakry:2017utr,Bakry:2018kpn,Bakry:2020ebo}.

The resultant minimum $\chi^{2}$ and the corresponding fit parameters $\alpha$ and $R_{0}^{2}$ from the fits to the lattice data Eq.\eqref{chisquared} are collected in Table.~\ref{Fits}. 
The first column of the table highlights the five string models defined by Eqs.\eqref{fitmod1} - \eqref{fitmod_3}. 
That is, in the first two rows, the fit parameter $R_{0}^2$ (related to the UV cutoff) returned from the fit to the width at both the LO and the NLO mean-square width of the NG string Eq.\eqref{fitmod1} and   \eqref{fitmod2}, respectively. However, in the next entries the table includes the values of both $R_{0}^2$ and the rigidity $\alpha_r$ parameters returned from the fits to the mean-square width for selected models of the PK string, Eqs.\eqref{fitmod_1}-\eqref{fitmod_3}.

\begin{figure*}[t]
\begin{center}						
	\subfigure[]{\includegraphics[scale=0.37]{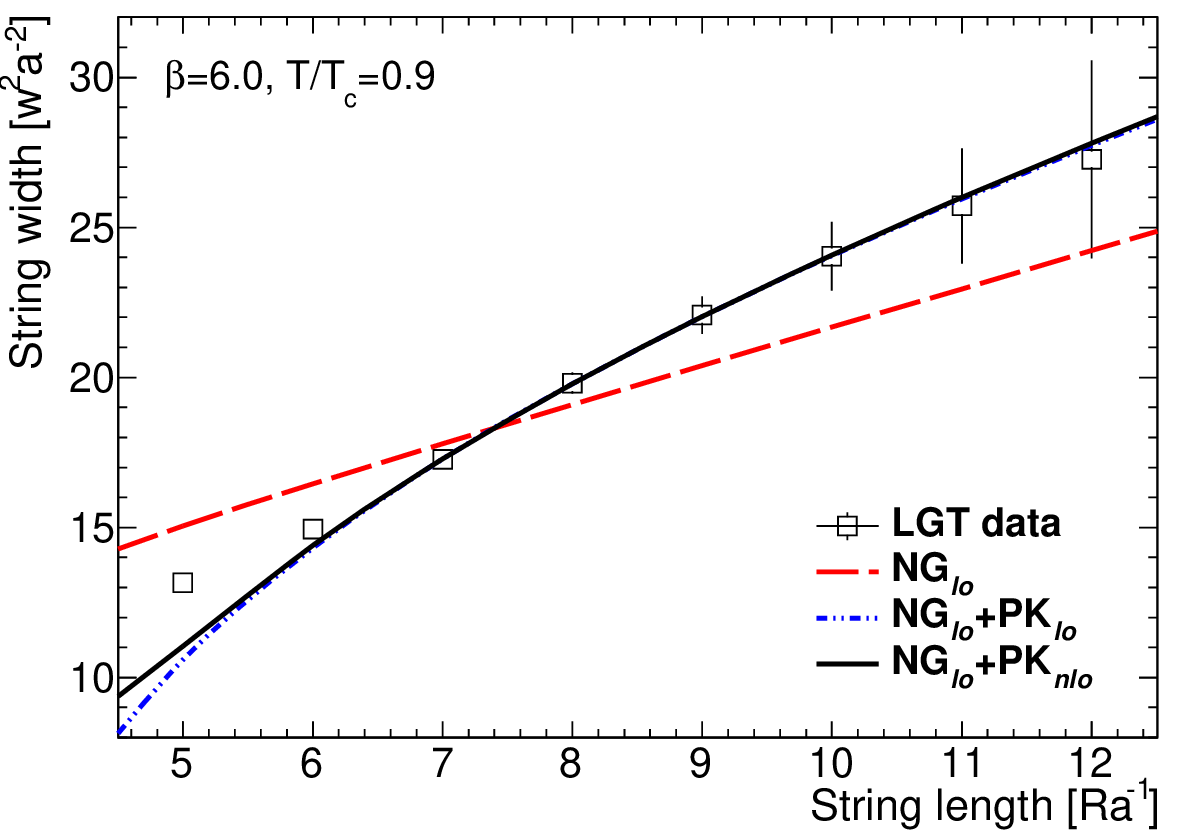}}
	\subfigure[]{\includegraphics[scale=0.37]{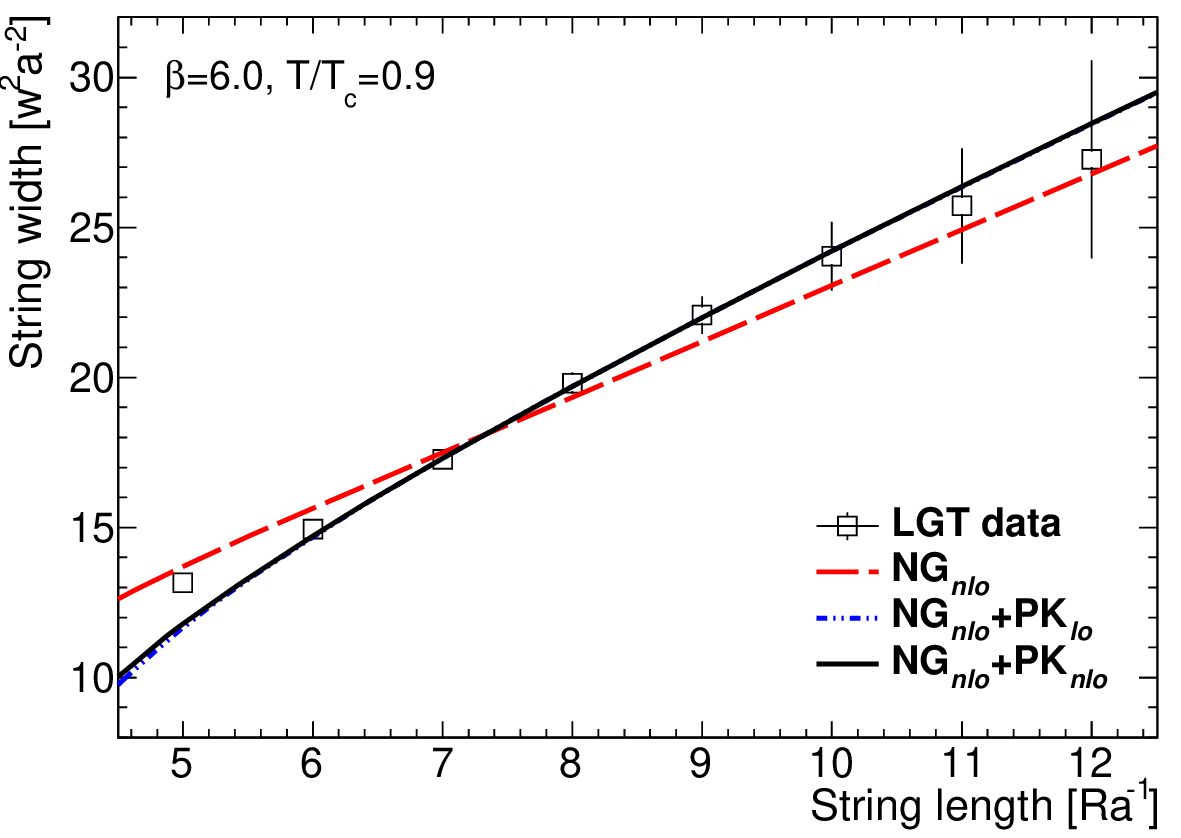}}  
	\caption{Same as Fig.~\ref{LatticeData}, the mean-square width of the string $W^2(R/2)$ versus $Q\bar{Q}$ separations measured at the middle plane $R/2$ at $T /T_{c} = 0.9$, however, the fit range involves larger source separation $R\in[0.7,1.2]$ fm. 
	(a) The fits of the LO width square of NG string, Eq.~(\ref{NGLED}) and combination of rigidity contributions given by PK string models Eqs.~\eqref{fitmod_1}-\eqref{fitmod_3}. (b) The fits of the NLO width of NG string, Eq.~\eqref{NGNLO}, and combination of rigidity contributions given by PK string models Eqs.~\eqref{fitmod_1}-\eqref{fitmod_3}.}
\label{LatticeData2}	
\end{center}								
\end{figure*}

The returned $\chi^{2}$ values, in Table.\ref{Fits}, imply that the fits are better with respect to the pure NG string (\ref{fitmod2}) at NLO than with the LO approximation \eqref{fitmod1}. Nonetheless, the NLO approximation does not produce an optimal match with the numerical data until large source separations $R \geq 0.8$ fm. 

Further examination of $\chi^2$ values from the fit of $W^{2}_{model_3}$ of the PK string \eqref{fitmod_1} reveals an improvement over fits of $W^{2}_{model_1}$ pertinent to the free-string model \eqref{fitmod1}. This PK string model, in which the LO rigidity corrections appear to play a similar role to the string self-interaction, yields acceptable $\chi^2$ values over a fit interval of $R \in [0.6,1.2]$ fm. 

Despite the reduction in $\chi^2$ values when considering either the PK string's rigidity or the NG string's self-interaction, we find that combining both corrections yields better values of $\chi^{2}_{\rm{d.o.f.}}=10.91/6$ over the entire source separation interval $ R \in [0.5,1.2]$ fm.

When the fit behavior of the LO and the NLO corrections of smooth PK string models [Eq.(\ref{fitmod_2}) and \eqref{fitmod_3}] is opposed, the values of the returned $\chi^2$ are found to differ slightly. The corrections acquired from the NLO stiffness terms are relatively minor and within the measurement uncertainties. The projected improvements from these NLO terms can be significant if smaller length scales, finer lattices, or alternate gauge models are considered, which are beyond the scope of this study. Nonetheless, the fit findings show that the NLO perturbative modifications are consistent with the lattice data at typical values of the rigidity parameter $\alpha_r$.

Figs.\ref{LatticeData} and \ref{LatticeData2} reflect the above discussed points, including the $\chi^2$ values in Table.\ref{Fits}, in particular the improvement in the fit with respect to the rigid PK string compared to that obtained merely on the basis of the pure NG string. The plots in Figs.\ref{LatticeData} and \ref{LatticeData2} are for fits to Eqs.(\ref{NGLED}) and \eqref{NGNLO} in addition to Eqs.\eqref{fitmod_1}-\eqref{fitmod_3}. The fits are over a subset of the intermediate source separation region $R\in[0.5,0.7]$ fm  Fig.\ref{LatticeData}. The Fig.\ref{LatticeData2} renders curves from the fits over the subset $R\in[0.7,1.2]$ fm at the end.

The graphs in Figs.\ref{LatticeData} and \ref{LatticeData2} show the mean-square width as calculated using the theoretical models of Eqs.\eqref{fitmod1}-\eqref{fitmod_3}. Fig.\ref{LatticeData} depicts fitted curves over source separation $R \in [0.5,0.7]$, whereas Fig.~\ref{LatticeData2} displays the fitted width square over $R \in [0.7,1.2]$. Both figures, as well as the $\chi^2$ values in Table \ref{Fits} reflect the arguments highlighted above, specifically the improvement in fit with respect to the stiff PK string compared to that produced purely on the basis of the NG string. 

For source separation range $R\in[0.5,0.7]$ fm, the rendering in Fig.\ref{LatticeData} of the fitted width of the pure NG string at either order exhibits significant deviations from the lattice data compared to the fits in Fig.\ref{LatticeData2} over the fit interval $R\in[0.7,1.2]$ fm. The plots in Fig.\ref{LatticeData} suggest the incompetence of the pure NG string as a physical description integrating the subtle features of the QCD flux tubes. However, for both fit intervals, the improvements with respect to the rigid PK string models [Eqs.\eqref{fitmod_1}-\eqref{fitmod_3}] are clearly evident in Figs.\ref{LatticeData} and \ref{LatticeData2}.

In regard to the stability of the fits, we find the values of $\alpha_r$ to be equal (within the uncertainties) and stable for all fit intervals commencing from $R_{min}=0.5$ fm ($R \in [R_{min},R_{max}] $) up to $R_{max}=1.2$ fm. Moreover, we report the same value of $\alpha_r$ for the fit intervals commencing from $R_{min}=0.4$ fm as well~\cite{Bakry:2020ebo}.

It should be noted that the small values of the returned $\chi^{2}_{\rm{d.o.f.}} < 1$, for the fit of rigid models $model_3-model_5$ on the other intervals such as $R \in [0.7,1.2]$, are indicating the growing width of the error bars at a large distance. However, stability and consistency with the returned values in the above can be checked by substituting the returned values of $\alpha_r$ on any of the former intervals within the latter, which produces fits with acceptable values of $\chi^{2}_{\rm{d.o.f.}}$ and thereof proves the stability of the fits. In addition, we find the returned value of $\alpha_r$ from the fits of the highly accurate $Q\bar{Q}$ potential data to rigid-string model on these intervals reproduces the values $\alpha_r$ within the uncertainties \cite{Bakry:2020flt}.

The properties of the energy profile of the smooth QCD string ought to be discerned in the light of the counterpart IR observable, namely, the string tension and the $Q\bar{Q}$ potential. 
Our up-to-date lattice simulation and fit analysis of the static $Q\bar{Q}$ data have been reported in detail in Ref.~\cite{Bakry:2020flt} (see also Refs.~\cite{Bakry:2017utr,Bakry:2018kpn,Bakry:2020ebo}).

The pure NG string at the NLO does not provide a precise match to the numerical data of the static $Q\bar{Q}$ potential either~\cite{Bakry:2020flt}. Fits to the PK string at leading order, on the other hand, return a stable minimum in the $\chi^{2}$ for a rigidity factor of roughly $\alpha_{r} \approx 0.3(2)$ on the interval $R \in [0.6,1.1]$ fm (without boundary corrections)~\cite{Bakry:2020flt}. Remarkably, by the inclusion of the rigidity terms in $Q\bar{Q}$ string potential, the accurate value of the string tension of $\sigma a^2=0.044$ ~\cite{Koma:2017hcm} is reproduced when taken as a free fit parameter. We plug both values of fit parameters $\sigma$ and $\alpha_r$ in Eq.(\ref{fitmod_1}) for the fit to the mean-square width of flux tube as a consistency check, and we attain good $\chi^2$ values in the fit range $R \in [0.6, 1.2]$ fm. 

The close fit produced from the two probes of the confining string, namely, the energy profile and the $Q\bar{Q}$ potential, possibly supports the faithfulness of the physics of the smooth QCD string rather than being a mere artifact of the model's extended parametric space.

\section{Summary and Conclusion}
\label{sec:Conclusions}

In summary, we have considered the geometrical (PK) string proposed by Polyakov~\cite{POLYAKOV19} and Kleinert~\cite{Kleinert:1986bk}, in which the surface representation in the string action depends on the area as well as the geometrical configuration of the embedded world sheet. This is embodied in the effective bosonic string action by introducing terms proportional to the extrinsic-curvature as a next-order operator after the NG action~\cite{POLYAKOV19,Kleinert:1986bk}. The curvature term with a positive-valued rigidity parameter favors the smooth string's world sheet configurations over the sharply creased surfaces.

In this work, we have derived the width of the energy profile of PK rigid strings in $D$ dimension at both low and high temperatures. The perturbative arguments regarding string's rigidity have been implemented  as an effective low-energy parameter expansion around the encompassed NG string action. That is, the perturbative expansion of the extrinsic-curvature term is taken around the free NG string action.

The technique of the $\zeta$ function is incorporated in the regularization of the divergent sums appearing in the corresponding expectation values of the mean-square width of the string. The resultant Green's function integrals have been evaluated with an open-string propagator on a cylinder, i.e., the Dirichlet boundary condition. Our main results are given by Eq.\eqref{WExt} for the width corrections at the LO level and Eq.\eqref{Ext2loopT} for the NLO corrections, which lay out a general expression valid at any temperature and dimension.

The LO rigid-string correction to the mean-square width Eq.\eqref{WExt} indicates an inverse decrease with the square of the length scale $R$. As expected, this entails rigidity effects that are outstanding near the intermediate and short string length scale. The next-to-leading expression of rigid-string correction Eq.\eqref{Ext2loopT}, introduces delicate corrections of the mean-square width over short distances, which would extend the applicability of the derived formula.

We have examined the width corrections induced by the smooth PK string  with respect to the numerical lattice data for a high temperature close to the deconfinement point $T/T_c = 0.9$. We consider lattice data corresponding to the Monte Carlo simulations of the flux tube profile in the middle plane in pure $SU(3)$ Yang-Mills theory in four dimensions~\cite{Bakry:2017utr,Bakry:2018kpn,Bakry:2020ebo}. The fits of the mean-square width considering the rigidity and self-interaction corrections are in a good match with the lattice Monte Carlo data of QCD flux tube at both intermediate and large color source separation distances.

Nevertheless, it remains demanding to layout the width profile taking into account the Lorentz-invariant boundary action~\cite{Billo:2012da,Bakry}, exploring the full effective string actions in the numerical simulation of $SU(3)$ Yang-Mills theory~\cite{Bakry:2020ebo} and other gauge models. The energy ~\cite{Bicudo:2018jbb} profile of the excited meson is a very relevant system to implement such a program as well. In addition to that, it would be interesting to consider other boundary conditions involving open strings; such as the mixed Dirichlet-Neumann boundary condition, which generalizes the discussed theoretical considerations to the baryonic configurations~\cite{Jahn2004700,deForcrand:2005vv,Pfeuffer,Bakry:2016aod,Bakry:2014gea}. This is the target of our next reports.

%%%%%%%%%%%%%%%%%%%%%%%%%%%%%%%%%%%%%
%%%%%%%%%%%%%%%%%%%%%%%%%%%%%%%%%%%%% 
\begin{acknowledgments}
We thank B. Brandt, M.Caselle, F. Gliozzi, M. Pepe, and U. Wiese for useful comments and discussions.
We thank the Armed Forces of Ukraine for providing security to perform this work. This work has become possible only because of the resilience and courage of the Ukrainian Army. This work has been funded by the Chinese Academy of Sciences President's International Fellowship Initiative Grants No.2015PM062 and No.2016PM043, the Polish National Science Centre (NCN) Grant No. 2016/23/B/ST2/00692, the Recruitment Program of Foreign Experts, NSFC Grants No.~11035006,~11175215,~11175220, and the Hundred Talent Program of the Chinese Academy of Sciences (Y101020BR0).
\end{acknowledgments}
%---------------------------------------------

%%%%%%%%%%%%%%%%%%%%%%%%%%%%%%%%%%%%%
%%%%%%%%%%%%%%%%%%%%%%%%%%%%%%%%%%%%% 
%\section{appendix}

%%%%%%%%%%%%%%%%%%%%%%%%%%%%%%%%%%%%%%%%%%%%%%%%%%%%%
%%%%%%%%%%%%%%%%%%%%%%%%%%%%%%%%%%%%%%%%%%%%%%%%%%%%%
\appendix
\section{Functions and Identities}
\label{Appendix_I}  

The following outlines functions and identities~\cite{WhittakerWatson,abramowitz,Erdelyi} that will be used along with the calculations of perturbative corrections to the mean-square width:

\begin{itemize}
	\item
	The Dedekind $\eta$ function over the complex plane is defined as
	\begin{align}		
		\eta(\tau) &=q^{\frac{1}{24}} \prod _{k=1}^{\infty } \left(1-q^k\right)\\
		&=e^{\frac{ i\pi   \tau }{12}} \prod _{k=1}^{\infty } \left(1-e^{2 i \pi   k \tau }\right) \nonumber,
	\end{align}		
	with  $q=e^{2 i\pi\tau}$ and $\tau=\frac{i L_{T}}{2 R}$. The identity	
	\begin{equation}
		\sum _{j=1}^{\infty } \log \left(1-q^j\right)=\log \left(q^{-\frac{1}{24}} \eta (\tau )\right),
	\end{equation}
	also follows from the above definition.
	
	\item 
	The Eisenstein series over the complex plane is defined as
	\begin{equation}
		E_{2k}(\tau)=1+(-1)^{k} \frac{4k}{B_k} \sum_{n=1}^{\infty} \frac{n^{2k-1}}{1-q^n},
		\label{eq:Eisenstein_Series}
	\end{equation}
	where $B_{k}$ are Bernoulli numbers defined as
	\begin{equation}
		\mathbf{\zeta}(2k)=\frac{B_k (2 \pi)^{2k}}{2k!},~~~~~~~~~\mathbf{\zeta}(1-2k)=(-1)^{k}\frac{B_k}{2k!}.
		\label{eq:Bernoulli_Numbers}	
	\end{equation}
	for even and odd numbers respectively.
	
	In particular, $E_{2}$ and $E_{4}$ are given by
	\begin{align}		
		E_{2}(\tau) =& 1 -24 \sum_{n=1}^{\infty} \frac{n}{1-q^n},\\
		E_{4}(\tau) =& 1 +240 \sum_{n=1}^{\infty} \frac{n^{3}}{1-q^n}.
		\label{E2E4}
	\end{align}
	The modular transform $\tau \rightarrow \frac{-1}{\tau}$ of the Dedekind $\eta$ function and the Eisenstein series $E_{2k}$ is given by
	\begin{align}
		\eta({\tau}) & = \frac{1}{\sqrt{i \tau}}\eta\left(\frac{-1}{\tau}\right),\\
		E_{2}(\tau) & = \frac{1}{ \tau^{2 }}E_{2}\left(\frac{-1}{\tau}\right)-\frac{1}{6 i \pi \tau},
		\label{eq:modularTransf_EisenSeries}          
	\end{align}
	respectively.

	\item \underline{Identity(I):}\\
	The sum $\sum_n \frac{n \left(q^n+1\right)}{q^n-1}$ can be expressed in the form of the Eisenstein series  
	\begin{align}
		\sum_n \frac{n \left(q^n+1\right)}{q^n-1}&=\sum_n\frac{n}{q^n-1}-\sum_n \frac{n q^n}{1-q^n} \\
		&=\sum_n \frac{n}{q^n-1}-\sum_n \frac{n q^n}{1-q^n}\nonumber \\
		&=\sum_n \frac{n}{q^n-1}-\frac{1}{24} (1-E_2(\tau ))\nonumber \\
		&=\sum_n \frac{n \left(q^n-q^n+1\right)}{q^n-1}-\frac{1}{24} (1-E_2(\tau ))\nonumber \\
		&= -\frac{2}{24} (1-E_2(\tau ))-\sum_n n \nonumber \\
		&=\frac{1}{12}-\frac{2}{24} (1-E_2(\tau ))\nonumber\\
		&=\frac{1}{12} E_2(\tau).\nonumber 
	\end{align}
	
	\item \underline{Identity(II)} The product representation of the Jacobi elliptic theta function of the first kind $\vartheta_1(z,q)$ is defined as 
	\begin{equation}
		\scalemath{0.85}{ \vartheta_{1}(z,q)=2q^{-\frac{1}{4}} \sin(z) \prod_{k=1}^{\infty} \left(1-q^{2k}\right)\left(-2q^{2k}cos(2z)+q^{4k}+1\right) }
		\label{IdentityII}  
	\end{equation}
	
	\item \underline{Identity(III)} The logarithmic derivatives of $\vartheta_1(z,q)$ can be expressed as a series sum of hyperbolic functions as
	\begin{equation}
		\sum _{k=-\infty}^{\infty } \coth(\pi  k \lambda +z)=i\frac{\vartheta _1^{\prime}\left(i z,e^{-\pi  \lambda }\right)}{\vartheta _1\left(i z,e^{-\pi  \lambda }\right)}.
		\label{IdentityIII}
	\end{equation}
	
	\item \underline{Identity(IV)} The modular $\pi \tau$ increment in the argument of the $\vartheta _1$ function has the property
	\begin{equation}
		\vartheta _1(\pi  \tau +z,q)=-\frac{e^{-2 i z} }{q} \vartheta _1(z,q).
		\label{IdentityIV:eq1}
	\end{equation}

	The generic formula for $m\in \mathbf{Z}$ for the modular increment of $z$,
	\begin{equation}
		\vartheta _1(z+m \pi  \tau ,q)= \frac{(-1)^m}{q^{m}} \vartheta _1(z,q) e^{-i (\pi  (m-1) m \tau +2 m z)}.\\
		\label{IdentityIV:eq2}
	\end{equation}
	
	The modular increase of the logarithmic derivative can be deduced from properties of the $\vartheta_1$ function; for $m=-1$ we obtain
	\begin{equation}
		\frac{\vartheta _1^{\prime }(z-\pi  \tau ,q)}{\vartheta _1(z-\pi  \tau ,q)}=\frac{\vartheta _1^{\prime }(z,q)}{\vartheta _1(z,q)}-2 i,
		\label{IdentityIV:eq3}
	\end{equation}
	and
	\begin{equation}
		\frac{\vartheta _1^{\prime \prime}(z-\pi  \tau ,q)}{\vartheta _1(z-\pi  \tau ,q)}=\frac{\vartheta^{\prime\prime} \left(1,z,q \right)}{\vartheta _1\left(z,q \right)}+\frac{4 i \vartheta _1^{\prime }\left(z,q \right)}{\vartheta _1\left(z,q \right)}-4.
		\label{IdentityIV:eq4}
	\end{equation}
	
	\item \underline{Identity(V)} The square of the logarithmic derivative~\cite{Zhao,SHEN1993299} of the theta function $\vartheta_1$ is given by  
	\begin{equation}
		\left(\frac{\vartheta _1^{\prime }(z,q)}{\vartheta _1(z,q)}\right)^2=\frac{\vartheta _1^{\prime \prime}(z,q)}{\vartheta _1(z,q)}-\left(\frac{\vartheta _1^{\prime }(z,q)}{\vartheta _1(z,q)}\right)^{\prime}.
		\label{IdentityV}
	\end{equation}
	
	\item \underline{Identity(VI):}
	%\numparts
	\begin{align}
		\sum _{k=1}^{\infty } \frac{k^3}{1-q^k}&=\frac{k^3 \left(q^k-q^k+1\right)}{1-q^k} \nonumber\\
		&=\sum _{k=1}^{\infty } \frac{k^3 q^k}{1-q^k}+\sum _{k=1}^{\infty } k^3\nonumber\\
		&=\frac{1}{240} (E_4(\tau)-1)+\mathbf{\zeta}(-3).  \label{IdentityVI}  
	\end{align}
	%\endnumparts
	
	\item \underline{Identity(VII)} The following quotient of the second derivative of the elliptic $\theta_1$ function has the series representation~\cite{Zhao,SHEN1993299} given by        
	\begin{equation}
		\scalemath{0.7}{ \frac{\vartheta _1^{\prime \prime}\left(z,e^{-\frac{L \pi }{2 R}}\right)}{\vartheta _1\left(z,e^{-\frac{L \pi }{2 R}}\right)}=16 \sum _{n=1}^{\infty } \frac{e^{-\frac{(\pi  L) n}{2 R}} \cos (2 n z)}{\left(1-e^{-\frac{(\pi  L) n}{2 R}}\right)^2}+8 \sum _{n=1}^{\infty } \frac{n e^{-\frac{(\pi  L) n}{2 R}}}{1-e^{-\frac{(\pi  L) n}{2 R}}}-1 }.
		\label{IdentityVII}  
	\end{equation}.
	
	\item \underline{Identity(VIII)} The Jacobi elliptic theta function $\vartheta_4(z,q)$ is defined as an infinite product such that
	\begin{equation} 
		\scalemath{0.9}{ \vartheta _4(z,q)=\prod _{k=1}^{\infty } \left(1-q^{2 k}\right) \left(-2 q^{2 k-1} \cos (2 z)+q^{4 k-2}+1\right) }.
		\label{IdentityVIII}  
	\end{equation}
	
	\item \underline{Identity(IX)} The modular increment of the argument of the Jacobi elliptic $\vartheta_4$ is equivalent to
	\begin{equation} 
		\scalemath{0.9}{ \vartheta _4(z+i m \log (q),q)=(-1)^m q^{-m^2} e^{2 i m z} \vartheta _4(z,q); m\in \mathbf{Z} }
		\label{IdentityVIII}  
	\end{equation}
	
	\item \underline{Identity(X)} The products that appear in Eq.~\eqref{Prod} as a result of the algebraic manipulations of the sums are related to the elliptic $\theta_4$. For instance, the following shows explicitly this relation for the product:\\        
	\begin{align}
		\mathcal{P}_1& \scalemath{0.85}{ = -\prod _{k=1}^{\infty } \left(1-e^{-\frac{\pi  k L_{T}}{2 R}+\frac{\pi  L_{T}}{R}}\right) \left(1-e^{-\frac{\pi  k L_{T}}{2 R}-\frac{\pi  L_{T}}{2 R} }\right) }\\
		&\scalemath{0.85}{ = \prod _{k=1}^{\infty } \Big(e^{-\frac{\pi  k L_{T}}{2 R}+\frac{\pi  L_{T}}{2 R}  }+e^{-\frac{\pi  k L_{T}}{2 R}-\frac{\pi  L_{T}}{R} }-e^{-\frac{\pi  k L_{T}}{R}-\frac{\pi  L_{T}}{2 R}}-1 \Big) } \nonumber \\
		&\scalemath{0.7}{ =\prod _{k=1}^{\infty } \Bigg( e^{-\frac{1}{2} \left(-\frac{\pi  k L_{T}}{R}-\frac{\pi  L_{T}}{2 R}\right)} \left(e^{-\frac{\pi  k L_{T}}{2 R}+\frac{\pi  L_{T}}{2 R}  }+e^{-\frac{\pi  k L_{T}}{2 R}-\frac{\pi  L_{T}}{R}  }\right)/e^{-\frac{1}{2} \left(-\frac{\pi  k L_{T}}{R}-\frac{\pi  L_{T}}{2 R}\right)} } \nonumber \\
		&~~~~~~~~~~~~~~~~~~~~~~~~~~~~~~~~~~~~~~~~~~~~~~ \scalemath{0.85}{ -1-e^{-\frac{\pi  k L_{T}}{R}-\frac{\pi  L_{T}}{2 R}} \Bigg) } \nonumber \\
		&\scalemath{0.85}{ = \prod _{k=1}^{\infty } \Bigg( -1-e^{-\frac{\pi  k L_{T}}{R}}+2 e^{\frac{1}{2} \left(-\frac{\pi  k L_{T}}{R}-\frac{\pi  L_{T}}{2 R}\right)} \cosh \left(\frac{3 \pi  L_{T}}{4 R} \right) \Bigg) } \nonumber \\
		&\scalemath{0.85}{ =\frac{1}{G_{1}} \vartheta _4\left(\frac{i (3 L_{T} \pi )}{8 R},e^{-\frac{L_{T} \pi }{4 R}}\right) }; \nonumber 
		\label{IdentityIX:eq1}
	\end{align}
	with $G_{1}=\prod _{k=1}^{\infty } \left(1-e^{-\frac{\pi  k L_{T}}{2 R}}\right) $

	\item \underline{Identity(XI):}\\
	The logarithmic series 
	\begin{align}
		\scalemath{0.85}{ \sum _{j=0}^{\infty } \log \left(1-q^j\right) } & \scalemath{0.85}{ =\sum _{j=1}^{\infty } \log \left(1-q^j\right)-\underset{\epsilon \to 0}{\text{lim}} \sum _{k=1}^{\infty } \frac{q^{k \epsilon }}{k} } \\
		&\scalemath{0.85}{ =\sum _{j=1}^{\infty } \log \left(1-q^j\right)-\lim_{\epsilon \to 0} \, \frac{1}{2} (\mathbf{\zeta}(1-\epsilon )+\mathbf{\zeta}(\epsilon +1)) }, \nonumber
	\end{align}        
	owing to the harmonic series defined as,        
	\begin{equation}
		\zeta(k)=\sum_{n=1}^{\infty}\frac{1}{n^{k}},
	\end{equation}
	diverges at $k=1$. The principle value of $\mathbf{\zeta}(1)$ is $\gamma$, where $\gamma=0.5772$ is the  Euler-Mascheroni constant.\\

	\item \underline{Zeta Regularization}\\
	In the following,  we quote the partition of some basic properties of the zeta-regularized products~\cite{Yoshimoto2002}.
	Let us consider the convergence index, which is the least integer $h$ such that the series for $Z[h+ 1]$ converges absolutely.\\	
	One-partition property.\\	
	Let $\{\varphi^1 \}_k$ and $\{\varphi ^2\}_k$ be $\mathbf{\zeta}$- and let regularizable sequences $\{\varphi \}_k=\{\varphi^1\}_k \cup  \{\varphi ^2\}_k$ disjoint union then
	\begin{equation}
		\prod_{k} \varphi_k =\prod_{k} \{ \varphi^1 \}_k \prod_{k} \{\varphi ^2\}_k.
	\end{equation}
	Two-splitting property.\\
	If $\{\varphi _k\}$ is zeta-regularizable and $\prod_k a_k $ is convergent absolutely, then
	\begin{equation}
		\prod_{k} a_k \varphi_k= \prod_k a_k \prod_{k} \varphi_k.
	\end{equation}
	
\end{itemize}

%%%%%%%%%%%%%%%%%%%%%%%%%%%%%%%%%%%%%%%%%%%%%%%%%%%%%
%%%%%%%%%%%%%%%%%%%%%%%%%%%%%%%%%%%%%%%%%%%%%%%%%%%%%
\section{The free-string propagator and width on cylinder} 
\label{Appendix_II}

The Green's correlator can be expressed in terms of Jacobi theta functions
\begin{align}
	G(\mathbf{\zeta;\zeta'})& \scalemath{0.9}{ =  \frac{1}{\pi  \sigma }\sum _{n=1}^{\infty } \frac{1}{n \left(1-q^n\right)}\sin \left(\frac{\pi  n \zeta}{R}\right) \sin \left(\frac{\pi  n \zeta'}{R}\right) } \\		
	&~~~~~~~~~~~~~~ \scalemath{0.9}{ \times \left(q^n e^{\frac{\pi  n (\zeta_0-\zeta^{'}_{0})}{R}}+e^{-\frac{\pi n (\zeta_0-\zeta^{'}_{0})}{R}}\right) }, \nonumber
	\label{eq:GreenFuncCorrJacobiTheta}
\end{align}
where $q=e^{-\pi \frac{L_{T}}{R} }$  is the complementary nome.

Let us define the following auxiliary identities:
\begin{align}
	\kappa_{1}&= (\zeta_{0}^{'}-\zeta_0-i \zeta -i \zeta'), \\
	\kappa_{2}&= (\zeta_{0}^{'}-\zeta_0+i \zeta -i \zeta'), \\ 
	\kappa_{3}&= (\zeta_{0}^{'}-\zeta_0-i \zeta +i \zeta'), \\ 
	\kappa_{4}&= (\zeta_{0}^{'}-\zeta_0+i \zeta +i \zeta'). 
	\label{eq:kappa_1234}
\end{align}
With these identities, the Green's function from  Eq.\eqref{eq:GreenFuncCorrJacobiTheta} can be recast as
\begin{align}
	\label{eq:GreenFunc_4Terms}	
	&G(\kappa)=\frac{1}{4 \pi  \sigma } \sum _{n=1}^{\infty } \frac{1}{n \left(1-q^n\right)} \\
	&~~~~~~~ \scalemath{0.9}{ \times\Bigg[ -q^ne^{-\frac{\pi n \kappa_1}{R}}+q^n e^{-\frac{\pi n \kappa_2}{R}}+q^n e^{-\frac{\pi n \kappa_3}{R}}-q^n e^{-\frac{\pi n \kappa_4}{R}} } \nonumber \\
	&~~~~~~~~~~~~~~~~~~~~~~~~~~~~~~~~~ \scalemath{0.9}{ -e^{\frac{\pi n \kappa_1}{R}}+e^{\frac{\pi n \kappa_2}{R}}
		+e^{\frac{\pi n \kappa_3}{R}}-e^{\frac{\pi n \kappa_4}{R}} \Bigg]  }\nonumber \\
	&=\frac{1}{4 \pi  \sigma } \sum _{n=1}^{\infty } \frac{1}{n \left(1-q^n\right)}\nonumber \\
	&~~~~~~~\scalemath{0.9}{ \times\Bigg[ 
		-\left(q^ne^{-\frac{\pi n \kappa_1}{R}} + e^{\frac{\pi n \kappa_1}{R}} \right)
		+\left(q^n e^{-\frac{\pi n \kappa_2}{R}} + e^{\frac{\pi n \kappa_2}{R}} \right) } \nonumber \\
	&~~~~~~~~~~~~~~~~~~~~~~~~~~~~~~~~~~~~~~~ \scalemath{0.9}{ +\left(q^n e^{-\frac{\pi n \kappa_3}{R}} + e^{\frac{\pi n \kappa_3}{R}} \right) } \nonumber \\
	&~~~~~~~~~~~~~~~~~~~~~~~~~~~~~~~~~~~~~~~ \scalemath{0.9}{ -\left(q^n e^{-\frac{\pi n \kappa_4}{R}} + e^{\frac{\pi n \kappa_4}{R}}  \right) \Bigg] }.\nonumber
\end{align}
Considering the general term in the sum
\begin{align}
	S(\kappa)&=\sum _{n=1}^{\infty } \frac{1}{n \left(1-q^n\right)} \left( q^n e^{-\frac{(\pi  n) \kappa}{R}}+e^{\frac{(\pi  n) \kappa}{R}} \right) \label{eq:General_sumTerm} \\
	&=\sum _{n=1}^{\infty } \left(\sum _{k=1}^{\infty } \frac{q^{k n} e^{-\frac{(\pi  n) \kappa}{R}}}{n}+\sum _{k=0}^{\infty } \frac{q^{k n} e^{\frac{(\pi  n) \kappa}{R}}}{n}\right) \nonumber \\
	&=-\sum _{k=0}^{\infty } \log \left(1-q^k e^{\frac{\pi  \kappa}{R}}\right)-\sum _{k=1}^{\infty } \log \left(1-q^k e^{-\frac{\pi  \kappa}{R}}\right) \nonumber \\
	&=-\log \left[ \prod _{k=0}^{\infty } \left(1-q^k e^{\frac{\pi  \kappa}{R}}\right) \prod _{k=1}^{\infty } \left(1-q^k e^{-\frac{\pi  \kappa}{R}}\right)\right] \nonumber \\
	&=-\log \left[\prod _{k=1}^{\infty } \left(1-q^{k-1} e^{\frac{\pi  \kappa}{R}}\right) \left(1-q^k e^{-\frac{\pi  \kappa}{R}}\right)\right] \nonumber \\ 
	&=\log \left[\prod _{k=1}^{\infty } \left(1-q^k\right)\right] \nonumber \\
	& -\log \left[\prod _{k=1}^{\infty } \left(1-q^k\right) \left(1-q^{k-1} e^{\frac{\pi  \kappa}{R}}\right) \left(1-q^k e^{-\frac{\pi  \kappa}{R}}\right)\right].\nonumber 		
\end{align}

Using the results for the general term in $G(\kappa_{i})$ and with some algebraic manipulations, the propagator in Eq.\eqref{eq:GreenFunc_4Terms} becomes
\begin{equation}
	G(\kappa)= \frac{1}{4 \pi  \sigma} \log \Bigg[
	\frac{
		A_{\kappa_1} A_{\kappa_4} } { 
		A_{\kappa_2} A_{\kappa_3} }\Bigg],
	\label{eq:GreenFunc_kappa}
\end{equation}
where
\begin{align}
	& A_{\kappa_i} = \prod _{n=1}^{\infty } \left(1-q^n\right) \left(1-q^{n-1} e^{\frac{\pi  \kappa_i}{R}}\right) \left(1-q^n e^{-\frac{\pi  \kappa_i}{R}}\right),\\
	& i=1,...,4.  \nonumber
	\label{eq:GreenFunc_redefinition}
\end{align}

In terms of Jacobi theta functions, the propagator in Eq.\eqref{eq:GreenFunc_kappa} reads
\begin{align}
	& \scalemath{0.75}{ G(\kappa ) } \\
	& \scalemath{0.75}{ =\frac{1}{4 \pi  \sigma}\log \Bigg[ \frac{\vartheta _4\left(-\frac{(i \pi ) \kappa_1}{2 R}-\frac{i \pi  L_{T}}{4 R},e^{-\frac{\pi  L_{T}}{2 R}}\right) \vartheta _4\left(-\frac{(i \pi ) \kappa_4}{2 R}-\frac{i \pi  L_{T}}{4 R},e^{-\frac{\pi  L_{T}}{2 R}}\right)}{\vartheta _4\left(-\frac{(i \pi ) \kappa_2}{2 R}-\frac{i \pi  L_{T}}{4 R},e^{-\frac{\pi  L}{2 R}}\right) \vartheta _4\left(-\frac{(i \pi ) \kappa_3}{2 R}-\frac{i \pi  L_{T}}{4 R},e^{-\frac{\pi  L_{T}}{2 R}}\right)} \Bigg] }. \nonumber
\end{align}
Using the following transformation $\xi _i=\frac{ -i \pi}{2}  \left(\frac{\kappa_i}{R}+\frac{L_{T}}{2 R}\right)$ 
the propagator can be recast into the following closed form:
\begin{eqnarray}
	G(\xi)=\frac{1}{4 \pi \sigma } \log \left[\frac{\theta _4\left(\xi _1,\sqrt{q}\right)\theta _4\left(\xi _4,\sqrt{q}\right)}{\theta _4\left(\xi _2,\sqrt{q}\right)\theta _4\left(\xi _3,\sqrt{q}\right)}\right].
	\label{eq:GreenPropagator_compact}
\end{eqnarray}
%%%%%%%%%%%%%%%%%%%%%%%%%%%%%%%%%%%%%%%%%%%%%%%%

%%%%%%%%%%%%%%%%%%%%%%%%%%%%%%%%%%%%%%%%%%%%%%%%%%%%%
%%%%%%%%%%%%%%%%%%%%%%%%%%%%%%%%%%%%%%%%%%%%%%%%%%%%%
\section{Contribution of LO term of extrinsic curvature}
\label{Appendix_III}

The next term in Eq.\eqref{33} representing the LO perturbation due to the instantaneous string in terms of the corresponding Green's functions is
\begin{align}
	\label{eq:Appx_Integral}
	&W^2_{\rm{Ext_{(\ell o)}}}\\
	&\scalemath{0.85}{= -(D-2)\alpha_{r} \lim_{\substack{\epsilon' \to 0\\ \epsilon \to 0}} \int_{0}^{R}\int_{0}^{L_T} d\zeta' d\zeta'_{0}  \Big[ G(\mathbf{\zeta;\zeta'})\partial_{\alpha}^{2} \partial_{\alpha'}^{2} G(\mathbf{\zeta';\zeta}) }  \nonumber \\
	&~~~~~~~~~~~~~~~~~~~~~~~~~~~~~~~~~~~~~~ \scalemath{0.85}{ +\partial_{\alpha}^{2} G(\mathbf{\zeta;\zeta'}) \ \partial_{\alpha'}^{2} G(\mathbf{\zeta';\zeta}) \Big] }. \nonumber
\end{align}

The correlator $G (\zeta', \zeta^{\prime}_{0}; \zeta , \zeta_{0} )\equiv  \langle X(\zeta,\zeta_{0}) X(\zeta', \zeta^{\prime}_{0}) \rangle$ on a cylinder of size $RL_{T}$ with fixed boundary conditions at $\zeta^{1} = 0$ and $\zeta^{1} = R$ and periodic boundary conditions in $\zeta_{0}$ with period $L_{T}$ is given by 
\begin{align}
	G(\mathbf{\zeta;\zeta'})& =  \frac{1}{\pi  \sigma }\sum _{n=1}^{\infty } \frac{1}{n \left(1-q^n\right)}\sin \left(\frac{\pi  n \zeta}{R}\right) \sin \left(\frac{\pi  n \zeta'}{R}\right)  \\
	&~~~~~~~~~~~~~\times \left(q^n e^{\frac{\pi  n (\zeta_0-\zeta^{'}_{0})}{R}}+e^{-\frac{\pi n (\zeta_0-\zeta^{'}_{0})}{R}}\right). \nonumber
	\label{eq:Gauss}
\end{align}

We split the integral Eq.(\ref{eq:Appx_Integral}) into two parts, $W^2_{\rm{Ext_{(\ell o)}}}=-(I^{(1)}_{1}+I^{(1)}_{2})$, and integrating. The first part of Eq.(\ref{eq:Appx_Integral}) yields a vanishing series 
\begin{equation}
	\begin{split}
		\label{eq:GreenFunc_AppIII_I1}	
		I^{(1)}_{1} & \scalemath{0.85}{ = (D-2) \lim_{\substack{\epsilon' \to 0\\ \epsilon \to 0}} \int_{0}^{R} d\zeta'\int_{0}^{L_T} d\zeta^{'}_{0} \bigg[G(\mathbf{\zeta;\zeta'})\, \partial_{\alpha}\partial_{\alpha} \partial_{\alpha'}\partial_{\alpha'} G(\mathbf{\zeta';\zeta})\bigg] } \\
		& \scalemath{0.85}{ =(D-2) \lim_{\substack{\epsilon' \to 0\\ \epsilon \to 0}} \sum _{m=1}^{\infty } \sum _{n=1}^{\infty}\frac{2 \pi  n^3}{m R^3 \sigma ^2 \left(m^2-n^2\right) \left(q^m-1\right) \left(q^n-1\right)} } \\
		& \times\Bigg[ e^{-\frac{\pi  (m-n) \epsilon '}{R}} \left(q^m+e^{\frac{2 \pi  m \epsilon '}{R}}\right) \sin \left(\frac{\pi  m (\zeta+\epsilon )}{R}\right) \\
		& ~~~~~~~~~~~~~~~ \times \left(q^n e^{-\frac{2 \pi  n \epsilon '}{R}}+1\right) \sin \left(\frac{\pi  n (\zeta+\epsilon )}{R}\right)\Bigg] \\
		& \times \left[ m \cos (\pi  m) \sin (\pi  n)-n \sin (\pi  m) \cos (\pi  n) \right]\\
		& =0 
	\end{split}	
\end{equation}
since the integrand is proportional to the free equation of motion (\ref{EOM}).

The second part of Eq.(\ref{eq:Integral}),
%---------------------------------------------
\begin{equation}
	\begin{split}
		\label{eq:GreenFunc_AppIII_I2}	
		&I^{(1)}_{2}(R,L_{T}) \\
		&\scalemath{0.85}{ =(D-2) \lim_{\substack{\epsilon' \to 0\\ \epsilon \to 0}} \int_0^R\int_0^{L_{T}}  d\zeta_{0} d\zeta \bigg[ \partial_{\alpha} \partial_{\alpha} G(\mathbf{\zeta;\zeta'}) \, \partial_{\alpha'} \partial_{\alpha'} G(\mathbf{\zeta';\zeta}) \bigg] }  \\
		&\scalemath{0.85}{ =(D-2) \lim_{\substack{\epsilon' \to 0\\ \epsilon \to 0}}
			\int_{0}^{R} \int_{0}^{L_T} d\zeta'  d\zeta_{0} \sum _{m=1}^{\infty } \sum _{n=1}^{\infty}  \frac{\pi ^2 m n}{R^4 \sigma ^2 \left(q^m-1\right) \left(q^n-1\right)} } \\
		&~~~~~~~ \scalemath{0.8}{ \times \left(q^n+e^{\frac{2 \pi  n \epsilon' }{R}}\right) \left(e^{\frac{2 \pi  m (-L_{T}+2 \zeta_{0}+\epsilon' )}{R}}+q^m\right) e^{-\frac{\pi  (-L_{T} m+\epsilon'  (m+n)+2 m \zeta_{0})}{R}} }  \\
		&~~~~~~~ \scalemath{0.8}{ \times \sin \left(\frac{\pi  m \zeta}{R}\right) \sin \left(\frac{\pi  n \zeta}{R}\right) \sin \left(\frac{\pi  m (\zeta+\epsilon )}{R}\right) \sin \left(\frac{\pi  n (\zeta+\epsilon )}{R}\right) } \\
		&\scalemath{0.8}{ =(D-2) \lim_{\epsilon' \to 0 } \int _0^{L_{T}} d\zeta_{0} \sum _{m=1}^{\infty } \sum _{n=1}^{\infty} \frac{  2 \pi   \left(n^2-m^2\right) (m (-2 \pi  n)) \left(q^n+e^{\frac{2 \pi  n \epsilon' }{R}}\right) }{8 R^3 \sigma ^2 (m-n) (m+n) \left(q^m-1\right) \left(q^n-1\right) } } \\
		&~~~~~~~ \scalemath{0.85}{ \times e^{-\frac{\pi  (-L_{T} m+\epsilon'  (m+n)+2 m \zeta_{0})}{R}} \left(e^{\frac{2 \pi  m (-L_{T}+2 \zeta_{0}+\epsilon' )}{R}}+q^m\right) } \\
		&\scalemath{0.85}{ =\frac{-\pi (D-2)}{4 R^2 \sigma ^2} \sum _{m=1}^{\infty } \sum _{n=1}^{\infty}  \frac{ q^{-m} \left(q^m+1\right)^2  \left(n \left(q^n+1\right)\right)}{\left(q^n-1\right)} }.  
	\end{split}	
\end{equation}

Rewriting the two sums separately,
\begin{equation}
	I^{(1)}_{2}=\frac{-\pi (D-2)}{4 R^2 \sigma ^2}S_{1}S_{2},
\end{equation}
such that
\begin{align}
	S_{1}&= \sum _{m=1}^{\infty } (1+q^m)^2q^{-m} \\
	&=\sum _{m=1}^{\infty }(2+q^{-m}+q^{m}) \nonumber\\
	&=2\mathbf{\zeta}(0)-1\nonumber\\
	&=-2,\nonumber
\end{align}
with Riemann zeta function regularization of the divergent series, and
\begin{align}
	S_{2}&=\sum _{n=1}^{\infty }\frac{n(1+q^n)}{(-1+q^{n})}\\
	&=\sum _{n=1}^{\infty }\left(-n-2\frac{nq^n}{1-q^n} \right).\nonumber
\end{align}
Making use of the Eisenstein series definition
\begin{equation}
	E_{l}(\tau)=C_{l} \sum _{k=1}^{\infty} \frac{k^{l-1} q^{k}}{1-q^{k}}+1,
	\label{eq:AppIII_EisensteinSeries}  
\end{equation}
and zeta regularization 
\begin{align}
	\sum _{n=1}^{\infty } n &= \mathbf{\zeta}(-1),\\
	\frac{E_2 (\tau)-1}{24}&=-\sum _{n=1}^{\infty } \frac{n q^n}{1-q^{n}},\nonumber
\end{align}
where $C_{2 n}=\frac{-4 n}{B_{2 n}}$ with $B_{2 n}$ representing the corresponding Bernoulli numbers, the integral $I^{(1)}_{2}$ from Eq.\eqref{eq:GreenFunc_AppIII_I2} becomes 
\begin{equation}
	I^{(1)}_{2}=\frac{\pi(D-2) \alpha_{r} }{24 R^2 \sigma ^2 } ~E_2 \left( \frac{-L_{T}}{2 R}\right).
	\label{A11}  
\end{equation}
Using the modular transformation of the Eisenstein series, Eq.\eqref{eq:modularTransf_EisenSeries},
\begin{equation}
	E_2\left( \frac{-L_{T}}{2 R} \right) = \frac{4R^{2}}{L_{T}^{2}} E_{2}\left(- \frac{2 R} {L_{T}} \right)+\frac{12 R}{\pi L_{T}},
\end{equation}  
the limit $R\gg L_{T}$ of the integral \eqref{A11} becomes
\begin{align}
	I^{(1)}_{2}& = \frac{(D-2) \alpha_{r}}{\sigma ^2} \left(\frac{\pi}{6L_{T}^2} +\frac{1}{2RL_{T}}\right)\\ 
	&=\frac{(D-2)\alpha_{r}} {2 \sigma^2 R L_{T}} + C(L_{T}), \nonumber
\end{align}
where $C(L_{T})$ is a function of the thermodynamic scale $L_{T}$.

%%%%%%%%%%%%%%%%%%%%%%%%%%%%%%%%%%%%%%%%%%%%%%%%
%%%%%%%%%%%%%%%%%%%%%%%%%%%%%%%%%%%%%%%%%%%%%%%%
\section{Detailed Calculations of NLO Contractions}
\label{Appendix_IV} 
The second loop contractions Eqs.\eqref{tbyt},\eqref{AG},\eqref{BG},\eqref{CG}, \eqref{DG}, and \eqref{EG} involve 16 integrals over products  of Green’s functions, which are listed below.

The two nonvanishing integrals contributing to the pure NG string Eqs.\eqref{tbyt} and \eqref{AG}, are given by
\begin{align}		
	\label{eq:GreenString11_01}	
	&\scalemath{0.85}{ I^{(2)}_{1}(R,L_{T}) }\\
	&\scalemath{0.7}{ =\lim_{\substack{\epsilon' \to 0\\ \epsilon \to 0}} \int _0^R\int _0^{L_{T}}  d\zeta d\zeta_{0} \Big[ \partial_{\alpha'} G\left(\zeta,\zeta_{0};\zeta',\zeta'_{0}\right)  \partial_{\beta'}\partial_{\beta''} G\left(\zeta',\zeta'_{0};\zeta'',\zeta''_{0}\right)   \partial_{\alpha''} G\left(\zeta'',\zeta''_{0};\zeta,\zeta_{0}\right)  \Big] } \nonumber 
\end{align}
and  
\begin{align}		
	\label{eq:GreenString12_02}	
	&\scalemath{0.85}{I^{(2)}_{2}(R,L_{T}) } \\
	&\scalemath{0.7}{=\lim_{\substack{\epsilon' \to 0\\ \epsilon \to 0}} \int _0^R\int _0^{L_{T}}  d\zeta d\zeta_{0}  \Big [\partial_{\alpha'}G\left(\zeta,\zeta_{0};\zeta',\zeta^{\prime}_{0}\right)   \partial_{\alpha'} \partial_{\beta''} G\left(\zeta',\zeta^{\prime}_{0};\zeta'',\zeta_{0}''\right)  \partial_{\beta''} G\left(\zeta'',\zeta''_{0};\zeta,\zeta_{0} \right) \Big] } \nonumber 
\end{align}
which have been evaluated in detail in Ref.~\cite{Gliozzi:2010zt}.

In the following, we will evaluate each of the second-order Green's integrals using substitutions~\cite{Gliozzi:2010zt}, which is consistent with the cylindrical boundary condition, $\zeta'_{0}=L_{T}-\zeta_{0}$ and $ \zeta''=\zeta'+\epsilon ,\zeta''_{0}=\zeta'_{0}+\epsilon'$.

%%%%%%%%%%%%%%%%%%%%%%%%%%%%%%%%%%%%%%%%%%%%%%%%
%%%%%%%%%%%%%%%%%%%%%%%%%%%%%%%%%%%%%%%%%%%%%%%%
\newpage
\begin{itemize}
	\item \underline{\bf{Integral $I^{(2)}_{3}(R,L_{T})$}}: 
\end{itemize}	
\begin{align}
	\label{eq:GreenString_03}	
	&I^{(2)}_{3}(R,L_{T}) \\
	& \scalemath{0.65}{=\lim_{\substack{\epsilon' \to 0\\ \epsilon \to 0}} \int _0^R\int _0^{L_{T}}  d\zeta d\zeta_{0}  \Big[\partial_{\alpha} \partial_{\alpha'} G\left(\zeta,\zeta_{0};\zeta',\zeta_{0}^{\prime}\right) \partial_{\alpha} \partial_{\beta''} G\left(\zeta'',\zeta^{\prime\prime}_0;\zeta,\zeta_0 \right) \partial_{\beta''} \partial_{\alpha'} G\left(\zeta',\zeta_{0}^{\prime};\zeta'',\zeta^{\prime\prime}_{0}\right)\Big] } \nonumber\\		
	&\scalemath{0.8}{ = \lim_{\substack{\epsilon' \to 0\\ \epsilon \to 0}} \int _0^R\int _0^{L_{T}} d\zeta d\zeta_{0}
		\sum _{m=1}^{\infty }\sum _{n=1}^{\infty }\sum _{k=1}^{\infty } \frac{\pi  m}{R^4 \sigma ^3 \left(q^k-1\right) \left(q^m-1\right) \left(q^n-1\right)} }  \nonumber\\
	&~~~~~~~ \scalemath{0.8}{  \times \left(q^m e^{-\frac{2 \pi  m \epsilon '}{R}}-1\right) } \nonumber\\
	&~~~~~~~ \scalemath{0.8}{  \times \exp \left(-\frac{\pi  \left((k+m) (L_{T}-\zeta_{0})+\zeta_{0} (k+n)+(n-m) \left(L_{T}-\zeta_{0}+\epsilon '\right)\right)}{R}\right) } \nonumber\\			
	&~~~~~~~ \scalemath{0.8}{ \times \Bigg\{\bigg[\left(q^k e^{\frac{2 \pi  k \zeta_{0}}{R}}-e^{\frac{2 \pi  k (L_{T}-\zeta_{0})}{R}}\right) \left(q^n e^{\frac{2 \pi  n \left(L_{T}-\zeta_{0}+\epsilon '\right)}{R}}+e^{\frac{2 \pi  n \zeta_{0}}{R}}\right) } \nonumber\\
	&~~~~~~~ \scalemath{0.8}{ \times\sin \left(\frac{\pi  k \zeta}{R}\right) \sin \left(\frac{\pi  k (\zeta-R)}{R}\right)\sin \left(\frac{\pi  m (\zeta-R)}{R}\right) } \nonumber\\
	&~~~~~~~ \scalemath{0.8}{ \times\sin \left(\frac{\pi  n \zeta}{R}\right) \cos \left(\frac{\pi  m (-R+\zeta+\epsilon )}{R}\right) \cos \left(\frac{\pi  n (-R+\zeta+\epsilon )}{R}\right)\bigg] } \nonumber\\
	&~~~~~~~\scalemath{0.8}{  +\bigg[\left(e^{\frac{2 \pi  k (L_{T}-\zeta_{0})}{R}}+q^k e^{\frac{2 \pi  k \zeta_{0}}{R}}\right)\left(e^{\frac{2 \pi  n \zeta_{0}}{R}}-q^n e^{\frac{2 \pi  n \left(L_{T}-\zeta_{0}+\epsilon '\right)}{R}}\right) } \nonumber\\
	&~~~~~~~ \scalemath{0.8}{ \times\sin \left(\frac{\pi  k \zeta}{R}\right) \cos \left(\frac{\pi  k (\zeta-R)}{R}\right) \cos \left(\frac{\pi  m (\zeta-R)}{R}\right) } \nonumber\\
	&~~~~~~~\scalemath{0.8}{  \times\sin \left(\frac{\pi  n \zeta}{R}\right) \sin \left(\frac{\pi  m (-R+\zeta+\epsilon )}{R}\right) \sin \left(\frac{\pi  n (-R+\zeta+\epsilon )}{R}\right) \bigg] \Bigg\}  }.\nonumber
\end{align}

The integrals over the spatial coordinates $\zeta$ yield a vanishing result
\begin{align}
	\scalemath{0.65}{f_1(R)} & \scalemath{0.65}{ = \int_0^R \sin \left(\frac{\pi  k \zeta}{R}\right) \sin \left(\frac{2 \pi  m (R-\zeta)}{R}\right) \sin \left(\frac{\pi  n \zeta}{R}\right) \sin \left(\frac{\pi  (k-n) (R-\zeta)}{R}\right) \, d\zeta }\\
	& \scalemath{0.65}{ =0 }. \nonumber \\	
	\scalemath{0.65}{f_2(R)} & \scalemath{0.65}{ =\int_0^R \sin \left(\frac{\pi  k \zeta}{R}\right) \sin \left(\frac{2 \pi  m (R-\zeta)}{R}\right) \sin \left(\frac{\pi  n \zeta}{R}\right) \sin \left(\frac{\pi  (k+n) (R-\zeta)}{R}\right) \, d\zeta}\\
	& \scalemath{0.65}{ =0 }. \nonumber 
\end{align}

%%%%%%%%%%%%%%%%%%%%%%%%%%%%%%%%%%%%%%%%%%%%%%%%
%%%%%%%%%%%%%%%%%%%%%%%%%%%%%%%%%%%%%%%%%%%%%%%%
\newpage
\begin{itemize}
	\item \underline{\bf{Integral $I^{(2)}_{4}(R,L_{T})$}}:
\end{itemize}
\begin{align}
	\label{eq:GreenString_04}	
	&I^{(2)}_{4}(R,L_{T}) \\
	& \scalemath{0.7}{ =\lim_{\substack{\epsilon' \to 0\\ \epsilon \to 0}} \int _0^R\int _0^{L_{T}}  d\zeta d\zeta_{0} \Big[ G\left(\zeta,\zeta_{0};\zeta',\zeta^{\prime}_{0}\right)  \partial_{\beta} \partial_{\alpha''} G\left(\zeta'',\zeta^{\prime\prime}_{0};\zeta,\zeta_{0}\right) \partial_{\beta''}\partial_{\alpha '} G\left(\zeta',\zeta^{\prime}_{0};\zeta'',\zeta^{\prime\prime}_{0}\right)\Big] } \nonumber\\
	& \scalemath{0.7}{ =\int _0^R\int _0^{L_{T}}   d\zeta d\zeta_{0}
		\sum _{m=1}^{\infty }\sum _{n=1}^{\infty }\sum _{k=1}^{\infty } \frac{\pi^2 m n}{\pi  k \sigma  \left(1-q^k\right)   2R^4 \sigma^2 \left(q^n-1\right) } } \nonumber\\
	&~~~~~~~~~ \scalemath{0.7}{ \times \left(q^k e^{\frac{\pi  k (2 \zeta_0-L_{T})}{R}}+e^{-\frac{\pi  k (2 \zeta_0-L_{T})}{R}}\right) \left( e^{-\frac{\pi  n (\zeta_0+2 \zeta_0)}{R}} \left(q^n e^{\frac{2 \pi  L_{T} n}{R}}-e^{\frac{4 \pi  n \zeta_0}{R}}\right)\right) } \nonumber\\
	&~~~~~~~~~ \scalemath{0.7}{ \times\sin \left(\frac{\pi  k \zeta}{R}\right) \sin \left(\frac{\pi  k (\zeta-R)}{R}\right) \sin \left(\frac{2 \pi  m (R-\zeta)}{R}\right) \sin \left(\frac{\pi  n (R-2 \zeta)}{R}\right) }.\nonumber
\end{align}
The integral $I^{(2)}_{4}(R,L_{T})$ has a zero value since the spatial integrals vanishes,  
\begin{align}
	\scalemath{0.9}{ f(R)} &\scalemath{0.9}{=\int_0^R  \, d\zeta  \sin \left(\frac{\pi  k \zeta}{R}\right) \sin \left(\frac{\pi  k (\zeta-R)}{R}\right) }\\
	&~~~~~~~~~ \scalemath{0.9}{ \times \sin \left(\frac{2 \pi  m (R-\zeta)}{R}\right) \sin \left(\frac{\pi  n (R-2 \zeta)}{R}\right) } \nonumber\\
	&=0.\nonumber
\end{align}

%%%%%%%%%%%%%%%%%%%%%%%%%%%%%%%%%%%%%%%%%%%%%%%%
%%%%%%%%%%%%%%%%%%%%%%%%%%%%%%%%%%%%%%%%%%%%%%%%
\begin{itemize}
	\item \underline{\bf{Integral $I^{(2)}_{5}(R,L_{T})$}}:
\end{itemize}

The off-diagonal terms of the integral  
\begin{align}
	\label{eq:GreenString_05}	
	&I^{(2)}_{5}(R,L_{T})\\
	&\scalemath{0.7}{ =\lim_{\substack{\epsilon' \to 0\\ \epsilon \to 0}} \int _0^R\int _0^{L_{T}}  d\zeta d\zeta_{0} ~[\partial_{\alpha'}^{2} G\left(\zeta,\zeta_{0};\zeta',\zeta^{\prime}_{0}\right)  \partial_{\beta}^{2} G\left(\zeta'',\zeta^{\prime\prime}_{0};\zeta,\zeta_{0} \right) \partial_{\alpha''} \partial_{\alpha'} G\left(\zeta',\zeta^{\prime}_{0};\zeta'',\zeta^{\prime\prime}_{0} \right)] } \nonumber\\   
	&\scalemath{0.7}{ =\frac{\pi^{3}}{8R^{5}\sigma^{3}} \int _0^{L_{T}}d\zeta_{0} \sum_{m=1}^{\infty}\sum_{n=1}^{\infty}\sum_{k=1}^{\infty} \frac{k}{(q^{k}-1)}\frac{m}{(q^{m}-1)}\frac{n}{(q^{n}-1)} } \nonumber\\
	&~~~~~~~\scalemath{0.7}{ \times\Bigg\{
		\left(q^{k}+1 \right)
		\left(q^{m}e^{-\frac{\pi m(-2\zeta_{0})}{R}} + e^{\frac{\pi m(-2\zeta_{0})}{R}} \right)
		\left(q^{-n}e^{-\frac{\pi n(2\zeta_{0}-L_{T})}{R}} + q^{2n}e^{\frac{\pi n2\zeta_{0}}{R}} \right) } \nonumber\\
	&~~~~~~~ \scalemath{0.7}{ -
		\left(q^{k}+1 \right)
		\left(q^{m}e^{-\frac{\pi m(-2\zeta_{0})}{R}} + e^{\frac{\pi m(-2\zeta_{0})}{R}} \right)
		\left(q^{-n}e^{-\frac{\pi n(2\zeta_{0}-L_{T})}{R}} + q^{2n}e^{\frac{\pi n2\zeta_{0}}{R}} \right)\Bigg\}  } \nonumber\\
	&=0.\nonumber
\end{align}
Form the equation of motion \eqref{EOM}. It is obvious that the integral vanishes for all terms $\alpha$ and $\beta$.

%%%%%%%%%%%%%%%%%%%%%%%%%%%%%%%%%%%%%%%%%%%%%%%%
%%%%%%%%%%%%%%%%%%%%%%%%%%%%%%%%%%%%%%%%%%%%%%%%
\newpage
\begin{itemize}
	\item \underline{\bf{Integral $I^{(2)}_{6}(R,L_{T})$}}:
\end{itemize}

Similarly, the integrals over all terms $\alpha$ and $\beta$ of 
\begin{align}
	\label{eq:GreenString_06}
	&I^{(2)}_{6}(R,L_{T}) \\
	&\scalemath{0.7}{=\lim_{\substack{\epsilon' \to 0\\ \epsilon \to 0}} \int _0^R\int _0^{L_{T}}  d\zeta d\zeta_{0}  \Big[ G\left(\zeta,\zeta_{0};\zeta',\zeta^{\prime}_{0}\right)   \partial_{\alpha'} \partial_{\alpha ''}  G\left(\zeta'',\zeta^{\prime\prime}_{0};\zeta,\zeta_{0} \right)   \partial_{\alpha ''}^{2} \partial_{\beta }^2 G\left(\zeta',\zeta^{\prime}_{0};\zeta'',\zeta^{\prime\prime}_{0} \right) \Big] },  \nonumber
\end{align}
are trivial from the equation of motion \eqref{EOM}.

%%%%%%%%%%%%%%%%%%%%%%%%%%%%%%%%%%%%%%%%%%%%%%%%
%%%%%%%%%%%%%%%%%%%%%%%%%%%%%%%%%%%%%%%%%%%%%%%%
\begin{itemize}
	\item \underline{\bf{Integral   $I^{(2)}_{7}(R,L_{T})$}}:
\end{itemize}

The sums of the diagonal $\alpha=\beta$ elements and off-diagonal elements $\alpha \neq \beta$ of the following integral are identical from the equation of motion; considering $\alpha \neq \beta$ the Green's integral is
\begin{align}
	\label{eq:GreenString_07}	
	& I^{(2)}_{7}(R,L_{T}) \\
	& \scalemath{0.7}{ =\lim_{\substack{\epsilon' \to 0\\ \epsilon \to 0}} \int _0^R\int _0^{L_{T}}  d\zeta d\zeta_{0} \Big[ \partial_\beta G\left(\zeta,\zeta_{0};\zeta',\zeta^{\prime}_{0}\right)  \partial_{\beta''} G\left(\zeta'',\zeta^{\prime\prime}_{0};\zeta',\zeta^{\prime}_{0}\right)  \partial_{\alpha ''}^{2} \partial_{\alpha '}^{2}  G\left(\zeta',\zeta^{\prime}_{0};\zeta'',\zeta^{\prime\prime}_{0}\right) \Big] } \nonumber\\
	& \scalemath{0.7}{ =\lim_{\epsilon^{\prime}\rightarrow0} \sum_{k=1}^{\infty}
		\frac{\pi^{3} k^{3}}{R^{6}\sigma^{3}(1-q^{k})}\left(q^{k}e^{-\frac{\pi k\epsilon^{\prime} }{R}} + e^{\frac{\pi k\epsilon^{\prime} }{R}} \right) \int_{0}^{L_{T}}d\zeta_{0}\sum_{m=1}^{\infty}\sum_{n=1}^{\infty}
		\Bigg\{\bigg[
		\frac{1}{(q^{m}-1)}\frac{1}{(q^{n}-1)} } \nonumber\\
	&~~~~~ \scalemath{0.8}{ \times\left(q^{m} e^{\frac{\pi m (L_{T}-2\zeta_{0}+\epsilon^{\prime})}{R}} + e^{-\frac{\pi m (L_{T}-2\zeta_{0}+\epsilon^{\prime})}{R}} \right)} \nonumber\\		
	&~~~~~ \scalemath{0.8}{ \times\left(q^{n} e^{\frac{\pi n (2\zeta_{0}-L_{T})}{R}} + e^{-\frac{\pi n (2\zeta_{0}-L_{T})}{R}} \right) } \nonumber\\
	&~~~~~ \scalemath{0.8}{ \times
		\lim_{\epsilon\rightarrow0} \int_{0}^{R} d\zeta
		{\rm{sin}}\left(\frac{\pi m \zeta}{R}\right)
		{\rm{cos}}\left(\frac{\pi n \zeta}{R}\right)
		{\rm{sin}}\left(\frac{\pi k (\zeta-R)}{R}\right)
	} \nonumber\\
	&~~~~~ \scalemath{0.8}{ \times 
		{\rm{sin}}\left(\frac{\pi n (\zeta-R)}{R}\right)
		{\rm{sin}}\left(\frac{\pi k (\zeta-R+\epsilon)}{R}\right)
		{\rm{cos}}\left(\frac{\pi n (\zeta-R+\epsilon)}{R}\right)\bigg] } \nonumber\\
	&~~~~~ \scalemath{0.8}{ +\Bigg[
		\left(q^{n}e^{\frac{\pi n (2\zeta_{0}-L_{T})}{R}} - e^{-\frac{\pi n (2\zeta_{0}-L_{T})}{R}} \right) } \nonumber\\
	&~~~~~ \scalemath{0.8}{ + \left(q^{m}e^{\frac{\pi m (L_{T}-2\zeta_{0}+\epsilon^{\prime})}{R}} - e^{-\frac{\pi m (L_{T}-2\zeta_{0}+\epsilon^{\prime})}{R}} \right) } \nonumber\\
	&~~~~~ \scalemath{0.8}{ \times
		\lim_{\epsilon\rightarrow0}\int_{0}^{R}d\zeta 
		{\rm{sin}}\left(\frac{\pi m \zeta}{R}\right)
		{\rm{cos}}\left(\frac{\pi n \zeta}{R}\right)
		{\rm{sin}}\left(\frac{\pi k (\zeta-R)}{R}\right)
	} \nonumber\\
	&~~~~~~~\scalemath{0.8}{ \times
		{\rm{sin}}\left(\frac{\pi n (\zeta-R)}{R}\right)
		{\rm{sin}}\left(\frac{\pi k (\zeta-R+\epsilon)}{R}\right)
		{\rm{sin}}\left(\frac{\pi m (\zeta-R+\epsilon)}{R}\right)\Bigg]\Bigg\} }. \nonumber
\end{align}

After integrating over $\zeta^{1}$, the integral assumes the form
\begin{equation}
	\begin{split}
		I^{(2)}_{7}(R,L_{T})&  \scalemath{0.8}{ =\frac{\pi ^3}{8 R^5 \sigma ^3}\sum _{m=1}^{\infty } \frac{1}{q^m-1}\sum _{n=1}^{\infty } \frac{1}{q^n-1} \sum _{k=1}^{\infty } \frac{k^3}{1-q^k} } \\
		&~~~ \scalemath{0.8}{ \times \int _0^{L_{T}} \underset{\epsilon '\to 0}{\text{lim}} \left(q^{-n} e^{-\frac{\pi  n (2 \zeta_{0})}{R}}-q^{2 n} e^{\frac{\pi  n (2 \zeta_{0})}{R}}\right) }\\
		&~~~ \scalemath{0.8}{ ~~~~~\times \left(e^{\frac{\pi  m \left(\epsilon '-2 \zeta_{0}\right)}{R}}-q^m e^{-\frac{\pi  m \left(\epsilon '-2 \zeta_{0}\right)}{R}}\right) d\zeta_{0} }.
		\label{eq:GreenString_07time}
	\end{split}
\end{equation}

The direct integration of $\zeta^{0}$ yields a mixed series over the running $n$ and $m$, which complicates the algebraic manipulations. In the following, we proceed first by encapsulating the three sums appearing in the above equation \eqref{eq:GreenString_07time} in closed form in terms of the standard functions before integration. The sum over $n$ reads
\begin{equation}
	\mathcal{S}_1=\sum _{n=1}^{\infty } \frac{q^{2 n} e^{\frac{\pi  n (2 \zeta^{0})}{R}}-q^{-n} e^{-\frac{\pi  n (2 \zeta^{0})}{R}}}{q^n-1}.   
	\label{Sum:S1}
\end{equation}
Expanding the denominator
\begin{equation}
	\scalemath{0.8}{ \mathcal{S}_1=\sum_{j=0}^{\infty}\sum _{n=1}^{\infty } -e^{-\frac{\pi  j L_{T} n}{R}} \left(e^{\frac{\pi  L_{T} n}{R}} \left(-e^{-\frac{\pi  n (2 \zeta^{0})}{R}}\right)+e^{-\frac{2 \pi  L_{T} n}{R}} e^{\frac{\pi  n (2 \zeta^{0})}{R}}\right) }, 
\end{equation}
and summing over $n$ 
\begin{equation}
	\scalemath{0.9}{ \mathcal{S}_1=\sum_{j=0}^{\infty} \frac{1}{1-e^{\frac{\pi  ((j+2) L_{T}-2 \zeta^{0})}{R}}}+ \sum_{j=0}^{\infty} \frac{1}{e^{\frac{\pi  ((j-1) L_{T}+2 \zeta^{0})}{R}}-1} }.
	\label{Sum:S1_2}
\end{equation}
After rearranging the terms, in hyperbolic form of the sum becomes
\begin{align}
	\mathcal{S}_1& =-\frac{1}{2} \sum _{j=0}^{\infty } \coth \left(-\frac{\pi  j L_{T}}{2 R}+\frac{\pi  L_{T}}{2 R}-\frac{\pi  \zeta^{0}}{R}\right) \\
	&~~~~ -\frac{1}{2} \sum _{j=0}^{\infty } \coth \left(\frac{\pi  j L_{T}}{2 R}+\frac{\pi  L_{T}}{R}-\frac{\pi  \zeta^{0}}{R}\right). \nonumber
\end{align}
Redefining the sum index over the first term in Eq.(\ref{Sum:S1_2}) $j\to j-1$ followed by sign inversion $j \to -j$, yields
\begin{align}
	\scalemath{0.8}{ \mathcal{S}_1 } & \scalemath{0.7}{ =-\frac{1}{2} \sum _{j=1}^{\infty } \coth \left(-\frac{\pi  j L_{T}}{2 R}+\frac{\pi  L_{T}}{R}-\frac{\pi  \zeta^{0}}{R}\right) 
		-\frac{1}{2} \sum _{j=0}^{\infty } \coth \left(\frac{\pi  j L_{T}}{2 R}+\frac{\pi  L_{T}}{R}-\frac{\pi  \zeta^{0}}{R}\right) } \nonumber \\
	& \scalemath{0.7}{ =-\frac{1}{2} \sum _{j=-1}^{-\infty } \coth \left(\frac{\pi  j L_{T}}{2 R}+\frac{\pi  L_{T}}{R}-\frac{\pi  \zeta^{0}}{R}\right) -\frac{1}{2} \sum _{j=0}^{\infty } \coth \left(\frac{\pi  j L_{T}}{2 R}+\frac{\pi  L_{T}}{R}-\frac{\pi  \zeta^{0}}{R}\right) } \nonumber \\
	& \scalemath{0.7}{ =-\frac{1}{2} \sum _{j=-\infty }^{\infty } \coth \left(\frac{\pi  j L_{T}}{2 R}+\frac{\pi  L_{T}}{R}-\frac{\pi  \zeta^{0}}{R}\right) }.
\end{align}

With the use of Eqs.\eqref{IdentityII} and \eqref{IdentityIII} of Identities (II) and (III), the sum over the hyperbolic function with running $j$ can be expressed in closed form with the use of the Jacobi elliptic $\vartheta_{1}$ function such that
\begin{equation}
	\mathcal{S}_1=-\frac{i}{2} \left(\frac{\vartheta_{1}'(\frac{-\pi i \zeta^{0}}{R}+\frac{\pi i L_{T}}{R},e^{-\frac{L_{T}\pi}{2R}})}{\vartheta_{1}(\frac{-\pi i \zeta^{0}}{R}+\frac{\pi i L_{T}}{R},e^{-\frac{L_{T}\pi}{2R}})}\right).
\end{equation}

From Eq.\eqref{IdentityIV:eq2} of Identity (IV) this reduces to
\begin{equation}
	\mathcal{S}_1=\frac{i}{2} \left(\frac{\vartheta_{1}'(\frac{\pi i \zeta^{0}}{R},e^{-\frac{L_{T}\pi}{2R}})}{\vartheta_{1}(\frac{\pi i \zeta^{0}}{R},e^{-\frac{L_{T}\pi}{2R}})}\right)-2.
\end{equation}

The second product is a series over the running index $m$,   
\begin{equation}
	\mathcal{S}_2=\sum _{m=1}^{\infty } \frac{e^{\frac{\pi  m (-2 \zeta^{0})}{R}}-q^{m} e^{-\frac{\pi  m (-2 \zeta^{0})}{R}}}{q^m-1}.
	\label{Sum:S2} 
\end{equation}
This series can be manipulated in a similar way to Eq.\eqref{Sum:S1}. Expanding the denominator of Eq.~\eqref{Sum:S2} such that
\begin{equation}
	\scalemath{0.8}{ \mathcal{S}_2=  -\sum_{l=0}^{\infty}\sum _{m=1}^{\infty } e^{\frac{ -(\pi m l) L_{T}}{R}} \left(e^{\frac{-\pi  L_{T} m}{R}} \left(-e^{-\frac{\pi  m (-2 \zeta^{0})}{R}}\right)+e^{\frac{\pi  m (-2 \zeta^{0})}{R}}\right) },
\end{equation}
and summing over the running $m$ yields
\begin{equation}
	\begin{split}
		\mathcal{S}_2 & \scalemath{0.8}{ = -\sum _{j=0}^{\infty}\frac{e^{-\frac{\pi  ((j+1) L_{T}-2 \zeta_{0})}{2 R}}}{e^{-\frac{\pi  ((j+1) L_{T}-2 \zeta_{0})}{2 R}}-e^{\frac{\pi  ((j+1) L_{T}-2 \zeta_{0})}{2 R}}}+\frac{e^{-\frac{\pi  (j L_{T}+2 \zeta_{0})}{2 R}}}{e^{\frac{\pi  (j L_{T}+2 \zeta_{0})}{2 R}}-e^{-\frac{\pi  (j L_{T}+2 \zeta_{0})}{2 R}}} } \\
		& \scalemath{0.65}{ =-\sum _{j=0}^{\infty }\left[ \frac{1}{2} \left(\coth \left(\frac{\pi  (j L_{T}+2 \zeta_0)}{2 R}\right)-1\right)+\frac{1}{2} \left(1-\coth \left(\frac{\pi  (j L_{T}+L_{T}-2 \zeta_{0})}{2 R}\right)\right)\right] }.\\
	\end{split}
\end{equation}
Similar to Eq.~\eqref{Sum:S1}, we accordingly redefined the summation index running over the first term,
\begin{equation}
	\begin{split}
		\mathcal{S}_{2}& \scalemath{0.7}{ =-\sum_{j=1}^{\infty } \frac{1}{2} \left[\coth \left(\frac{\pi ((j-1) L_{T}+2 \zeta_{0})}{2 R}\right)\right] -\sum_{j=0}^{\infty } \frac{1}{2} \left[\coth \left(\frac{\pi  (-j L_{T}-L_{T}+2 \zeta_{0})}{2 R}\right)\right] } \\
		&\scalemath{0.7}{ =-\sum_{j=-1}^{-\infty } \frac{1}{2} \left[\coth \left(\frac{\pi  ((-j-1) L_{T}+2 \zeta_{0})}{2 R}\right)\right] -\sum_{j=0}^{\infty } \frac{1}{2} \left[\coth \left(\frac{\pi ((-j-1)L_{T}+2 \zeta_{0})}{2 R}\right)\right] } \\		
		&\scalemath{0.7}{ =-\sum _{j=-\infty}^{\infty} \frac{1}{2} \coth \left[\frac{\pi (-(j+1)L_{T}+2 \zeta_{0})}{2 R}\right] }.
	\end{split}
\end{equation}

The series sum $S_{2}$ is over hyperbolic functions, which in closed form would read 
\begin{equation}
	\begin{split} 
		\mathcal{S}_2&=\frac{i}{2}\left(\frac{\vartheta_{1}'(\frac{-\pi i \zeta^{0}}{R}+\frac{\pi i L_{T}}{2R},e^{-\frac{L_{T}\pi}{2R}})}{\vartheta_{1}(\frac{-\pi i \zeta^{0}}{R}+\frac{\pi i L_{T}}{2R},e^{-\frac{L_{T}\pi}{2R}})}\right) \\
		&=\frac{-i}{2}\left(\frac{\vartheta_{1}'(\frac{-\pi i \zeta^{0}}{R} ,e^{-\frac{L_{T}\pi}{2R}})}{\vartheta_{1}(\frac{-\pi i \zeta^{0}}{R} ,e^{-\frac{L_{T}\pi}{2R}})}\right) +1
	\end{split}   
\end{equation}

Making use of Eq.\eqref{IdentityVI}-Identity (VI), the third sum in Eq.\eqref{eq:GreenString_07time} can be regularized with the $\mathbf{\zeta}$ function such that 
\begin{equation}
	\begin{split}
		\mathcal{S}_3&=\sum _{k=1}^{\infty } \frac{k^3}{1-q^k}\\
		&=\frac{1}{240}(E_4(\tau )-1)+\mathbf{\zeta}(-3).
	\end{split}
\end{equation}

The integral over $\zeta^{0}$ then reads
\begin{equation}
	\begin{split}
		&\scalemath{0.7}{ I^{(2)}_{7}(R,L_{T})=\frac{\pi ^3}{8 R^5 \sigma ^3} \frac{(E_4(\tau )+1)}{240} } \\
		&\scalemath{0.7}{ \times \int_{0}^{L_T} d\zeta^{0} \left\{\frac{1}{4}\left(\frac{\vartheta_{1}'(\frac{-i \pi \zeta^{0}}{R},e^{-\frac{L_{T}\pi}{2R}})}{\vartheta_{1}(\frac{-i \pi \zeta^{0}}{R},e^{-\frac{L_{T}\pi}{2R}}))}\right)^{2}+\frac{3i}{2}\left(\frac{\vartheta_{1}'(\frac{-i \pi \zeta^{0}}{R},e^{-\frac{L_{T}\pi}{2R}})}{\vartheta_{1}(\frac{-i \pi \zeta^{0}}{R},e^{-\frac{L_{T}\pi}{2R}}))}\right)-2\right\}  }.
	\end{split}
	\label{I7}
\end{equation}

The square term of the logarithmic derivative [Eq.\eqref{IdentityVII}-Identity (VII)] in the integrand can be recast into the form  
\begin{equation}
	\scalemath{0.9}{ \mathcal{I}_a=\frac{1}{4} \int_{0}^{L_T} d\zeta^{0} \left[ \frac{\vartheta _1^{\prime \prime}(\frac{\pi \zeta^{0}}{R},e^{-\frac{L_{T}\pi}{2R}} )}{\vartheta _1(\frac{\pi \zeta^{0}}{R},e^{-\frac{L_{T}\pi}{2R}} )}-\left(\frac{\vartheta _1^{\prime }( \frac{\pi \zeta^{0}}{R},e^{-\frac{L_{T}\pi}{2R}} )}{\vartheta _1(\frac{\pi \zeta^{0}}{R},e^{-\frac{L_{T}\pi}{2R}}  )}\right)^{'}\right] }.
	\label{Ia}  
\end{equation}

Let $z=\frac{\pi \zeta^{0}}{R}$ and $q_{1}=e^{\frac{-\pi L_{T}}{2 R}}$, employing the series form for the quotient of double derivative of elliptic $\vartheta_1$ Eq.\eqref{IdentityVI} of Identity (VI), the integral Eq.~\eqref{Ia} accordingly decomposes to  
\begin{equation}
	\begin{split}
		&\scalemath{0.8}{ \int_{0}^{L_{T}}  \frac{\vartheta _1^{\prime\prime}(z,q_{1})}{\vartheta _1(z,q)} dt } \\
		&\scalemath{0.8}{ =\int_{0}^{L_T} d\zeta^{0} \left(16 \sum _{n=1}^{\infty } \frac{e^{-\frac{\pi L_{T}n}{2 R}} \cos \left(2 i \pi  n \left(-\frac{\zeta^{0}}{R}\right)\right)}{\left(1-e^{-\frac{\pi L_{T}n}{2 R}}\right)^2}+8 \sum _{n=1}^{\infty } \frac{n e^{-\frac{\pi L_{T}n}{2 R}}}{1-e^{-\frac{\pi L_{T}n}{2 R}}}-1\right) }\\
		& \scalemath{0.8}{ = \frac{8 R}{\pi} \sum _{n=1}^{\infty } \frac{e^{\frac{\pi  L_{T} n}{2 R}} \sinh \left(\frac{2 \pi  L_{T} n}{R}\right)}{n \left(e^{\frac{\pi  L_{T} n}{2 R}}-1\right)^2}+L_{T} \left(8 \sum _{n=1}^{\infty } \frac{n e^{-\frac{\pi  L_{T} n}{2 R}}}{1-e^{-\frac{\pi  L_{T} n}{2 R}}}-1\right) } \\
		&\scalemath{0.8}{ = \frac{8 R}{\pi} \sum _{n=1}^{\infty } \frac{e^{\frac{\pi  L_{T} n}{2 R}} \sinh \left(\frac{2 \pi  L_{T} n}{R}\right)}{n \left(e^{\frac{\pi  L_{T} n}{2 R}}-1\right)^2}+L_{T} \left(\frac{8}{24} \left(1-E_2\left(i\frac{\pi  L_{T}}{4 R}\right)\right)-1\right) }.
	\end{split}
	\label{theta:pp}
\end{equation}

Expanding the denominator and summing over the index $n$, the series of the first term in Eq.~\eqref{theta:pp} would read
\begin{equation}
	\label{Prod}	
	\begin{split}
		&\scalemath{0.8}{ \frac{8 R}{\pi} \sum _{n=1}^{\infty } \frac{e^{\frac{\pi  L_{T} n}{2 R}} \sinh \left(\frac{2 \pi  L_{T} n}{R}\right)}{n \left(e^{\frac{\pi  L_{T} n}{2 R}}-1\right)^2} }\\
		&\scalemath{0.8}{ =\frac{4 R}{\pi} \sum _{n=1}^{\infty } \frac{e^{-\frac{3 \pi  L_{T} n}{2 R}} \left(e^{\frac{\pi  L_{T} n}{2 R}}+1\right) \left(e^{\frac{\pi  L_{T} n}{R}}+1\right) \left(e^{\frac{2 \pi  L_{T} n}{R}}+1\right)}{n \left(e^{\frac{\pi  L_{T} n}{2 R}}-1\right)} } \\
		&\scalemath{0.8}{ =\frac{4 R}{\pi} \sum _{n=1}^{\infty } \frac{1}{n}e^{-\frac{3 \pi  L_{T} n}{2 R}} \left(e^{\frac{\pi  L_{T} n}{2 R}}+1\right) \left(e^{\frac{\pi  L_{T} n}{R}}+1\right)\left(e^{\frac{2 \pi  L_{T} n}{R}}+1\right)\sum _{k=1}^{\infty } e^{-\frac{k (\pi  L_{T} n)}{2 R}} } \\
		& =\frac{-4 R}{\pi} \Bigg\{ \log \left[ \prod_{k=1}^{\infty} \left(1-e^{-\frac{\pi  (k-4) L_{T}}{2 R}}\right)\left(1-e^{-\frac{\pi  (k+3) L_{T}}{2 R}}\right)\right]  \\ 
		&~~~~~~~~~~ +\log \left[\prod_{k=1}^{\infty}  \left(1-e^{-\frac{\pi  (k-3) L_{T}}{2 R}}\right)\left(1-e^{-\frac{\pi  (k+2) L_{T}}{2 R}}\right)\right]\\
		&~~~~~~~~~~ +\log \left[\prod_{k=1}^{\infty}  \left(1-e^{-\frac{\pi  (k-2) L_{T}}{2 R}}\right)\left(1-e^{-\frac{\pi  (k+1) L_{T}}{2 R}}\right)\right]\\
		&~~~~~~~~~~ +\log \left[\prod_{k=1}^{\infty}                    \left(1-e^{-\frac{\pi  (k-1) L_{T}}{2 R}}\right)\left(1-e^{-\frac{\pi  k L_{T}}{2 R}}\right)\right]\\
		&~~~~~~~~~~ -4 \log \left[\prod _{k=1}^{\infty } \left(1-e^{-\frac{\pi  L_{T}}{2 R}}\right)\right] \Bigg\}.
	\end{split}
\end{equation}

Employing Identities (VII-X) which defines the Jacobi elliptic theta $\vartheta_4$ and its relevant properties, the series sum becomes
\begin{equation}
	\begin{split}
		& \frac{8 R}{\pi} \sum _{n=1}^{\infty } \frac{e^{\frac{\pi  L_{T} n}{2 R}} \sinh \left(\frac{2 \pi  L_{T} n}{R}\right)}{n \left(e^{\frac{\pi  L_{T} n}{2 R}}-1\right)^2}\\
		&\scalemath{0.8}{ =\frac{4 R}{\pi} \Bigg\{ 4 \log \left[\prod _{k=1}^{\infty } \left(1-e^{-\frac{\pi  L_{T}}{2 R}}\right)\right]-\log \left[\vartheta _4\left(-\frac{i (L_{T} \pi )}{8 R},e^{-\frac{L_{T} \pi }{4 R}}\right)\right] } \\
		&\scalemath{0.8}{ -\log \left[\vartheta _4\left(-\frac{i (3 L_{T} \pi )}{8 R},e^{-\frac{L_{T} \pi }{4 R}}\right)\right]-\log \left[\vartheta _4\left(-\frac{i (5 L_{T} \pi )}{8 R},e^{-\frac{L_{T} \pi }{4 R}}\right)\right] }\\
		&\scalemath{0.8}{ -\log \left[\vartheta _4\left(-\frac{i (7 L_{T} \pi )}{8 R},e^{-\frac{L_{T} \pi }{4 R}}\right)\right]\Bigg\} }\\
		&\scalemath{0.8}{ =\frac{4 R}{\pi}\Bigg[ 4\log \left( e^{\frac{-L_{T}\pi}{48 R}} \eta(\frac{i L_{T}}{4 R})\right)+\gamma-\log\left(e^{\frac{-9 \pi L_{T}}{4 R}} \right) \Bigg] }.
	\end{split}
	\label{sum_n}
\end{equation}
Owing to the vanishing of the four elliptic theta functions involved in Eq.~\eqref{sum_n}, this produces a logarithmic divergent term. Taking into account the sum representation of Jacobi elliptic functions ~\cite{WhittakerWatson,abramowitz}, the whole series within the logarithm can conveniently~\cite{Nesterenko:1997ku} be taken proportional to the Euler-Mascheroni constant $\gamma$ as defined in Identity (XI). The last log term in Eq.\eqref{sum_n} is a phase term arising from the transformation of the elliptic theta functions (Identity (IX)) chosen to ensure the convergence in the limit $L_{T} \rightarrow \infty$.

Substituting the above expression into Eq.~\eqref{theta:pp}, the integral is then   
\begin{equation}
	\begin{split}
		\scalemath{0.9}{ \int_{0}^{L_{T}}  \frac{\vartheta _1^{\prime\prime}(z,q)}{\vartheta _1(z,q)} d\zeta^{0} }&\scalemath{0.8}{ =\frac{4 R}{\pi} \left[\log \left( e^{\frac{-L_{T}\pi}{48 R}} \eta(\frac{i L_{T}}{4 R})\right)+\gamma + \log\left(e^{\frac{9 L_{T}\pi}{4 R}} \right)\right]}\\
		&\scalemath{0.8}{~~~~ +L_{T} \left(\frac{8}{24} \left(1-E_2\left(i\frac{\pi  L_{T}}{4 R}\right)\right)-1\right) }.
	\end{split}
\end{equation}

Evaluating the limits of the second part of integral ~\eqref{Ia} yields
\begin{equation}
	\begin{split}
		\scalemath{0.75}{ \frac{R}{ i \pi } \int_{0}^{L_{T}}\left[ \frac{\partial}{\partial z} \frac{\vartheta _1^{\prime }\left(z,e^{-\frac{L_{T} \pi }{2 R}}\right)}{\vartheta _1\left(z,e^{-\frac{L_{T} \pi }{2 R}}\right)}\right]\, dz } & \scalemath{0.75}{ = \frac{R}{ i \pi } \left[ \frac{\vartheta _1^{\prime }\left(\frac{i (\pi  L_{T})}{R},e^{-\frac{L_{T} \pi }{2 R}}\right)}{\vartheta _1\left(\frac{i (\pi  L_{T})}{R},e^{-\frac{L_{T} \pi }{2 R}}\right)}-\frac{\vartheta _1^{\prime }\left(0,e^{-\frac{L_{T} \pi }{2 R}}\right)}{\vartheta _1\left(0,e^{-\frac{L_{T} \pi }{2 R}}\right)}\right] }\\
		&\scalemath{0.75}{ =  -\frac{4 R}{\pi} }.
	\end{split}
\end{equation}

The second part of the integral in Eq.~\eqref{I7} is a straightforward to evaluate; substituting the calculus of Eq.~\eqref{Ia} back into the integral \eqref{I7}, the total integral over $\zeta^{0}$ then reads
\begin{equation}
	\begin{split}
		I^{(2)}_{7}(R,L_{T})&\scalemath{0.8}{ =\frac{\pi ^3}{8 R^5 \sigma ^3} \frac{(E_4(\tau )+1)}{240} } \\
		& \scalemath{0.8}{ \times \Bigg[\frac{4 R}{\pi} \left(\log \left( \eta \left(\frac{i L_{T}}{4 R}\right)\right)+\gamma \right) -\frac{L_{T}}{12} \left(2-E_{2}\left(i\frac{\pi L_{T}}{4 R} \right) \right)\Bigg] }. 
	\end{split}
	\label{I7}
\end{equation}

%%%%%%%%%%%%%%%%%%%%%%%%%%%%%%%%%%%%%%%%%%%%%%%%
%%%%%%%%%%%%%%%%%%%%%%%%%%%%%%%%%%%%%%%%%%%%%%%%
\newpage
\begin{itemize}
	\item \underline{\bf{Integral $I^{(2)}_{8}(R,L_{T})$}}
\end{itemize}

\begin{equation}
	\begin{split}
		&I^{(2)}_{8}(R,L_{T})\\
		&\scalemath{0.7}{ =\lim_{\substack{\epsilon' \to 0\\
					\epsilon \to 0}} \int _0^R\int _0^{L_{T}}  d\zeta d\zeta_{0} ~[ G\left(\zeta,\zeta_{0};\zeta',\zeta^{\prime}_{0}\right) \partial_{\beta} \partial_{\alpha''}^{2}  G\left(\zeta'',\zeta^{\prime\prime}_{0};\zeta',\zeta^{\prime}_{0}\right) \partial_{\beta''} \partial_{\alpha'}^{2} G\left(\zeta',\zeta^{\prime}_{0};\zeta'',\zeta^{\prime\prime}_{0}\right)]} \\ 
		&\scalemath{0.7}{ =\lim_{\substack{\epsilon \to 0}} \int_{0}^{R}\int_{0}^{L_{T}}d\zeta d\zeta_{0}\sum_{m=1}^{\infty}\sum_{n=1}^{\infty}\sum_{k=1}^{\infty}
			\frac{\pi^{3} k^{2}}{R^{6}\sigma^{3}(q^{k}-1)}\frac{m^{2}(-1)^{m+n+1}}{(q^{m}-1)}\frac{1}{n(q^{n}-1)}
			e^{-\frac{\pi (L_{T}-2\zeta_{0})(m+n)}{R}} }\\
		&~~~~~~\scalemath{0.7}{ \times \left( e^{\frac{2\pi n(L_{T}-2\zeta_{0})}{R}} + q^{n} \right)
			\left(q^{k+m} e^{\frac{2\pi m (L_{T}-2\zeta_{0})}{R}} + 1 \right)
			{\rm{cos}}\left(\frac{\pi(k(\zeta+\epsilon)- m\zeta)}{R}\right) }\\
		&~~~~~~\scalemath{0.8}{  + \left(q^{k} + q^{m}e^{\frac{2\pi m(L_{T}-2\zeta_{0})}{R}} \right)
			{\rm{cos}}\left(\frac{\pi(k(\zeta+\epsilon)+ m\zeta)}{R}\right) }\\
		&~~~~~~~~~~\scalemath{0.8}{ \times {\rm{sin}}^{2}\left(\frac{\pi k \zeta}{R}\right)
			{\rm{sin}}^{2}\left(\frac{\pi n \zeta}{R}\right)
			{\rm{sin}}^{2}\left(\frac{\pi m (\zeta+\epsilon)}{R}\right) }\\
		&\scalemath{0.8}{ =\int_{0}^{L_{T}} d\zeta_{0}\sum_{m=1}^{\infty}\sum_{n=1}^{\infty}\sum_{k=1}^{\infty}
			\frac{\pi^{3} k^{2}}{8R^{5}\sigma^{3}}\frac{m^{2}(-1)^{m+n}}{(q^{m}-1)}\frac{1}{n(q^{n}-1)} }
		\\
		&\scalemath{0.8}{ \times{\rm{exp}}\left( -\frac{\pi (L_{T}-2\zeta_{0})(m+n)}{R} \right)
			\left(q^{m} e^{\frac{2\pi m(L_{T}-2\zeta_{0})}{R}} - 1 \right)
			\left(e^{\frac{2\pi n (L_{T}-2\zeta_{0})}{R}} + q^{n} \right) }\\
		&\scalemath{0.8}{ =-\frac{\pi^{3}}{8R^{5}\sigma^{3}} \int_{0}^{L_{T}} d\zeta_{0}   \sum_{m=1}^{\infty}
			\frac{m^{2}(-1)^{m}\left(q^{m} e^{\frac{2\pi m(L_{T}-2\zeta_{0})}{R}} - 1 \right)}{(q^{m}-1)}} \\
		&~~~~~~~~~~~~~~~~~\scalemath{0.8}{ \times \sum_{n=1}^{\infty} \frac{(-1)^{n}\left(q^{n}e^{-\frac{\pi n (L_{T}-2\zeta_{0})}{R}} + e^{\frac{\pi n (L_{T}-2\zeta_{0})}{R}} \right)}{n(q^{n}-1)}
			\sum_{k=1}^{\infty}k^{2} }\\
		&=0.
		\label{eq:GreenString_08}
	\end{split}
\end{equation} 

The integral vanishes by virtue of the regularization of the divergent sum $\sum_{k=1}^{\infty} k^{2}=\mathbf{\zeta}(-2)=0$.\\

%%%%%%%%%%%%%%%%%%%%%%%%%%%%%%%%%%%%%%%%%%%%%%%%
%%%%%%%%%%%%%%%%%%%%%%%%%%%%%%%%%%%%%%%%%%%%%%%%
\begin{itemize}
	\item \underline{\bf{Integral $I^{(2)}_{9}(R,L_{T})$}}
\end{itemize}

Considering the sums of integrals over the off-diagonal elements $\alpha \neq \beta$ of the integral 
\begin{equation}
	\begin{split}
		&I^{(2)}_{9}(R,L_{T})\\
		&\scalemath{0.7}{ =\lim_{\substack{\epsilon' \to 0\\ \epsilon \to 0}} \int _0^R\int _0^{L_{T}}  d\zeta d\zeta_{0}  ~[G\left(\zeta,\zeta_{0};\zeta',\zeta^{\prime}_{0}\right) \partial_{\beta} \partial_{\beta''}^{2}  G\left(\zeta'',\zeta^{\prime\prime}_{0};\zeta',\zeta^{\prime}_{0}\right) \partial_{\alpha'} \partial^2_{\alpha''\beta''} G\left(\zeta',\zeta^{\prime}_{0};\zeta'',\zeta^{\prime\prime}_{0}\right)]}\\
		&\scalemath{0.9}{=\lim_{\substack{\epsilon' \to 0}} \int _0^{L_{T}} d\zeta_{0} \sum_{m=1}^{\infty}\sum_{n=1}^{\infty}\sum_{k=1}^{\infty} \frac{\pi^{3} k^{2}}{8R^{5}\sigma^{3}(q^{k}-1)}\frac{m^{2}}{(q^{m}-1)}\frac{1}{n(q^{n}-1)} }\\
		&~~~~~~\scalemath{0.9}{ \times \left(1 - q^{k} \right) \left(q^{m}e^{\frac{\pi m(L_{T}-2\zeta_{0}+\epsilon^{\prime})}{R}} - e^{-\frac{\pi m(L_{T}-2\zeta_{0}+\epsilon^{\prime})}{R}}\right) }\\
		&~~~~~~~~~~~~~~~~~\scalemath{0.9}{ \times	\left(q^{n}e^{\frac{\pi n(2\zeta_{0}-L_{T})}{R}} + e^{-\frac{\pi n(2\zeta_{0}-L_{T})}{R}}\right) }\\
		&=0,
		\label{eq:GreenString_09}
	\end{split}
\end{equation}
which yields since the $k$-indexed summation $\sum _{k=1}^{\infty } k^2=0$ vanishes with the employment of the $\zeta$ function regularization.

The sum of the integrals corresponding to the diagonal components assumes the form
\begin{equation}
	\begin{split}
		\label{eq:GreenString_09Dia}
		&I^{(2)}_{9}(R,L_{T})\\
		&\scalemath{0.7}{ =\lim_{\substack{\epsilon' \to 0\\
					\epsilon \to 0}} \int _0^R\int _0^{L_{T}}  d\zeta d\zeta_{0}  ~[G\left(\zeta,\zeta_{0};\zeta',\zeta^{\prime}_{0}\right) \partial_\alpha \partial_{\alpha''}^{2}  G\left(\zeta'',\zeta^{\prime\prime}_{0};\zeta,\zeta_{0}\right) \partial_{\alpha''}^{2}  \partial_{\alpha'} G\left(\zeta',\zeta^{\prime}_{0};\zeta'',\zeta^{\prime\prime}_{0}\right)] }\\
		&\scalemath{0.7}{ =\lim_{\epsilon^{\prime}\rightarrow0}\int_{0}^{L_{T}} d\zeta_{0}\sum_{m=1}^{\infty}\sum_{n=1}^{\infty}\sum_{k=1}^{\infty}
			\frac{\pi^{3} (-1)^{k+n}}{8R^{5}\sigma^{3}k(q^{k}-1)}\frac{m^{2}}{(q^{m}-1)}\frac{n^{2}}{(q^{n}-1)}} \\
		&~~~~~~\scalemath{0.7}{ \times{\rm{exp}}\left(-\frac{\pi\left(k(L_{T}-2\zeta_{0})+n(L_{T}+2\zeta_{0})+(m+n)\epsilon^{\prime}) \right) }{R}\right) }\\
		&~~~~~~\scalemath{0.7}{ \times 
			\left(e^{\frac{2\pi k (L_{T}-2\zeta_{0})}{R}} + q^{k} \right)
			\left(e^{\frac{2\pi m \epsilon^{\prime}}{R}} - q^{m} \right)
			\left(e^{\frac{4\pi n \zeta_{0}}{R}} -q^{n} e^{\frac{2\pi n (L_{T}-\epsilon^{\prime})}{R}} \right) }\\
		&\scalemath{0.7}{ =\frac{\pi^{2}}{16R^{4}\sigma^{3}} \sum_{m=1}^{\infty}\sum_{n=1}^{\infty}\sum_{k=1}^{\infty}
			\frac{nm^{2}(q^{n}+1)}{(q^{n}-1)}
			\frac{(-1)^{k+n}}{k(q^{k}-1)}
			\frac{(q^{k+n}-1)}{q^{k+n}}
			(q^{k}+1)
			(q^{k}-q^{n}) }\\
		&=0.
	\end{split}
\end{equation}
The sum of the integral is zero owing to the zero value of the regularized divergent sum  
\begin{equation}
	\sum_{m=1}^{\infty}m^{2}={\zeta}(-2)=0.
\end{equation}~~~\\

%%%%%%%%%%%%%%%%%%%%%%%%%%%%%%%%%%%%%%%%%%%%%%%%
%%%%%%%%%%%%%%%%%%%%%%%%%%%%%%%%%%%%%%%%%%%%%%%%
\newpage
\begin{itemize}
	\item \underline{\bf{Integral; $I^{(2)}_{10}(R,L_{T})$}}
\end{itemize}

The sum over the diagonal components, $\alpha=\beta$, of the integral is given by  
\begin{equation}
	\begin{split}
		&I^{(2)}_{10}(R,L_{T})\\
		&\scalemath{0.7}{ =\lim_{\substack{\epsilon' \to 0\\
					\epsilon \to 0}}  \int _0^R\int _0^{L_{T}}  d\zeta d\zeta_{0}  ~[\partial_{\alpha' } G\left( \mathbf{\zeta;\zeta'} \right) \partial_{\alpha'}\partial_{\beta''}^{2}G\left(\zeta',\zeta^{\prime}_{0};\zeta'',\zeta^{\prime\prime}_{0}\right) \partial_{\beta}^{2} G\left(\zeta'',\zeta^{\prime\prime}_{0};\zeta',\zeta^{\prime}_{0}\right)
			] }\\
		&\scalemath{0.7}{ =\lim_{\substack{\epsilon' \to 0\\
					\epsilon \to 0}}\int _0^{L_{T}}  d\zeta_{0}  \sum_{m=1}^{\infty}\sum_{n=1}^{\infty}\sum_{k=1}^{\infty} \frac{\pi ^3 k^2 m }{R^6 \sigma ^3 \left(1-q^k\right) \left(1-q^m\right) \left(1-q^n\right)} }\\						
		&\scalemath{0.7}{ \times \Bigg(\left(q^k e^{-\frac{\pi  k \epsilon '}{R}}+e^{\frac{\pi  k \epsilon '}{R}}\right) \left(q^n e^{\frac{\pi  n (2 \zeta_{0}-L_{T})}{R}}+e^{-\frac{\pi  n (2 \zeta_{0}-L_{T})}{R}}\right) } \\						
		&~~~~~~~~~~~~~~~~~~~~~~~~~~~\scalemath{0.8}{ \times \left(q^m e^{\frac{\pi  m \left(L_{T}-2 \zeta_{0}+\epsilon '\right)}{R}}+e^{-\frac{\pi  m \left(L_{T}-2 \zeta_{0}+\epsilon '\right)}{R}}\right)\Bigg) }\\						
		&\scalemath{0.8}{  \times \int_{0}^{R} d\zeta \Bigg[ \cos \left(\frac{\pi  k (\zeta-R)}{R}\right) \sin \left(\frac{\pi n \zeta}{R}\right) }\\			
		&~~~~~~~~~~~~~~~~~~~~~~~~~~~ \scalemath{0.8}{\times \cos \left(\frac{\pi  n (\zeta-R)}{R}\right) \sin \left(\frac{\pi k (-R+\zeta+\epsilon )}{R}\right) }\\				
		&~~~~~~~~~~~~~~~~~~~~~~~~~~~ \scalemath{0.8}{ \times \sin \left(\frac{\pi  m \zeta}{R}\right) \sin \left(\frac{\pi  m (-R+\zeta+\epsilon )}{R}\right)\Bigg]  } \\
		&\scalemath{0.8}{  + \frac{1}{\pi^3 k m n \sigma ^3 \left(1-q^k\right) \left(1-q^m\right) \left(1-q^n\right)}  
			\left(\frac{\pi^3 k^3 q^k e^{-\frac{\pi  k \epsilon '}{R}}}{R^3}-\frac{\pi ^3 k^3 e^{\frac{\pi  k \epsilon '}{R}}}{R^3}\right) }\\  
		&\scalemath{0.8}{ \times \left(\frac{\pi^2 m^2 q^m e^{\frac{\pi  m \left(L_{T}-2 t+\epsilon '\right)}{R}}}{R^2}+\frac{\pi^2 m^2 e^{-\frac{\pi  m \left(L_{T}-2 t+\epsilon '\right)}{R}}}{R^2}\right) } \\
		&\scalemath{0.7}{ \times \left(\frac{\pi  n e^{-\frac{\pi  n (2 \zeta_{0}-L_{T})}{R}}}{R}-\frac{\pi  n q^n e^{\frac{\pi  n (2 \zeta_{0}-L_{T})}{R}}}{R}\right) }\\
		&\scalemath{0.8}{ \times \int_{0}^{R} d\zeta \Bigg[ \sin\left(\frac{\pi  k (\zeta-R)}{R}\right) \sin \left(\frac{\pi  m \zeta}{R}\right) \sin \left(\frac{\pi  n \zeta}{R}\right) }\\
		&\scalemath{0.8}{ \times \sin \left(\frac{\pi  n (\zeta-R)}{R}\right) \sin \left(\frac{\pi  k (-R+\zeta+\epsilon )}{R}\right) \sin \left(\frac{\pi  m (-R+\zeta+\epsilon )}{R}\right)\Bigg]} .
		\label{eq:GreenString_10}
	\end{split}
\end{equation}
Evaluating the integrals over $\zeta$ yields
\begin{equation}
	\begin{split}
		f_{1}(R)&\scalemath{0.8}{ =\int_0^R \bigg[\cos \left(\frac{\pi k (\zeta-R)}{R}\right) \sin \left(\frac{\pi m \zeta}{R}\right) \sin \left(\frac{\pi n \zeta}{R}\right) \cos \left(\frac{\pi  n (\zeta-R)}{R}\right) }\\
		&\scalemath{0.8}{\times \sin \left(\frac{\pi k(-R+\zeta+\epsilon )}{R}\right) \sin \left(\frac{\pi  m (-R+\zeta+\epsilon )}{R}\right)\bigg] \,d\zeta }\\
		&=0. 
	\end{split} 
\end{equation}
and
\begin{equation}
	\begin{split}
		f_{2}(R)&\scalemath{0.8}{ =\int_0^R \bigg[\sin \left(\frac{\pi k (\zeta-R)}{R}\right) \sin \left(\frac{\pi m \zeta}{R}\right) \sin \left(\frac{\pi n \zeta}{R}\right) \sin \left(\frac{\pi  n (\zeta-R)}{R}\right)} \\
		&\scalemath{0.8}{\times \sin \left(\frac{\pi k (-R+\zeta+\epsilon )}{R}\right) \sin \left(\frac{\pi  m (-R+\zeta+\epsilon )}{R}\right)\bigg] \, d\zeta }\\
		&\scalemath{0.8}{ =\frac{1}{8} R \sec (\pi  (m+n)) \cos \left(\frac{\pi k \epsilon }{R}\right) \cos \left(\frac{\pi m \epsilon }{R}\right)} ,
	\end{split} 
\end{equation}
that is,
\begin{equation}
	\begin{split}
		I_{10}^{(2)}&\scalemath{0.8}{ = \sum_{m=1}^{\infty}\sum_{n=1}^{\infty} \sum_{k=1}^{\infty} \frac{\pi ^3 k^2 m\left(q^m e^{\frac{2 \pi  m (L_{T}-2 \zeta_{0})}{R}}+1\right) \left(e^{\frac{2 \pi  n (L_{T}-2 \zeta_{0})}{R}}-q^n\right)}{R^6 \sigma ^3 \left(q^k-1\right) \left(q^m-1\right) \left(q^n-1\right)} }\\  
		&\scalemath{0.8}{ \times \frac{e^{-\frac{\pi  (L_{T}-2 \zeta_{0}) (m+n)}{R}} \left(1-q^k\right)}{R^6 \sigma ^3 \left(q^k-1\right) \left(q^m-1\right) \left(q^n-1\right)} } \\   
		&\scalemath{0.8}{  =\sum_{m=1}^{\infty}\sum_{n=1}^{\infty} \sum_{k=1}^{\infty} \frac{\pi ^2 k^2 m e^{-\frac{\pi  L_{T} (m+n)}{R}} }{2 R^5 \sigma ^3 (m-n) (m+n) \left(q^m-1\right) \left(q^n-1\right)} }\\
		&\scalemath{0.7}{  \times \Bigg( (m+n) e^{\frac{2 \pi  L_{T} m}{R}} \left(q^{m+n}-1\right)-(m+n) e^{\frac{2 \pi  L_{T} n}{R}} \left(q^{m+n}-1\right) }\\
		&\scalemath{0.8}{  +(m-n) \left(q^m-q^n\right) \left(-e^{\frac{2 \pi  L_{T} (m+n)}{R}}\right)+(m-n) \left(q^m-q^n\right)\Bigg) }.
	\end{split}
\end{equation}

Similar to integrals over Green's functions in Eqs.\eqref{eq:GreenString_09}, the integral vanishes by the virtue of $\zeta$ function regularization of the $\sum _{k=1}^{\infty} k^2=0$.

On the other hand, the sums over the off-diagonal elements $\alpha \neq \beta$ are given by the integral
\begin{equation}
	\begin{split}
		&I^{(2)}_{10}(R,L_{T})\\
		&\scalemath{0.7}{ =\lim_{\substack{\epsilon' \to 0\\
					\epsilon \to 0}}  \int _0^R\int _0^{L_{T}}  d\zeta d\zeta_{0}  ~[\partial_{\alpha' } G\left( \zeta,\zeta_{0};\zeta',\zeta^{\prime}_{0} \right) \partial_{\alpha'}\partial_{\beta''}^{2}G\left(\zeta',\zeta^{\prime}_{0};\zeta'',\zeta^{\prime\prime}_{0}\right) \partial_{\beta}^{2} G\left(\zeta'',\zeta^{\prime\prime}_{0};\zeta',\zeta^{\prime}_{0}\right)
			]} \\
		&\scalemath{0.7}{ =\lim_{\substack{\epsilon' \to 0\\
					\epsilon \to 0}}\int _0^{L_{T}}  d\zeta_{0}  \sum_{m=1}^{\infty}\sum_{n=1}^{\infty}\sum_{k=1}^{\infty}\frac{1}{\pi m R^2 \sigma ^3 \left(1-q^k\right) \left(1-q^m\right) \left(1-q^n\right)} } \\
		&\scalemath{0.7}{ \times \left(\frac{\pi k^2 q^k e^{-\frac{\pi  k \epsilon '}{R}}}{R^2}+ \frac{\pi ^2 k^2 e^{\frac{\pi  k \epsilon '}{R}}}{R^2}\right)  \left(q^n e^{\frac{\pi  n (2 \zeta_{0}-L_{T})}{R}}+e^{-\frac{\pi  n (2 \zeta_{0}-L_{T})}{R}}\right) } \\
		&\scalemath{0.7}{ \times \left(\frac{\pi ^2 m^2 q^m e^{\frac{\pi  m \left(L_{T}-2 \zeta_{0}+\epsilon '\right)}{R}}}{R^2}+\frac{\pi ^2 m^2 e^{-\frac{\pi  m \left(L_{T}-2 \zeta_{0}+\epsilon '\right)}{R}}}{R^2}\right) }\\
		&\scalemath{0.7}{ \times \int_{0}^{R} d\zeta\Bigg[ \cos\left(\frac{\pi  k (\zeta-R)}{R}\right) \sin \left(\frac{\pi  m \zeta}{R}\right) \sin \left(\frac{\pi  n \zeta}{R}\right) \cos \left(\frac{\pi  n (\zeta-R)}{R}\right) } \\
		&~~~~~~~~~~~~~~~~~~~~~~~~~~~ \scalemath{0.7}{ \times \sin \left(\frac{\pi  k (-R+\zeta+\epsilon )}{R}\right) \sin \left(\frac{\pi  m (-R+\zeta+\epsilon )}{R}\right)\Bigg] } \\
		&\scalemath{0.7}{ +\frac{k m \pi}{(n (1 - q^k) (1 - q^m) (1 - q^n) R^4 \sigma^3)}\left(\frac{\pi  k q^k e^{-\frac{\pi  k \epsilon '}{R}}}{R}-\frac{\pi  k e^{\frac{\pi  k \epsilon '}{R}}}{R}\right) } \\
		&\scalemath{0.7}{ \times \left(\frac{\pi  n e^{-\frac{\pi  n (2 \zeta_{0}-L_{T})}{R}}}{R}-\frac{\pi  n q^n e^{\frac{\pi  n (2 \zeta_{0}-L_{T})}{R}}}{R}\right)\left(q^m e^{\frac{\pi  m \left(L_{T}-2 \zeta_{0}+\epsilon '\right)}{R}}+e^{-\frac{\pi  m \left(L_{T}-2 \zeta_{0}+\epsilon '\right)}{R}}\right) }\\
		&\scalemath{0.7}{ \times \int_{0}^{R} d\zeta \Bigg[ \sin \left(\frac{\pi  k (\zeta-R)}{R}\right) \sin \left(\frac{\pi  m \zeta}{R}\right) \sin \left(\frac{\pi  n \zeta}{R}\right) \sin \left(\frac{\pi  n (\zeta-R)}{R}\right) } \\
		&\scalemath{0.7}{ \times \sin \left(\frac{\pi  k (-R+\zeta+\epsilon )}{R}\right) \sin \left(\frac{\pi  m (-R+\zeta+\epsilon )}{R}\right)\Bigg] }. \\
	\end{split}  
\end{equation}

Each of the two integrals involving $\zeta$ in the above equation evaluates to zero 
\begin{equation}
	\begin{split}
		f_{1}(R)&\scalemath{0.8}{ =\int_0^R \bigg[\cos \left(\frac{\pi k (\zeta-R)}{R}\right) \sin \left(\frac{\pi m \zeta}{R}\right) \sin \left(\frac{\pi n \zeta}{R}\right) \cos \left(\frac{\pi  n (\zeta-R)}{R}\right) }\\
		&\scalemath{0.8}{ \times \sin \left(\frac{\pi  k (-R+\zeta+\epsilon )}{R}\right) \sin \left(\frac{\pi  m (-R+\zeta+\epsilon )}{R}\right)\bigg] \, d\zeta } \\
		&=0, 
	\end{split} 
\end{equation}
and
\begin{equation}
	\begin{split}
		f_{2}(R)& \scalemath{0.8}{ = \int_0^R \bigg[\sin \left(\frac{\pi  k (\zeta-R)}{R}\right) \sin \left(\frac{\pi  m \zeta}{R}\right) \sin \left(\frac{\pi  n \zeta}{R}\right) \sin \left(\frac{\pi  n (\zeta-R)}{R}\right) } \\
		&\scalemath{0.8}{ \times \sin \left(\frac{\pi  k (-R+\zeta+\epsilon )}{R}\right) \sin \left(\frac{\pi  m (-R+\zeta+\epsilon )}{R}\right)\bigg] \, d\zeta }\\
		& =(-1)^{m+n}.
	\end{split}
\end{equation}

After integrating with respect to $\zeta_0$, we obtain the series sum
\begin{equation}
	\begin{split}
		I^{(2)}_{10}&\scalemath{0.9}{ =\sum_{m=1}^{\infty}\sum_{n=1}^{\infty}\sum_{k=1}^{\infty}\frac{(-1)^{m+n}  \pi ^2 k^2 m  e^{-\frac{\pi  L_{T} (m+n)}{R}}}{2 R^5 \sigma ^3 (m-n) (m+n) \left(q^m-1\right) \left(q^n-1\right)} }\\
		&\scalemath{0.9}{ \times \Bigg((m+n) e^{\frac{2 \pi  L_{T} m}{R}} \left(q^{m+n}-1\right) -(m+n)e^{\frac{2 \pi  L_{T} n}{R}} \left(q^{m+n}-1\right) }\\
		&\scalemath{0.9}{  +(m-n) \left(q^m-q^n\right) \left(-e^{\frac{2 \pi  L_{T} (m+n)}{R}}\right)+(m-n) \left(q^m-q^n\right)\Bigg) }\\
		&=0,
	\end{split} 
\end{equation}
which vanishes by virtue of the $\zeta$ function regularization of the sum $\sum _{k=1}^{\infty} k^2=0$.

%%%%%%%%%%%%%%%%%%%%%%%%%%%%%%%%%%%%%%%%%%%%%%%%
%%%%%%%%%%%%%%%%%%%%%%%%%%%%%%%%%%%%%%%%%%%%%%%%
\newpage
\begin{itemize}
	\item \underline{\bf{Integral $I^{(2)}_{11}(R,L_{T})$}}\\
\end{itemize}

The diagonal elements of $I^{(2)}_{11}(R,L_{T})$ coincide with that of $I^{(2)}_{10}(R,L_{T})$ given by Eq.~\eqref{eq:GreenString_10}. The following evaluates the off-diagonal components of the integral:
\begin{equation}
	\begin{split}
		& I^{(2)}_{11}(R,L_{T})\\
		&\scalemath{0.8}{ = \lim_{\substack{\epsilon' \to 0\\ \epsilon \to 0}} \int _0^R\int _0^{L_{T}}  d\zeta d\zeta_{0} \Big[ \partial_{\alpha'}^2 G\left(\zeta,\zeta_{0};\zeta',\zeta^{\prime}_{0} \right) } \\
		&~~~~~~~~~~~~~~~~~~~~~~~ \scalemath{0.8}{ \partial_{\alpha' }\partial _{\alpha'' }\partial _{\beta'' }  G\left(\zeta',\zeta^{\prime}_{0};\zeta'',\zeta^{\prime\prime}_{0} \right)			
			\partial _{\beta } G\left(\zeta'',\zeta^{\prime\prime}_{0};\zeta',\zeta^{\prime}_{0} \right) \Big] }\\						
		& \scalemath{0.8}{ = \lim_{\epsilon^{\prime}\rightarrow0}\int_{0}^{L_{T}} d\zeta_{0}\sum_{m=1}^{\infty}\sum_{n=1}^{\infty}\sum_{k=1}^{\infty} \frac{k } {m R^3 \sigma ^3 \left(1-q^k\right) \left(1-q^m\right) \left(1-q^n\right)} }\\ 
		&\scalemath{0.8}{ \times \left(\frac{\pi  k q^k e^{-\frac{\pi  k \epsilon '}{R}}}{R}-\frac{\pi  k e^{\frac{\pi  k \epsilon '}{R}}}{R}\right)
			\left(q^n e^{\frac{\pi  n (2 \zeta_{0}-L_{T})}{R}}+e^{-\frac{\pi  n (2 \zeta_{0}-L_{T})}{R}}\right) } \\
		&\scalemath{0.8}{ \times \left(\frac{\pi ^2 m^2 q^m e^{\frac{\pi  m \left(L_{T}-2 \zeta_{0}+\epsilon '\right)}{R}}}{R^2}+\frac{\pi ^2 m^2 e^{-\frac{\pi  m \left(L_{T}-2 t+\epsilon '\right)}{R}}}{R^2}\right) }\\
		&\scalemath{0.8}{ \times \int_{0}^{R}d\zeta_{1}\Bigg[ \sin \left(\frac{\pi  k (\zeta-R)}{R}\right) \sin \left(\frac{\pi  m \zeta}{R}\right) \sin \left(\frac{\pi  n \zeta}{R}\right) } \\
		&\scalemath{0.7}{ \times \cos \left(\frac{\pi  n (\zeta-R)}{R}\right)\sin \left(\frac{\pi  k (-R+\zeta+\epsilon )}{R}\right)  \sin \left(\frac{\pi  m (-R+\zeta+\epsilon )}{R}\right)\Bigg] } \\  
		&\scalemath{0.8}{ + \frac{k}{n R^3 \sigma ^3 \left(1-q^k\right) \left(1-q^m\right) \left(1-q^n\right)} \bigg(\frac{\pi ^2 k^2 q^k e^{-\frac{\pi  k \epsilon '}{R}}}{R^2} + \frac{\pi ^2 k^2 e^{\frac{\pi  k \epsilon '}{R}}}{R^2}\bigg) }\\   
		&\scalemath{0.7}{ \times \left(\frac{\pi  n e^{-\frac{\pi  n (2 \zeta_{0}-L_{T})}{R}}}{R}-\frac{\pi  n q^n e^{\frac{\pi  n (2 \zeta_{0}-L_{T})}{R}}}{R}\right) \left(q^m e^{\frac{\pi  m \left(L_{T}-2 \zeta_{0}+\epsilon '\right)}{R}}+e^{-\frac{\pi  m \left(L_{T}-2 t+\epsilon '\right)}{R}}\right) }\\
		&\scalemath{0.8}{ \times \int_{0}^{R}  d\zeta \Bigg[ \cos \left(\frac{\pi  k (\zeta-R)}{R}\right) \sin \left(\frac{\pi  m \zeta}{R}\right) \sin \left(\frac{\pi  n \zeta}{R}\right) }\\  
		&\scalemath{0.8}{ \times  \sin \left(\frac{\pi  n (\zeta-R)}{R}\right)\sin \left(\frac{\pi  k (-R+\zeta+\epsilon )}{R}\right) \sin \left(\frac{\pi  m (-R+\zeta+\epsilon )}{R}\right)\Bigg] }.
		\label{eq:GreenString_11}
	\end{split}
\end{equation}
It is easy to see that the integrals over $\zeta$ in the above equality are identically zero and hence
\begin{equation}
	I^{(2)}_{11}=0.
\end{equation}

%%%%%%%%%%%%%%%%%%%%%%%%%%%%%%%%%%%%%%%%%%%%%%%%
%%%%%%%%%%%%%%%%%%%%%%%%%%%%%%%%%%%%%%%%%%%%%%%%
%\newpage
\begin{itemize}
	\item \underline{\bf{Integral $I^{(2)}_{12}(R,L_{T})$}}\\
\end{itemize}

From the equation of motion \eqref{EOM}, the sum of the following integrals are trivial: 
\begin{equation}
	\begin{split}
		&I^{(2)}_{12}(R,L_{T})\\
		&\scalemath{0.7}{ =\lim_{\substack{\epsilon' \to 0\\
					\epsilon \to 0}}  \int _0^R\int _0^{L_{T}}  d\zeta d\zeta_{0} \Big[ \partial_{\alpha'} G\left(\zeta,\zeta_{0};\zeta',\zeta^{\prime}_{0}\right) \partial_{\alpha''} G\left(\zeta'',\zeta^{\prime\prime}_{0};\zeta',\zeta^{\prime}_{0}\right) \partial_{\beta'}^{2} \partial_{\alpha''}^{2} G\left(\zeta',\zeta^{\prime}_{0};\zeta'',\zeta^{\prime\prime}_{0}\right) \Big] }\\
		& =0.
	\end{split}
	\label{eq:GreenString_12}
\end{equation}

%%%%%%%%%%%%%%%%%%%%%%%%%%%%%%%%%%%%%%%%%%%%%%%%
%%%%%%%%%%%%%%%%%%%%%%%%%%%%%%%%%%%%%%%%%%%%%%%%
%\newpage
\begin{itemize}
	\item \underline{\bf{Integral $I^{(2)}_{13}(R,L_{T})$}}
\end{itemize}

The off-diagonal elements corresponding to $\alpha \neq \beta$ of the integral,
\begin{equation}
	\begin{split}
		&  I^{(2)}_{13}(R,L_{T})\\
		&\scalemath{0.7}{ =\lim_{\substack{\epsilon' \to 0\\
					\epsilon \to 0}}  \int _0^R\int _0^{L_{T}}  d\zeta d\zeta_{0} ~\Big[ \partial_{\beta } \partial_\alpha G\left(\zeta,\zeta_{0};\zeta',\zeta^{\prime}_{0}\right)  \partial_\beta \partial_{\alpha''} G\left(\zeta'',\zeta^{\prime\prime}_{0};\zeta,\zeta_{0}\right)  \partial_{\alpha''} \partial_{\alpha'} G\left(\zeta',\zeta^{\prime}_{0};\zeta'',\zeta^{\prime\prime}_{0}\right) \Big] }\\
		&\scalemath{0.9}{ = \lim_{\substack{\epsilon' \to 0\\
					\epsilon \to 0}} \int _0^{L_{T}} d\zeta_{0}~\sum_{m=1}^{\infty}\sum_{n=1}^{\infty}\sum_{k=1}^{\infty} \frac{\pi ^3 k m n(-1)^{k+n} }{2 R^6 \sigma ^3 \left(q^k-1\right) \left(q^m-1\right) \left(q^n-1\right)} } \\
		&\scalemath{0.9}{ \times \left(q^m+1\right)e^{-\frac{\pi  (k+n) (L_{T}-2 \zeta_{0})}{R}} 
			\left(q^k e^{\frac{2 \pi  k (L_{T}-2 \zeta_{0})}{R}}+1\right) \left(e^{\frac{2 \pi  n (L_{T}-2 \zeta_{0})}{R}}+q^n\right) } \\
		&\scalemath{0.9}{ \times \int _0^R d \zeta \sin ^2\left(\frac{\pi  k \zeta}{R}\right) \left[\cos \left(\frac{2 \pi  m \zeta}{R}\right)+\cos \left(\frac{2 \pi  n \zeta}{R}\right)\right] }, 
		\label{eq:GreenString_13}
	\end{split}
\end{equation}
yields a zero value for the integral $I^{(2)}_{13}=0$, which follows from the integral over $\zeta$,
\begin{equation}
	\begin{split}
		f_{1}(R)=& \scalemath{0.9}{ \int _0^R d\zeta  \Bigg[ \sin ^2\left(\frac{\pi  k \zeta}{R}\right) \left(\cos \left(\frac{2 \pi  m \zeta}{R}\right)+\cos \left(\frac{2 \pi  n \zeta}{R}\right)\right) \Bigg]} \\
		=&0.
	\end{split}
\end{equation}
The sum of the diagonal elements $\alpha=\beta$ of the integral,
\begin{equation}
	\begin{split}  
		& I^{(2)}_{13}(R,L_{T})\\
		&\scalemath{0.8}{ =\lim_{\substack{\epsilon' \to 0}}  \int _0^{L_{T}} d\zeta_{0} ~\frac{\pi ^3 k m n}{4 R^6 \sigma ^3 \left(q^k-1\right) \left(q^m-1\right) \left(q^n-1\right)}\left(e^{\frac{2 \pi  m \epsilon' }{R}}-q^m\right)} \\
		&\scalemath{0.8}{ \times \left(e^{\frac{2 \pi  n (L_{T}-2 \zeta_{0})}{R}}-q^n\right) \left(q^k e^{\frac{2 \pi  k (L_{T}-2 \zeta_{0}+\epsilon' )}{R}}+1\right) }\\     
		&\scalemath{0.8}{  \times \lim_{\substack{\epsilon \to 0}} \int_0^R  d\zeta \sin \left(\frac{\pi  k \zeta}{R}\right) \sin \left(\frac{\pi  k (-R+\zeta+\epsilon )}{R}\right) }\\
		&\scalemath{0.8}{  \times\Bigg[\cos \left(\frac{\pi  (m (2 R-2 \zeta-\epsilon )+n R)}{R}\right)-\cos \left(\frac{\pi  (m \epsilon -n R+2 n \zeta)}{R}\right) }\\
		&\scalemath{0.8}{ +\cos \left(\frac{\pi  (m \epsilon +n R-2 n \zeta)}{R}\right) -\cos \left(\frac{\pi  (m (-2 R+2 \zeta+\epsilon )+n R)}{R}\right)\Bigg] }, 
	\end{split}  
\end{equation}
gives a trivial contribution due to the vanishing of
\begin{equation}
	\begin{split}
		&\scalemath{0.9}{ f_{2}(R)= \lim_{\substack{\epsilon \to 0}} \int_0^R  \Bigg\{   \sin \left(\frac{\pi  k \zeta}{R}\right) \sin \left(\frac{\pi  k (-R+\zeta+\epsilon )}{R}\right) }\\
		&\scalemath{0.75}{ \times \Bigg[ \cos \left(\frac{\pi  (m (2 R-2 \zeta-\epsilon )+n R)}{R}\right) -\cos \left(\frac{\pi  (m \epsilon -n R+2 n \zeta)}{R}\right) }\\
		&\scalemath{0.75}{ +\cos \left(\frac{\pi  (m \epsilon +n R-2 n \zeta)}{R}\right) -\cos \left(\frac{\pi  (m (-2 R+2 \zeta+\epsilon )+n R)}{R}\right)\Bigg] \Bigg\} d\zeta }\\
		&=0.
	\end{split}
\end{equation}

%%%%%%%%%%%%%%%%%%%%%%%%%%%%%%%%%%%%%%%%%%%%%%%%
%%%%%%%%%%%%%%%%%%%%%%%%%%%%%%%%%%%%%%%%%%%%%%%%
\newpage
\begin{itemize}
	\item \underline{\bf{Integral $I^{(2)}_{14}(R,L_{T})$}}
\end{itemize}

\begin{equation}
	\begin{split}
		& I^{(2)}_{14}(R,L_{T})\\
		&\scalemath{0.7}{  =\lim_{\substack{\epsilon' \to 0\\
					\epsilon \to 0}} \int _0^R\int _0^{L_{T}}  d\zeta d\zeta_{0} \Big[\partial _{\alpha } G\left(\zeta,\zeta_{0};\zeta',\zeta^{\prime}_{0} \right) \partial _{\alpha ''}\partial _{\beta'' }G\left(\zeta',\zeta^{\prime}_{0};\zeta'',\zeta^{\prime\prime}_{0} \right)\partial_{\beta''} \partial_{\alpha}^{2}  G\left(\zeta'',\zeta^{\prime\prime}_{0};\zeta',\zeta^{\prime}_{0} \right) \Big] } \\	
		&\scalemath{0.7}{ =\lim_{\substack{\epsilon' \to 0\\
					\epsilon \to 0}}\int _0^{L_{T}} d\zeta_{0} \sum _{n=1}^{\infty } \sum _{m=1}^{\infty }\sum _{k=1}^{\infty }  \frac{\pi  k}{\pi  n R^2 \sigma ^3 \left(1-q^k\right) \left(1-q^m\right) \left(1-q^n\right)} }\\
		&\scalemath{0.7}{ \times \left(\frac{\pi  k q^k e^{-\frac{\pi  k \epsilon '}{R}}}{R}-\frac{\pi  k e^{\frac{\pi  k \epsilon '}{R}}}{R}\right) \left(q^n e^{\frac{\pi  n (2 \zeta_{0}-L_{T})}{R}}+e^{-\frac{\pi  n (2 \zeta_{0}-L_{T})}{R}}\right) } \\
		&\scalemath{0.7}{ \times \left(\frac{\pi  m q^m e^{\frac{\pi  m \left(L_{T}-2 \zeta_{0}+\epsilon '\right)}{R}}}{R}-\frac{\pi  m e^{-\frac{\pi  m \left(L_{T}-2 t+\epsilon '\right)}{R}}}{R}\right) }\\
		&\scalemath{0.7}{ \times \int _0^R \Bigg[ \sin \left(\frac{\pi  m \zeta}{R}\right)\sin \left(\frac{\pi  n \zeta}{R}\right) \sin \left(\frac{\pi  k (\zeta-R)}{R}\right) }\\
		&\scalemath{0.7}{ \times \sin \left(\frac{\pi  k (-R+\zeta+\epsilon )}{R}\right) \cos \left(\frac{\pi  n (\zeta-R)}{R}\right) \cos \left(\frac{\pi  m (-R+\zeta+\epsilon )}{R}\right)\Bigg] d\zeta }\\
		&\scalemath{0.7}{ +\frac{1}{\pi  n R^2 \sigma ^3 \left(1-q^k\right) \left(1-q^m\right) \left(1-q^n\right)} \left(\frac{\pi ^2 k^2 q^k e^{-\frac{\pi  k \epsilon '}{R}}}{R^2}+\frac{\pi ^2 k^2 e^{\frac{\pi  k \epsilon '}{R}}}{R^2}\right) } \\
		&\scalemath{0.7}{ \times \left(\frac{\pi  n e^{-\frac{\pi  n (2 \zeta_{0}-L_{T})}{R}}}{R}-\frac{\pi  n q^n e^{\frac{\pi  n (2 \zeta_{0}-L_{T})}{R}}}{R}\right) }\\
		&\scalemath{0.7}{ \times \left(\frac{\pi  m q^m e^{\frac{\pi  m \left(L_{T}-2 t+\epsilon '\right)}{R}}}{R}-\frac{\pi  m e^{-\frac{\pi  m \left(L_{T}-2 t+\epsilon '\right)}{R}}}{R}\right) }\\ 
		&\scalemath{0.7}{ \times \int_{0}^{R}\Bigg[ \cos \left(\frac{\pi  k (\zeta-R)}{R}\right) \sin \left(\frac{\pi  m \zeta}{R}\right) \sin \left(\frac{\pi  n \zeta}{R}\right) }\\
		&\scalemath{0.7}{ \times  \sin \left(\frac{\pi  n (\zeta-R)}{R}\right) \sin \left(\frac{\pi  k (-R+\zeta+\epsilon )}{R}\right) \cos \left(\frac{\pi  m (-R+\zeta+\epsilon )}{R}\right)\Bigg] d\zeta }.
		\label{eq:GreenString_14}
	\end{split}
\end{equation}

The integrals over $\zeta$ are
\begin{equation}
	\begin{split}
		f_{1}(R)&=\scalemath{0.7}{ \int_0^R \bigg[\sin \left(\frac{\pi  k (\zeta-R)}{R}\right) \sin \left(\frac{\pi  m \zeta}{R}\right) \sin \left(\frac{\pi  n \zeta}{R}\right) \cos \left(\frac{\pi  n (\zeta-R)}{R}\right) }\\
		&\scalemath{0.7}{ \times \sin \left(\frac{\pi  k (-R+\zeta+\epsilon )}{R}\right) \cos \left(\frac{\pi  m (-R+\zeta+\epsilon )}{R}\right)\bigg] \, d\zeta }\\
		&=0 
	\end{split} 
\end{equation}
and
\begin{equation}
	\begin{split}
		f_{2}(R)&\scalemath{0.7}{ = \int_0^R \bigg[\cos \left(\frac{\pi  k (\zeta-R)}{R}\right) \sin \left(\frac{\pi  m \zeta}{R}\right) \sin \left(\frac{\pi  n \zeta}{R}\right) \sin \left(\frac{\pi  n (\zeta-R)}{R}\right) }\\
		&\scalemath{0.7}{ \times \sin \left(\frac{\pi  k (-R+\zeta+\epsilon )}{R}\right) \cos \left(\frac{\pi  m (-R+\zeta+\epsilon )}{R}\right)\bigg] \, d\zeta }\\
		&\scalemath{0.7}{ =-\frac{1}{16} R \sec (\pi  (m+n)) \left[\cos \left(\frac{\pi  (k-m) \epsilon }{R}\right)+\cos \left(\frac{\pi  (k+m) \epsilon }{R}\right) \right] }.
	\end{split} 
\end{equation}

In the limit $\epsilon \to 0$, the sums over the running $n,m$ and $k$ diverge, 
\begin{equation}
	\begin{split}
		&I^{(2)}_{14}(R,L_{T})\\
		&\scalemath{0.7}{ =\frac{\pi^2}{32 R^4 \sigma^3}\lim_{\substack{\epsilon \to 0}} \sum_{n,m,k=1}^{\infty} \sec(\pi(m+n)) \left[\cos \left(\frac{\pi  (k-m) \epsilon }{R}\right)-\cos \left(\frac{\pi  (k+m) \epsilon }{R}\right) \right] }\\
		&\scalemath{0.7}{ = \frac{\pi^2}{32 R^4 \sigma^3} \lim_{\substack{\epsilon \to 0}} \sum_{n,m,k}(0) }\\
		&\scalemath{0.7}{ = \frac{\pi^2}{32 R^4 \sigma^3}  \sum_{n,m} (-1)^{m+n} }\\
		&\scalemath{0.7}{ = \frac{\pi^2}{32  R^4 \sigma^3} \times \frac{1}{4} }.  
	\end{split} 
\end{equation}

The divergent sums $\sum_{k} (0)$ and $\sum_{n}(-1)^{n}$ are evaluated as the $\zeta$ function and Grandi's series, respectively. 

%%%%%%%%%%%%%%%%%%%%%%%%%%%%%%%%%%%%%%%%%%%%%%%%
%%%%%%%%%%%%%%%%%%%%%%%%%%%%%%%%%%%%%%%%%%%%%%%%
%\newpage
\begin{itemize}
	\item \underline{\bf{Integral $I^{(2)}_{15}(R,L_{T})$}}\\
\end{itemize}

For $\alpha \neq \beta$ the sum of integral   
\begin{equation}
	\begin{split}
		&I^{(2)}_{15}(R,L_{T})\\
		&\scalemath{0.7}{ = \lim_{\substack{\epsilon' \to 0\\
					\epsilon \to 0}} \int _0^R\int _0^{L_{T}}  d\zeta d\zeta_{0}  ~\Big[G\left(\zeta,\zeta_{0};\zeta',\zeta^{\prime}_{0} \right) \partial _{\alpha' }\partial _{\beta'' } G\left(\zeta',\zeta^{\prime}_{0};\zeta'',\zeta^{\prime\prime}_{0} \right) \partial_{\alpha'' } \partial _{\beta''} \partial_{\alpha}^{2} G\left(\zeta'',\zeta^{\prime\prime}_{0};\zeta',\zeta^{\prime}_{0} \right)\Big] } \\
		&\scalemath{0.7}{ =\frac{(-1)^{m+n}\pi ^3 k^3 m }{n R^6 \sigma ^3 \left(q^k-1\right) \left(q^m-1\right) \left(q^n-1\right)} \left(e^{\frac{2 \pi  n (L_{T}-2 \zeta_{0})}{R}}+q^n\right) \left(q^m e^{\frac{2 \pi  m (L_{T}-2 \zeta_{0}+\epsilon' )}{R}}-1\right) }\\
		&\scalemath{0.7}{ \times \exp \left(-\frac{\pi  (\epsilon'  (k+m)+(L_{T}-2 \zeta_{0}) (m+n))}{R}\right) } \\
		&\scalemath{0.7}{ \times \left(e^{\frac{2 \pi  n (L_{T}-2 \zeta_{0})}{R}}+q^n\right) \left(q^m e^{\frac{2 \pi  m (L_{T}-2 \zeta_{0}+\epsilon' )}{R}}-1\right) }\\
		&\scalemath{0.7}{ \times \cos \left(\frac{\pi  k \zeta}{R}\right) \sin ^2\left(\frac{\pi  n \zeta}{R}\right) \sin \left(\frac{\pi  k (\zeta+\epsilon )}{R}\right) \sin \left(\frac{\pi  m (2 \zeta+\epsilon )}{R}\right)} ,
	\end{split}
\end{equation}
is trivial since the integration with respect to $\zeta$ yields a vanishing value,
\begin{equation}
	\begin{split}
		f(R)& \scalemath{0.7}{ = \int_0^R \cos \left(\frac{\pi  k \zeta}{R}\right) \sin ^2\left(\frac{\pi  n \zeta}{R}\right) \sin \left(\frac{\pi  k (\zeta+\epsilon )}{R}\right) \sin \left(\frac{\pi  m (2 \zeta+\epsilon )}{R}\right) \, d\zeta }\\
		&=0.
	\end{split}
\end{equation}
It is easy to show that the same result is valid for the diagonal components $\alpha=\beta$.

%%%%%%%%%%%%%%%%%%%%%%%%%%%%%%%%%%%%%%%%%%%%%%%%
%%%%%%%%%%%%%%%%%%%%%%%%%%%%%%%%%%%%%%%%%%%%%%%%
\newpage
\begin{itemize}
	\item \underline{\bf{Integral $I^{(2)}_{16}(R,L_{T})$}}\\
\end{itemize}

The diagonal components corresponding to $\alpha=\beta$ are given by
\begin{equation}
	\begin{split}
		&I^{(2)}_{16}(R,L_{T})\\
		&\scalemath{0.7}{  =\lim_{\substack{\epsilon' \to 0\\
					\epsilon \to 0}}  \int _0^R\int _0^{L_{T}}  d\zeta d\zeta_{0} \Big[ \partial_{\alpha'} G\left(\zeta,\zeta_{0};\zeta',\zeta^{\prime}_{0}\right) \partial_{\alpha''} G\left(\zeta'',\zeta^{\prime\prime}_{0};\zeta',\zeta^{\prime}_{0}\right) \partial_{\beta'}^{2} \partial_{\alpha''}^{2} G\left(\zeta',\zeta^{\prime}_{0};\zeta'',\zeta^{\prime\prime}_{0}\right)\Big] }\\
		&\scalemath{0.7}{ =\int_{0}^{L_{T}} d\zeta_{0}\lim_{\substack{\epsilon' \to 0}}\sum _{k=1}^{\infty }\sum _{m=1}^{\infty }\sum _{n=1}^{\infty }-\frac{\pi^{3}k^{3}}{R^{6}\sigma^{3}(1-q^{k})}\frac{1}{(1-q^{m})}\frac{1}{(1-q^{n})} }\\
		&~~\scalemath{0.7}{ \times
			\left(e^{\frac{\pi k \epsilon^{\prime}}{R}} + q^{k}e^{-\frac{\pi k \epsilon^{\prime}}{R}} \right)
			\left(q^{n}e^{\frac{\pi n (2\zeta_{0}-L_{T}) }{R}} + e^{-\frac{\pi n (2\zeta_{0}-L_{T}) }{R}}   \right)
			\left(e^{-\frac{\pi m \epsilon^{\prime}}{R}} + q^{m}e^{\frac{\pi m \epsilon^{\prime}}{R}} \right) }\\
		&~~\scalemath{0.7}{  \times\int_{0}^{R}{\rm{sin}}\left(\frac{\pi k (\zeta- R)}{R}\right)
			{\rm{sin}}\left(\frac{\pi m (\zeta-R)}{R}\right)
			{\rm{sin}}\left(\frac{\pi n \zeta}{R}\right) }\\
		&~~\scalemath{0.7}{ \times
			{\rm{cos}}\left(\frac{\pi n (\zeta - R)}{R}\right)
			{\rm{cos}}\left(\frac{\pi k (\zeta - R + \epsilon)}{R}\right)
			{\rm{sin}}\left(\frac{\pi m (\zeta - R + \epsilon)}{R}\right)d\zeta }\\
		&~~\scalemath{0.7}{ +\int_{0}^{L_{T}} d\zeta_{0}\lim_{\substack{\epsilon' \to 0}}\sum _{k=1}^{\infty }\sum _{m=1}^{\infty }\sum _{n=1}^{\infty } \frac{\pi^{3}k^{3}}{R^{6}\sigma^{3}(1-q^{k})}\frac{1}{(1-q^{m})}\frac{1}{(1-q^{n})} }\\
		&~~\scalemath{0.7}{ \times
			\left(e^{\frac{\pi k \epsilon^{\prime}}{R}} + q^{k}e^{-\frac{\pi k \epsilon^{\prime}}{R}} \right)
			\left(e^{-\frac{\pi n (2\zeta_{0}-L_{T}) }{R}}  - q^{n}e^{\frac{\pi n (2\zeta_{0}-L_{T}) }{R}} \right)
			\left(q^{m}e^{\frac{\pi m \epsilon^{\prime}}{R}} - e^{-\frac{\pi m \epsilon^{\prime}}{R}}  \right) }\\
		&~~\scalemath{0.7}{ \times\int_{0}^{R}{\rm{sin}}\left(\frac{\pi k (\zeta- R)}{R}\right)
			{\rm{sin}}\left(\frac{\pi m (\zeta-R)}{R}\right)
			{\rm{sin}}\left(\frac{\pi n \zeta}{R}\right) }\\
		&~~\scalemath{0.7}{ \times
			{\rm{sin}}\left(\frac{\pi n (\zeta - R)}{R}\right)
			{\rm{sin}}\left(\frac{\pi k (\zeta - R + \epsilon)}{R}\right)
			{\rm{sin}}\left(\frac{\pi m (\zeta - R + \epsilon)}{R}\right)d\zeta }\\
		&\scalemath{0.7}{ =\frac{\pi ^3}{8 R^5 \sigma ^3}\sum _{k=1}^{\infty } \frac{k^3 \left(q^k+1\right)}{q^k-1}\sum _{m=1}^{\infty } \frac{1}{q^m-1}\sum _{n=1}^{\infty } \frac{1}{q^n-1} }\\
		&~~\scalemath{0.7}{ \times \int _0^{L_{T}}\left(e^{\frac{\pi  m (-2 \zeta_{0})}{R}}-q^m e^{-\frac{\pi  m (-2 \zeta_{0})}{R}}\right) \left(q^{-n} e^{-\frac{\pi  n (2 \zeta_{0})}{R}}-q^{2 n} e^{\frac{\pi  n (2 \zeta_{0})}{R}}\right)d\zeta_{0} },
		\label{eq:GreenString_16}
	\end{split}
\end{equation}
and turn out to be nontrivial. Comparison of Eq.~\eqref{eq:GreenString_16} with Eq.~\eqref{eq:GreenString_07} shows a similar form; that is,

\begin{equation}
	I^{(2)}_{16}=I^{(2)}_{7}.
\end{equation}

The sum corresponding to the off-diagonal components $\alpha \neq \beta$ is given by
\begin{equation}
	\begin{split}
		&I^{(2)}_{16}(R,L_{T})\\
		&\scalemath{0.7}{ =\lim_{\substack{\epsilon' \to 0\\ \epsilon \to 0}} \int _0^R\int _0^{L_{T}}  d\zeta d\zeta_{0} \Big[\partial_{\beta} G\left(\zeta,\zeta_{0};\zeta',\zeta^{\prime}_{0}\right) \partial_{\alpha''} G\left(\zeta'',\zeta^{\prime\prime}_{0};\zeta',\zeta^{\prime}_{0}\right) \partial_{\alpha''}^{2} \partial_{\beta'} \partial_{\alpha'} G\left(\zeta',\zeta^{\prime}_{0};\zeta'',\zeta^{\prime\prime}_{0}\right)\Big] }.
	\end{split}
\end{equation}
with two vanishing spatial integrations 
\begin{equation}
	\begin{split}
		f_1(R)&\scalemath{0.8}{ =\underset{\epsilon \to 0}{\text{lim}}\int_0^R d\zeta \cos \left(\frac{\pi  k (\zeta-R)}{R}\right) \sin \left(\frac{\pi  m \zeta}{R}\right) \sin \left(\frac{\pi  n \zeta}{R}\right)} \\
		&\scalemath{0.8}{ \times \sin \left(\frac{\pi  n (\zeta-R)}{R}\right) \sin \left(\frac{\pi  k (-R+\zeta+\epsilon )}{R}\right) \cos \left(\frac{\pi  m (-R+\zeta+\epsilon )}{R}\right) } \\
		&=0,\\
		f_2(R)&\scalemath{0.8}{ =\underset{\epsilon \to 0}{\text{lim}}\int_0^R d\zeta \cos \left(\frac{\pi  k (\zeta-R)}{R}\right) \sin \left(\frac{\pi  m \zeta}{R}\right) \sin \left(\frac{\pi  n (\zeta-R)}{R}\right)} \\
		&\scalemath{0.8}{ \times \cos \left(\frac{\pi  n \zeta}{R}\right) \sin \left(\frac{\pi  k (-R+\zeta+\epsilon )}{R}\right) \sin \left(\frac{\pi  m (-R+\zeta+\epsilon )}{R}\right) }\\
		&=0.
	\end{split}
\end{equation}

%---------------------------------------------
%\section{references}
%---------------------------------------------
%Produces the bibliography via BibTeX.

%\end{appendix}

%\bibliography{DarkMatter_bibliography.bib}

% BibTeX users please use
% \bibliographystyle{}
% \bibliography{}
%
% Non-BibTeX users please use

\end{document}